\documentclass{jfm}
\usepackage{graphicx}
\usepackage{pgf}
\usepackage{subcaption}
\usepackage{caption}
\usepackage{epstopdf, epsfig}
\usepackage[export]{adjustbox}
\usepackage{multirow}

\usepackage{amsmath, bm, amssymb}

\usepackage[utf8]{inputenc}
\usepackage{pgfplots}
\DeclareUnicodeCharacter{2212}{−}
\usepgfplotslibrary{groupplots,dateplot}
\usetikzlibrary{patterns,shapes.arrows}
\pgfplotsset{compat=newest}

\graphicspath{{Figures/}}

\shorttitle{Optimal surface actuation for aerodynamic flows}
\shortauthor{Ernold Thompson, Andres Goza }

\title{Adjoint-based optimal actuation for separated flow past an airfoil }

\author{Ernold Thompson\aff{1}
  \corresp{\email{ernoldt2@illinois.edu}},
  Andres Goza\aff{1}
}

\affiliation{\aff{1}Department of Aerospace Engineering, University of Illinois at Urbana-Champaign, Urbana, Illinois, 61801, USA}

\begin{document}

\maketitle

\begin{abstract}
This study computes the optimal normal actuation 
on the surface of a NACA0012 airfoil at an angle of attack
of $\alpha = 15^{\circ}$ and a Reynolds number of $\Rey = 1000$, using costs defined for minimal drag and maximal lift. 
To allow for a general actuation profile, non-zero actuation is permissible on both the suction and pressure surfaces. This approach of optimal actuation along the full airfoil surface augments 
most other studies that have focused on parametrically varied open-loop control restricted to the suction surface.
The 
gradient-based optimization procedure requires the gradient of the cost functional with respect to the design variables,
which is
determined using the adjoint of the governing equations.
The optimal actuation profile for the two performance aims are compared. 
Where possible, similarities with commonly considered open-loop actuation in the form of backward traveling waves on the suction surface have been highlighted.
In addition, the key spatial locations on the airfoil surface
for the two optimal control strategies have been compared to earlier works 
where actuation has been limited to a sub-domain of the airfoil surface.
The flow features emerging from the optimal actuation variations, and their consequent influence on the instantaneous 
aerodynamic coefficients, 
have been analyzed.
To complement our findings with normal actuation, 
we also provide in the appendix: results for a more general form of actuation with independent $x$ and $y$ components, and for a different optimization window to assess the effect of this window parameter on the actuation profile and the flow features.

\end{abstract}

\begin{keywords}
\end{keywords}

\section{Introduction}

Active control of unsteady aerodynamic flows is of interest for obtaining performance benefits by modifying the surrounding flow field. 
Many flow control efforts have focused on mitigating stall.
For chord-based Reynolds numbers relevant to large-scale aerial craft, 
$\Rey = O(10^5)$--$O(10^6)$,
the thinner separated region in the presence of actuation leads to a higher suction peak (hence higher lift) as well as lower pressure drag due to reattachment of the separated flow
\citep{seifert1996delay, amitay2002role, benard2009lift, chapin2015active, Corke2002application, Corke2007, Corke2010, Cutler2005, Glezer2003, Crittenden2009}.
However, the optimal properties of actuation that provide these desired flow characteristics are not always known a-priori.
In the case of localized actuation technologies such as synthetic jets, 
parametric studies to answer questions regarding the optimal location of the actuator, its angle relative to the airfoil surface, the use of steady versus unsteady actuation, and the role of actuation magnitude, among others, have spanned several experimental and numerical investigations. 
In \cite{seifert1996delay}, an experimental comparison between steady and periodic blowing (with a steady component) at various locations on the suction surface suggested that periodic actuation is more efficient in terms of the magnitude of actuation required to achieve performance benefits.
While the non-dimensionalized actuation frequency in the above work was $O(1)$, in \cite{amitay2002role} actuation at a higher frequency of $O(10)$ was studied and found to be more effective at reattaching the separated shear layer than actuation at a frequency of $O(1)$.
The effectiveness of high-frequency actuation was also reported in \cite{chang1992forcing} and \cite{hsiao1990control}.
However, in \cite{raju2008dynamics}, who studied the role of actuation frequency and location at a lower Reynolds number of $\Rey=40$ $000$, actuation was most effective when the actuation frequency was around the natural separation bubble frequency as defined in \cite{kotapati2007numerical}.
Regarding the influence of actuation location, actuation slightly upstream of the locations of separation has generally been found to be effective \citep{amitay2001aerodynamic, raju2008dynamics}, though performance benefits from actuation at other locations has been found to be comparable in certain cases. 
For example, \cite{amitay2001aerodynamic} found actuation on the pressure side near the leading edge to yield lift improvements at an angle of attack of $\alpha = 0^{\circ}$.
In \cite{raju2008dynamics}, for one of the actuation frequencies considered, actuation on the airfoil surface at a location that coincides with the separation bubble was found to yield similar performance as actuation upstream of the separation bubble. 
The multivariate problem of choosing optimal parameters for localized actuation has also motivated the use of mathematical optimization \citep{duvigneau2006optimization, duvigneau2006simulation}.

Research into aerodynamic flow control has also been driven by new actuation opportunities made possible through advances in materials science. For example, actuation strategies involving the entire suction surface are now possible \citep{jones2018controlexp}.
In \citet{jones2018control}, the suction surface of an airfoil at $\Rey=50$ $000$ was actuated in the form of a standing wave. 
Among the frequencies considered, a non-dimensional frequency (non-dimensionalized by freestream velocity and chord length) of two was found to be most effective in reattaching the separated boundary layer. 
In a companion numerical study at the same Reynolds number \citep{jones2018control}, performance-beneficial actuation parameters gave rise to vortical structures. 
Among the frequencies considered, a non-dimensional frequency of two was found to be most beneficial (consistent with the experimental study). Actuation extending over the entire suction surface has also been studied in the form of traveling waves \citep{akbarzadeh2020controlling, akbarzadeh2019numerical, akbarzadeh2021role}.
In \cite{akbarzadeh2020controlling}, a comparison between forward traveling, backward traveling, and standing waves at $\Rey=50$ $000$ showed that backward traveling waves outperformed the other configurations.
Further, stall suppression was possible with some wave parameters.
A broader parametric exploration in \cite{akbarzadeh2021role} showed a non-monotonic dependence of performance on both frequency and amplitude.
In a related computational effort motivated by fish surface and body undulations, \cite{shukla2022hydrodynamics} assessed the effect of surface actuation on the thrust generated by an airfoil at $\alpha = 0^{\circ}$, $\Rey = 500$--$5000$.
The thrust-producing wake seen in the case of performance-beneficial surface undulations was linked to the vorticity generated by the backward moving undulations.
This observation is similar to the findings of \cite{akbarzadeh2019reducing}, where backward traveling waves in the form of surface undulations were considered for flow past an inclined plate at $\Rey=20$ $000$.
The performance of actuation was linked to the vortices generated and carried along the troughs of the surface undulations. 
\citet{thompson2022surface} considered normal actuation on the suction surface of a NACA0012 airfoil at $\Rey=1000$, which is relevant to insect flight \citep{shyy2008aerodynamics}.
The actuation was prescribed as a velocity boundary condition with no displacement of the airfoil surface. 
For the actuation parameters considered, the changes in the flow field were closely related to the actuation near the airfoil location of the maximum $y$-coordinate value. 
The wave parameters leading to maximum lift benefits had time scales (defined in terms of wavelength and wavespeed of the traveling wave) related to the advection time scale of the unactuated flow.
Further, the two time scales were of the same order of magnitude.
Since lift improvements resulted from an increased curvature of the streamlines near the location of the maximum $y$-coordinate and the concomitant increased pressure magnitude, lift-beneficial actuation was accompanied by a penalty in drag. 

The above studies demonstrate the potential of actuation to yield performance improvements across a wide range of Reynolds numbers, angles of attack and actuation configurations.
However, determining the optimal time scales and functional forms of actuation is not straightforward. In most cases, actuation is assumed to be periodic. Thus, the benefits and detriments of the positive and negative phases of actuation are tightly coupled.
Even when the actuation is periodic, the performance-beneficial actuation frequencies could be separated by orders of magnitude between which non-monotonic behavior of performance might be observed.
Further, the mechanisms by which actuation at the different frequency ranges lead to performance benefits can be non-intuitive. These mechanisms are different depending on whether one seeks to reduce drag or increase lift.
Finally, while a small number of studies considered surface actuation 
affecting the flow on both the pressure and suction sides of the body
\citep{akbarzadeh2019reducing,shukla2022hydrodynamics},
the majority of investigations have focused only on suction-surface actuation. 
The effect of actuation along the entire surface therefore remains incompletely understood.

Motivated by these challenges, in the current work we employ optimization to explore the actuation properties that lead to lift and drag benefits for the flow past a NACA0012 airfoil at a stalled angle of attack of $\alpha = 15^{\circ}$ and $\Rey=1000$.
The optimal actuation profile is determined as the solution of an optimization procedure where a cost functional (based on either minimizing time-integrated quantities involving drag or lift)  is minimized. 
The design variable in the iterative optimization problem is the time-varying normal actuation along the airfoil surface (including the pressure side) over a time window spanning about four vortex-shedding cycles of the unactuated flow.
Since lift-optimal actuation can lead to drag detriments \citep{thompson2022surface}, we optimize for lift and drag benefits separately. The optimal actuation when simultaneously optimizing for lift and drag benefits, for the same flow parameters as in our work, has been determined by \citet{paris2023reinforcement} using reinforcement-learning. The authors applied actuation on the suction surface near the point of maximum $y$-coordinate and utilized larger actuation magnitudes 
(non-dimensional surface velocities of roughly $15\%$) 
than we consider here. 
These larger actuation magnitudes are relevant for a number of actuation paradigms such as synthetic jets, and the joint consideration of lift and drag in the optimization is important for understanding efficient flight configurations. \citet{paris2023reinforcement} observed that the optimal actuation yields flow reattachment, and found the actuation profiles that produced this new flow state. We consider surface-distributed actuation and, motivated by surface actuation associated with small amplitudes, employ penalization to limit actuation magnitudes to approximately 
$3\%$ of the freestream velocity
(the same as the actuation amplitude in the open-loop control effort of \citep{thompson2022surface}). Also, distinctly optimizing for lift and drag, respectively, can help clarify relevant actuation strategies and associated physical mechanisms for cases where joint optimization of both criteria is not the key aim. For example, in insect flight the existence of the leading-edge vortex is known to have lift benefits \citep{dickinson1993unsteady} and the lift-optimal actuation studied here could inform MAV designs employing flapping as a means for lift generation.
On the other hand, the result of the drag-optimal actuation could have utility in vortex control during turning of aerodynamic vehicles.

The iterative optimization procedure makes use of gradients of the cost functional with respect to the design variable, which is efficiently computed using the adjoint of the governing equations \citep{flinois2015optimal, protas2002optimal, naderi2016numerical, jameson2003aerodynamic}.
The gradient of the cost functional is used to compute optimal actuation within an (iterative) nonlinear conjugate gradient algorithm. 
We discuss the interplay between the actuation profile on the suction and pressure sides of the airfoil and their respective roles in modifying the flow field for the lift- and drag-based costs. 
Physical insights are drawn by probing the adjoint field of the unactuated flow and by assessing changes to key vortex-shedding features relative to the unactuated case.
Based on our findings, we subsequently investigate the influence of additional optimization parameters and actuation forms. 
\section{Problem Formulation and Computational Methodology}
\label{sec:NM}
The airfoil considered is NACA0012 at an angle of attack of $\alpha = 15^{\circ}$, with the flow Reynolds number taken to be $\Rey = 1000$. Actuation is considered on both the suction as well as pressure surfaces of the airfoil and is taken to be along the local normal. Figure \ref{fig:Schematic} shows a schematic of the normal actuation on the airfoil surface at one instance of time (left figure; note that this actuation is generally not periodic despite that commonly assumed functional form). The normal actuation at each discretized body point is allowed to vary as a function of time as shown in figure \ref{fig:Schematic}, which shows an example of time varying actuation at one body point (right figure). The actuation on the airfoil surface is modeled as a velocity boundary condition, i.e., the surface is assumed to have zero displacement.  
While this representation is not equivalent to the surface morphing actuation that partly motivates this work, we note that for small surface deflections the leading order effect on the flow is due to the non-zero velocity condition and not the moving location at which that condition is imposed. Moreover, the results from this formulation, which is computationally simpler to implement, are also relevant to a variety of actuation strategies that apply momentum to the flow near the body's surface.

 The goal of the current work is to compute the actuation profile that yields optimal improvements in lift and drag. 
 Since actuation for lift improvement might not result in drag benefits, we consider actuation for the two performance imperatives of lift improvement and drag reduction separately. 
 This distinct exploration could also yield insights into the role of actuation for these   different aims.  
 For the rest of the paper, the discussion of aerodynamic performance is in terms of the coefficients of pressure, lift and drag defined respectively as: 
 \begin{equation}
    C_p=\frac{p-p_\infty}{\frac{1}{2}\rho U_\infty^2}, \quad C_l = \frac{F_y}{\frac{1}{2}\rho U_\infty^2 c}, \quad C_d = \frac{F_x}{\frac{1}{2}\rho U_\infty^2 c},
\end{equation}

where $\rho$ is the fluid density, $U_{\infty}$ is the freestream flow speed, $c$ is the chord length of the airfoil, $p$ is pressure, $p_{\infty}$ is the pressure in the freestream, and $F_x$ and $F_y$ are respectively the integrated force on the airfoil surface in the $x$ and $y$ directions. The actuation profile is determined by solving an optimization problem, where a cost functional based on lift or drag is minimized by varying the actuation. Since lift and drag are studied separately, two distinct cost functionals are considered:
\begin{figure}
    \centering
    \includegraphics[width=0.95\textwidth]{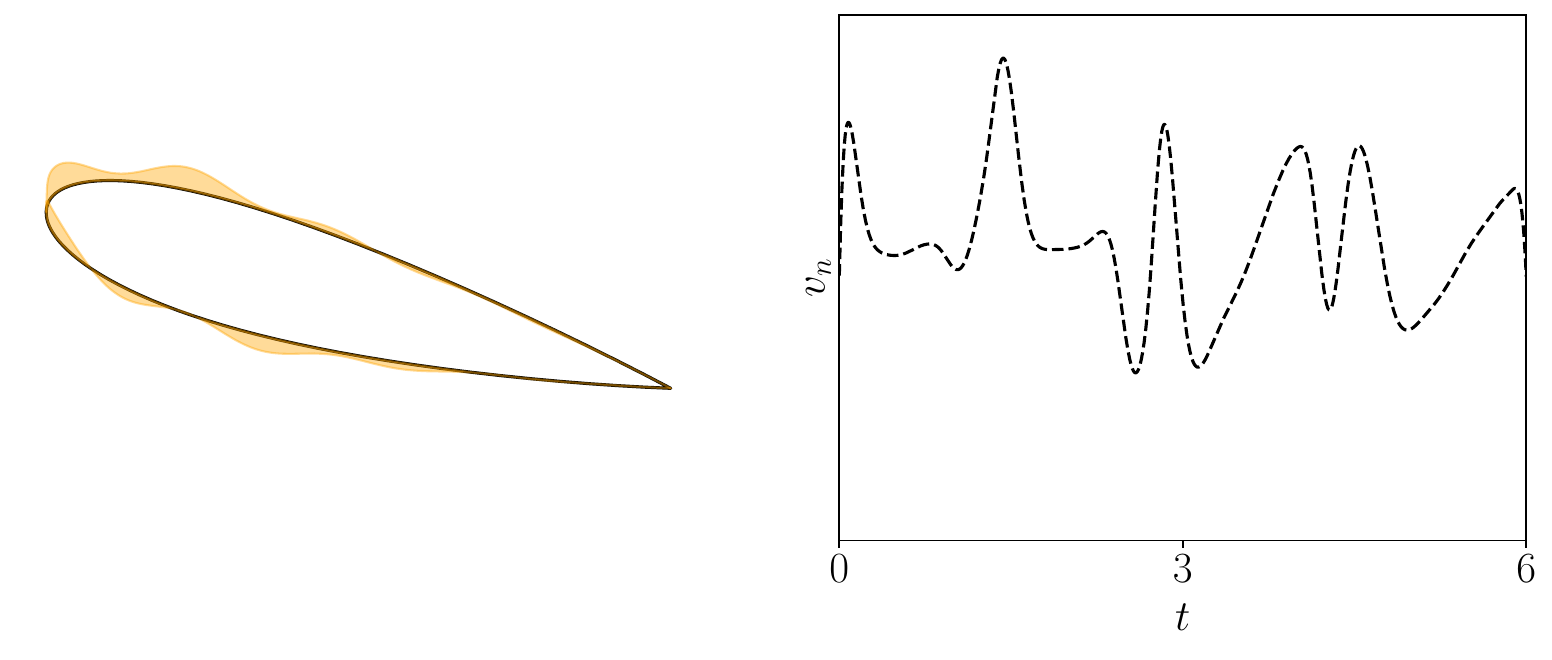}
    \caption{Schematic of airfoil with normal actuation. Left: actuation on the airfoil surface at one instance of time, right: time varying actuation at one point on the body.}
    \label{fig:Schematic}
\end{figure}
%
     \begin{align}
        \begin{split}
             \mathcal{J}_{C_{d}}(u_{act}) &= \frac{1}{2}\int_0^{T} \left[ (mC_d- {C_d}_{target})^2 + \phi \left(\int_\Gamma u_{act}^2(\bm{\xi}, t) d \bm{\xi} \right) \right] dt, \\
             \mathcal{J}_{C_{l}}(u_{act}) &= \frac{1}{2}\int_0^{T} \left[ ({C_l}_{target} - mC_l)^2 + \phi \left(\int_\Gamma u_{act}^2(\bm{\xi}, t) d \bm{\xi} \right) \right] dt ,  
        \end{split}
        \label{eq:cost}
    \end{align}

Each cost functional comprises of a contribution from the aerodynamic coefficients (which in turn depend on the actuation), a target value of the aerodynamic coefficient and a penalty term to limit the magnitude of actuation. Formulated in this way, both costs constitute a minimization problem. 
In the above equations, $\bm{\xi}$ represents the co-ordinate along the airfoil surface, $\Gamma$ is the spatial domain coinciding with the airfoil, $\phi$ is the penalty weight and $T$ is the optimization window. It was found that the inclusion of the target values led to smoother convergence of the optimization procedure. In the current work, the values of  ${C_l}_{target}$ and  ${C_d}_{target}$ were taken to be $10$ and $-5$ respectively. 
These target values were chosen to ensure that the quantities 
$({C_l}_{target} - mC_l)$ 
in the case of $\mathcal{J}_{C_l}$, and 
$(mC_d- {C_d}_{target})$ 
in the case of $\mathcal{J}_{C_d}$, are positive even in the presence of large transients at the start and the end of the optimization window. The weighting, $m$, is equal to $\Delta t / 2\Delta x$, with grid size, $\Delta x = 0.003571$ and time step, $\Delta t = 0.0004375$. These grid parameters were determined in the sub-optimal actuation study of \cite{thompson2022surface}, which contains details about the convergence properties of the simulations. 
The same grid parameters are used here as the optimal-actuation profile is demonstrated to have temporal and spatial scales within 
the parametric space considered there.
The penalty weight $\phi$ was chosen to ensure that the maximum magnitude of instantaneous actuation on the airfoil surface was about $0.03$. 
Since the $C_l$ and $C_d$ variations differ in magnitude, the penalty weights required to ensure acceptable magnitudes of actuation are different.
The cost functionals in equation (\ref{eq:cost}) also necessitate the definition of a time window, $T$. In our work, we fix this parameter to be $T=6$ which is about 4.25 vortex-shedding cycles of the unactuated flow. 
The length of the optimization window is limited by the accuracy of the adjoint gradients; the accuracy of the adjoint-based gradients deteriorates with the length of the optimization window \citep{flinois2015optimal}.

The simulations are performed with the immersed boundary method of \citet{colonius2008fast}, which solves the incompressible Navier-Stokes equations and models the influence of the body on the flow field as a forcing to the momentum equations. This method has been successfully utilized on a number of incompressible flow problems, including on a prior investigation of sub-optimal actuation on the same aerodynamic configuration of interest here \citep{thompson2022surface}. 

The optimal actuation profile is obtained using the nonlinear conjugate gradient method, which iteratively updates guesses for the space- and time-varying normal surface velocity using gradient information of the cost functional with respect to the surface actuation. The required gradients of the cost functionals are computed efficiently using the adjoint of the governing equations. The update for the actuation profile is determined in terms of the gradient at the current and prior iterations via the Polak-Ribiere formula \citep{polak1971computational}. The step size along this chosen direction is obtained via Brent's algorithm \citep{press2007numerical}. To avoid sharp variations in the actuation, smooth gradients are enforced  by introducing a small smoothing parameter and solving a Helmholtz equation \citep{jameson2003aerodynamic, bukshtynov2011optimal}.
More details on the adjoint equations that provide the gradient, the update to the search direction and step size, a validation of the computed gradients, and the smoothing procedure are provided in 
appendix \ref{sec:NM_details}. 
To indicate the sensitivity of the computed optimal results to the smoothing parameter, the dependence of the gradient field on the smoothing parameter is also indicated in section \ref{sec:gradients}.

\section{Performance of optimal actuation}\label{sec:Ap}
\begin{figure}
    \centering
\includegraphics[width=0.92\textwidth]{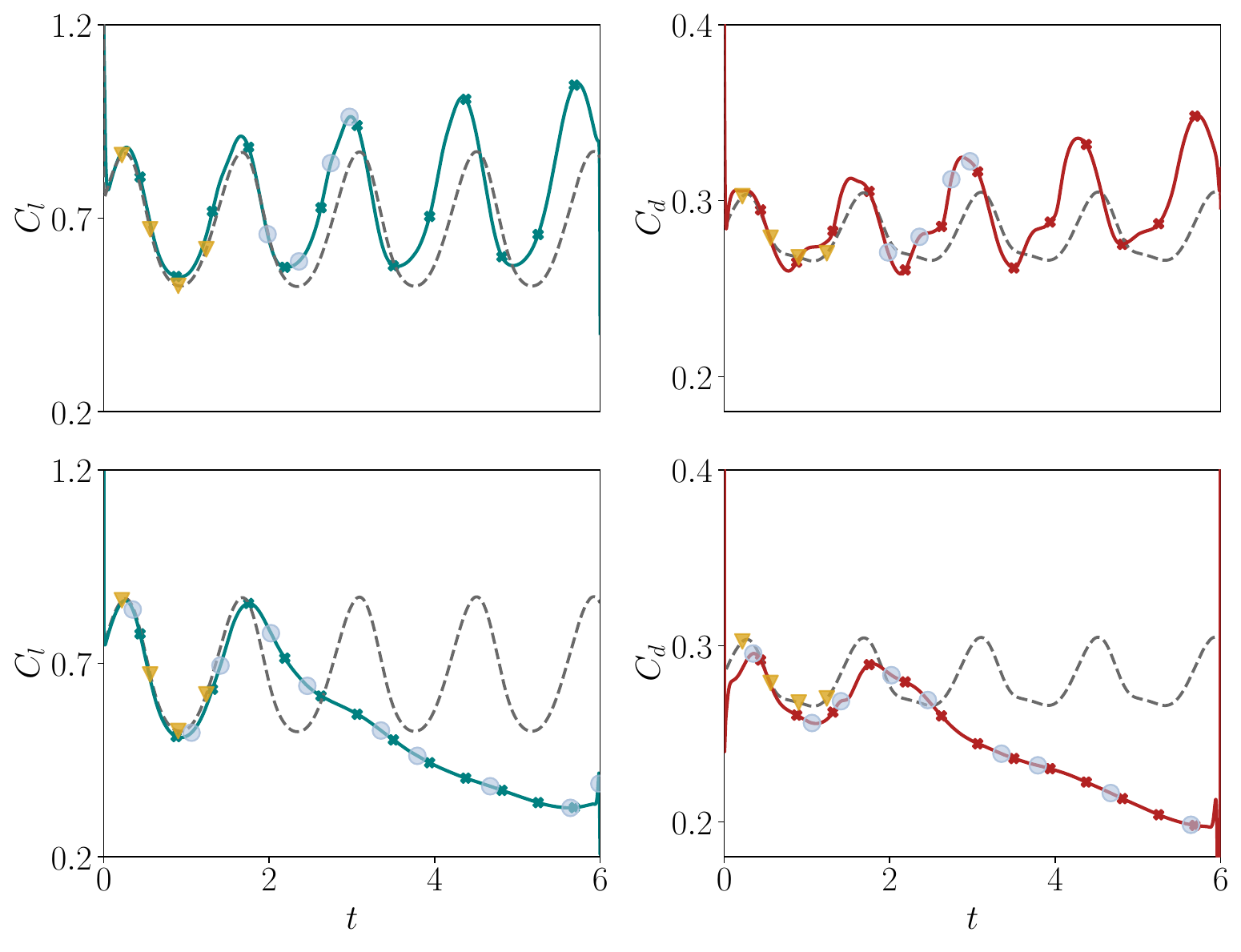}
    \caption{
    Time history of the aerodynamic forces with optimal actuation. Top row: $\mathcal{J}_{C_l}$, bottom row: $\mathcal{J}_{C_d}$. 
    Left column shows $C_l$ variation, right column shows $C_d$ variation.
    Dashed gray line in each plot represents the aerodynamic coefficient for the unactuated flow.
    Blue markers in the top row are at time instance of snapshots in figure \ref{fig:ClAP}, those in the bottom row are at snapshots in figures \ref{fig:DragFormation} and \ref{fig:DragAdvection}.
    Golden markers are at time instances of the snapshots in figure \ref{fig:BLSnaps}.
}
    \label{fig:TH}
\end{figure}

In this section we present the temporal variations of the coefficient of lift, $C_l$, and the coefficient of drag, $C_d$, resulting from the optimal actuation profiles for the two cost functionals. 
The optimization procedure is terminated when the mean aerodynamic coefficients vary less than 1\% between consecutive conjugate-gradient iterations. 
Figure \ref{fig:TH} shows the aerodynamic coefficients over the optimization window for the two cost functionals. 
With lift-optimal actuation, lift improvements can be achieved but with a penalty in drag (top row in figure \ref{fig:TH}).
Similarly, with drag-optimal actuation, the decrease in drag is accompanied by a drop in lift (bottom row in figure \ref{fig:TH}).
The more pronounced $C_l$ oscillations for lift-optimal actuation indicate that the temporal dynamics associated with vortex-shedding in the unactuated case (dashed gray curve) persist, with stronger vortical structures that yield an increase in both the minimum as well as maximum $C_l$ values of the vortex-shedding cycle. 
With drag-optimal actuation, the reduced oscillation magnitude in the $C_d$ and $C_l$ curves indicate a weakened vortex-shedding process, resulting in a $C_d$ of much smaller value than the minimum $C_d$ in the absence of actuation. 
The reduction of drag over the optimization window is reminiscent of the drag profile when actuation is employed for drag mitigation in flow past a cylinder \citep{flinois2015optimal}.
The observed variations of the aerodynamic coefficients 
with lift-optimal actuation suggest that performance improvements are not due to reattachment of the flow as in \cite{paris2023reinforcement}, where lift and drag benefits were simultaneously achieved, but by other means. 
With drag-optimal actuation, the decrease in fluctuations of the coefficients indicates a possible suppression of vortex shedding that will be explored in section \ref{sec:CdAnalysis}. 
The coefficients thus indicate that there are distinct mechanisms at play for the two cost functionals that will be explored in more detail.

Although the qualitative behavior of the $C_l$ variation with lift-optimal actuation is not drastically different from the unactuated case, the $C_d$ variation has noticeable differences near the local minima (top row in figure \ref{fig:TH}). 
While in the absence of actuation there is a single trough that is broader than the peak is sharp, with actuation two distinct local minima can be observed. The first minimum results in a drop in drag below the unactuated case (apart from the last trough) while the second has a drag value higher than the unactuated case.
The last trough in the actuated case has a smaller variation between the two local minima. 

\begin{table}
  \begin{center}
\def~{\hphantom{0}}
  \begin{tabular}{lcccc}
    Cost & $\% \Delta \overline{C_l}$ & $ \% \Delta \overline{C_d}$ & 
    $\% \Delta(\overline{C_l/C_d})$
    \\ 
    \hline\hline 
    $\mathcal{J}_{C_l}$ & 10.39 & 4.56 & 5.31\\
    $\mathcal{J}_{C_d}$ & -19.18 & -12.59 & -8.87  \\
  \end{tabular}
  \caption{Change in mean-aerodynamic coefficients, relative to the baseline unactuated case, due to optimal actuation}
  \label{tab:ClCd}
  \end{center}
\end{table}
\begin{table}
  \begin{center}
\def~{\hphantom{0}}
  \begin{tabular}{lccccccccc}
    \multirow{2}{7em}{Aerodynamic coefficient} & cycle & min & $t_{min}$ & $\Delta t_{min}$ & max & $t_{max}$ & $\Delta t_{max}$ & avg. \\    \\
    \hline\hline 
    \hspace{2em} \multirow{5}{2em}{$C_l$} 
    & 1 & 0.548 & 0.920 & - & 0.883 & 0.275 & - & 0.699 \\
     & 2 &0.573 & 2.206 & 1.286 & 0.912 & 1.658 & 1.383  & 0.730 \\
     & 3 &0.575& 3.541 & 1.335 & 0.963 & 2.984 & 1.326 & 0.752 \\
     & 4 &0.578 & 4.946 & 1.405 & 1.013& 4.338& 1.355 & 0.770 \\
     & 5 & \multicolumn{3}{c}{-----} & 1.048 & 5.726 & 1.388 & - \\
    & \textbf{ua} & {0.524} & {-} & {1.417} & {0.875} & {-} & {1.417} & {0.676} \\
    \hline
    \hspace{2em}\multirow{5}{2em}{$C_d$} 
    & 1 & 0.260 & 0.774 & - & 0.306 & 0.144 & - & 0.284 \\
     & 2 &0.258 & 2.134 & 1.360 & 0.313 & 1.528 & 1.384 & 0.288 \\
     & 3 &0.262& 3.498 & 1.364 & 0.325 & 2.876 & 1.348 & 0.294 \\
     & 4 &0.275 &4.803 & 1.306 & 0.335& 4.276 & 1.400 & 0.303 \\
     & 5 & \multicolumn{3}{c}{-----} & 0.349 & 5.718 & 1.442 & - \\
    & \textbf{ua} & {0.267} & {-} & {1.417} & {0.306} & {-} & {1.417} & {0.282} \\
    \end{tabular}
\caption{Metrics of $C_l, C_d$ over the optimization window with lift-optimal actuation (see the main text for a full description of the various columns). Last row in each half of the table corresponds shows values for the unactuated flow.}
  \label{tab:JCltemp}
    \end{center}
\end{table}
The change in the mean values of the aerodynamic coefficients for the two costs are provided in table \ref{tab:ClCd}. 
The mean values are computed by integrating the respective quantities over a time period excluding $0.1$ time units at the start and the end of the optimization window, to neglect the transients there.
Table \ref{tab:ClCd} also shows the change in the mean of 
${C_l}/{C_d}$ 
for the two costs. 
With the lift-optimal actuation, an increase in 
$\overline{C_l/C_d}$ 
is realized even though this quantity is not optimized for.
Even though the mean drag increases with lift-optimal actuation, the temporal variation of drag shows a decrease in the minimum drag 
for some of the vortex-shedding cycles (top right in figure \ref{fig:TH}).
To quantify the changes in the $C_l$ and $C_d$ variations resulting from lift-optimal actuation, metrics of their temporal behavior are tabulated in table \ref{tab:JCltemp} (a companion table is not shown for drag-optimal actuation, where the dynamics involve a gradual decay in both lift and drag rather than intricate dynamics with varied cycles).
A ``cycle'' is characterized by the appearance of a local maximum in the respective quantities, i.e., the start of a cycle coincides with the $t_{max}$ for either coefficient and is thus quantity dependent.  
The average value over a cycle for each coefficient is computed by integrating over the time window between two consecutive local maxima.
The quantity, $\Delta t_{max}$, of a cycle is the difference between the values of $t_{max}$ in the current cycle and the one prior.
Similarly, $\Delta t_{min}$ is the difference between the values of $t_{min}$ in the current cycle and the one before it.
Beyond the first minimum, the minima of $C_l$ do not vary substantially whereas the maxima do (see $C_l$ values in min and max columns). 
The time shift between the minima of $C_l$ increases monotonically while the shift between the maxima decreases from the first cycle to the second and increases monotonically thereafter.
On the other hand, the time shift between the minima of $C_d$ 
does not increase or decrease monotonically
while that between the maxima follows the same trend as the $C_l$ variation.
The last row in each half of the table shows the minimum and maximum value, and the time shift between them for the coefficients, of the unactuated flow.
While the shift in the minima and maxima for both $C_l$ and $C_d$ is the same in the absence of actuation, neither $\Delta t_{min}$ nor $\Delta t_{max}$ are the same for any of the cycles with actuation.
Further, the time shifts between the extrema of $C_l$ and $C_d$ are different for all cycles.
%
\section{Gradients of the unactuated flow}
\label{sec:gradients}
\begin{figure}
    \centering
    \subcaptionbox{\label{fig:airfoilsc}}{\hspace{0.12\textwidth}
    \includegraphics[width=0.62\textwidth]{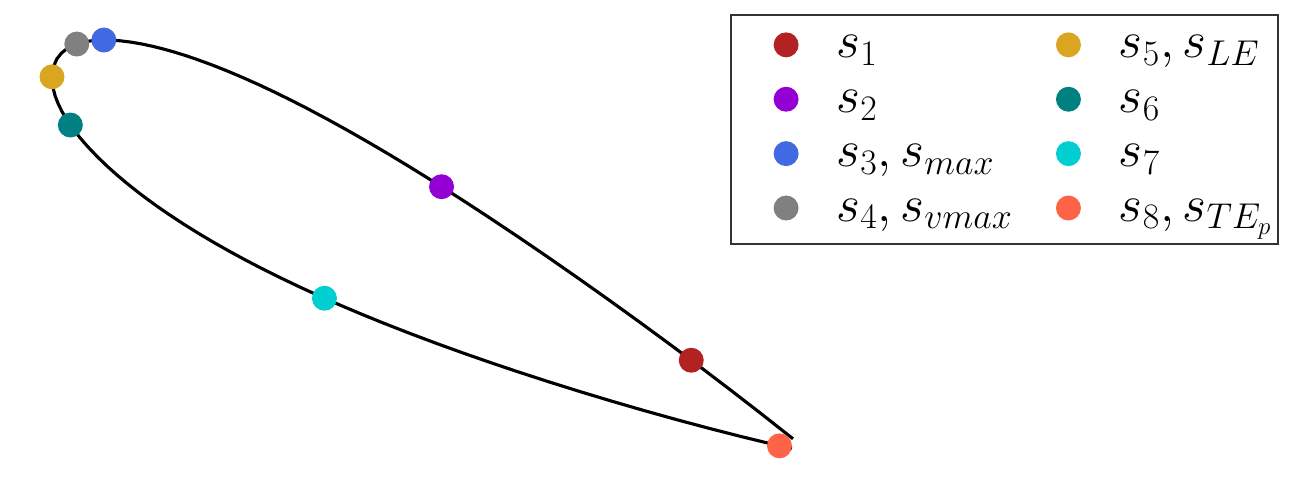}} 
    \\  \vspace{1.0em} 
    \subcaptionbox{\label{fig:gradients}}
    {\includegraphics[width=0.96\textwidth]{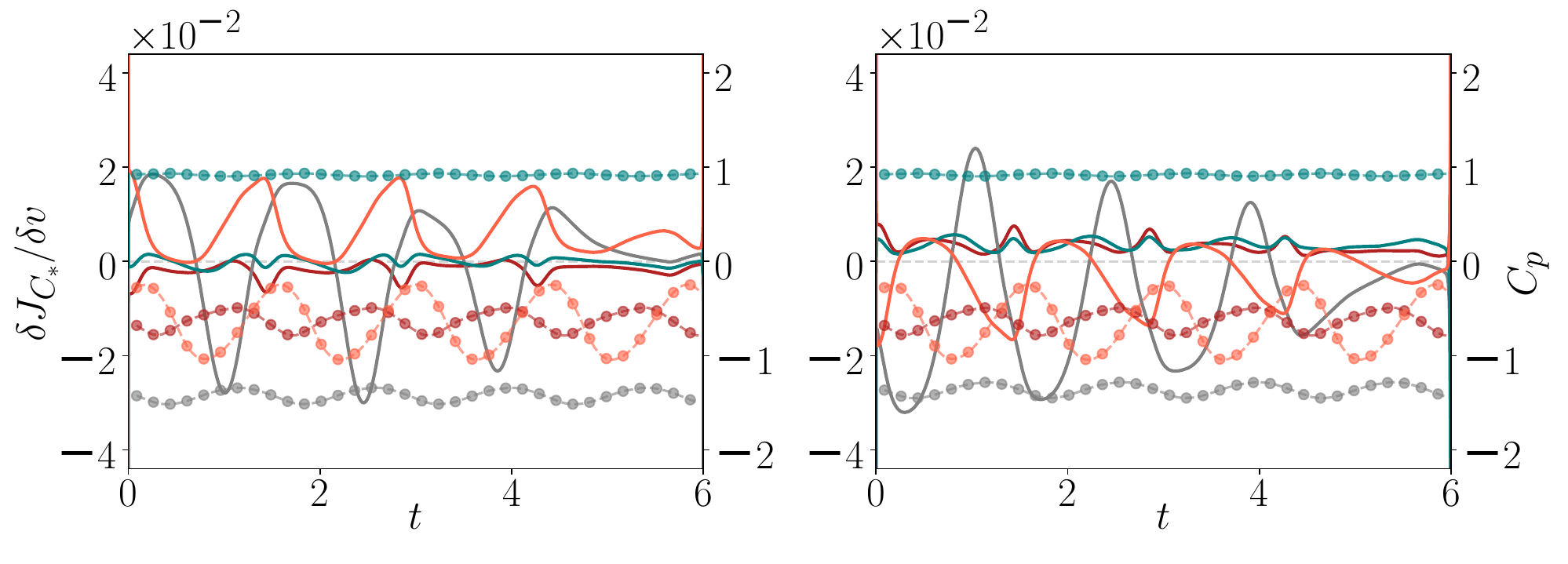}}  \\ \vspace{1.0em}
    \subcaptionbox{\label{fig:BLSnaps}}{\hspace{2.7em}\includegraphics[width=0.75\textwidth]%
    {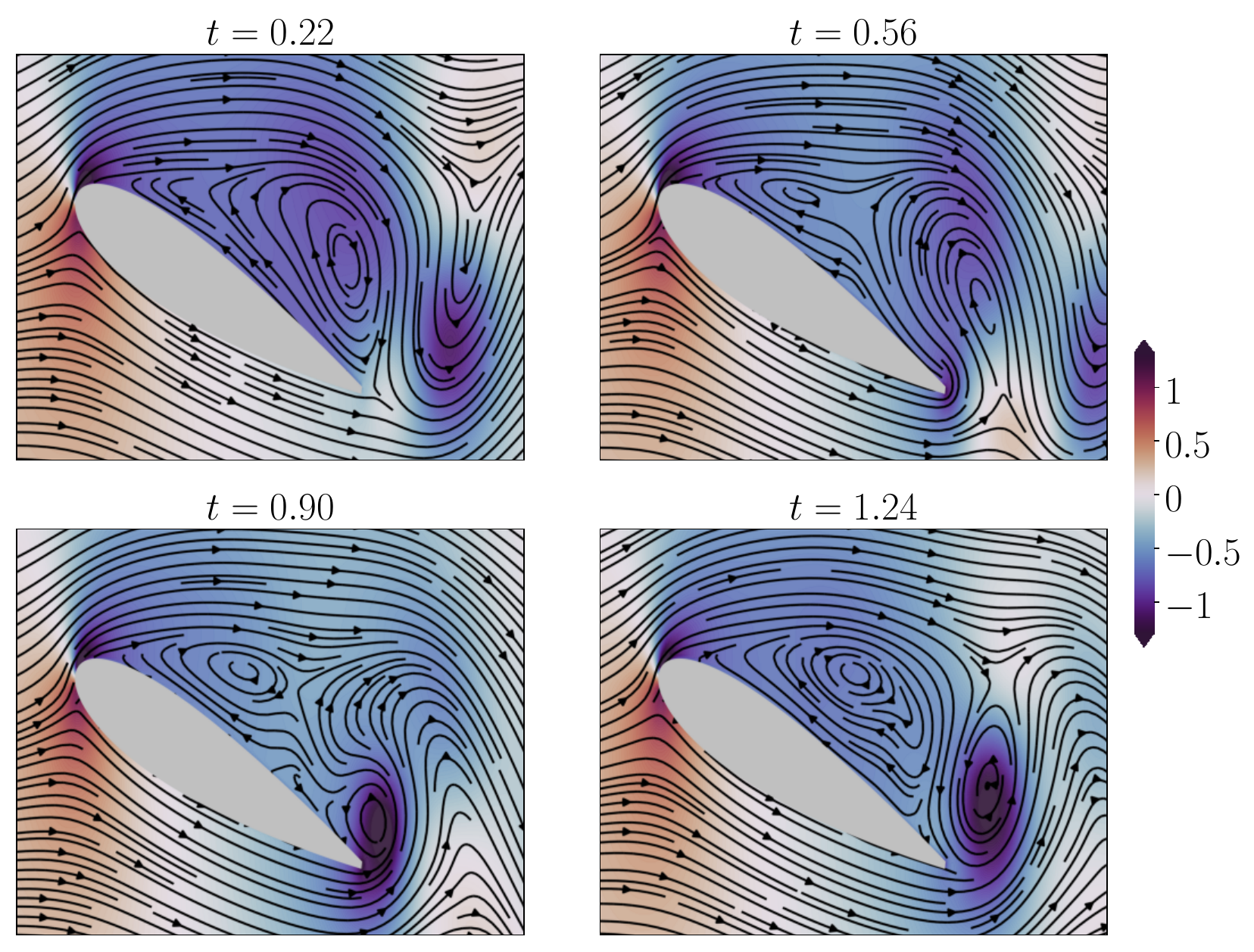}} \vspace{1.0em}
    \caption{(a) Airfoil schematic with markers at a subset of body points, 
    (b) Temporal variation of the gradients of the unactuated flow (solid line) and $C_p$ of the unactuated flow (dashed line with markers) at some of the points in (a); 
    left: 
    $\delta \mathcal{J}_{C_l}/ \delta v $, 
    right: 
    $\delta \mathcal{J}_{C_d}/ \delta v $;
    (c) Pressure contours and streamlines of the unactuated flow over one vortex-shedding cycle.}
    \label{fig:ClCdGrad}
\end{figure}
Since the optimal actuation achieved on convergence of the optimization procedure leads to distinct flow behavior from the unactuated case (as reflected in the modification of the aerodynamic coefficients shown in figure \ref{fig:TH}), analysis of the optimal actuation profile alone does not yield information
about the most suitable actuation for modifying the various flow phenomena of the unactuated vortex-shedding process.
The change in the vortex-shedding behavior brought about by optimal actuation also obscures a direct comparison between actuation for the separate performance goals of lift improvement and drag mitigation at the various time instances of the unactuated vortex-shedding process. 
Finally, it is unclear whether the cycle-to-cycle variation of the vortex-shedding process in the case of the lift-optimal actuation seen in figure \ref{fig:TH} is an artifact of the finite optimization window considered here.
To gain insights into what actuation might be beneficial for altering the unactuated flow, we first probe the gradients of the unactuated flow for the two performance goals using the adjoint flow field. 
The unactuated case exhibits limit-cycle oscillations so there are no cycle-to-cycle variations in vortex-shedding behavior over the optimization window. 

In this section, the term ``gradient'' refers to the direction of maximum decrease in the cost functional in response to a unit increase in the actuation at a location and time instance. 
This definition is in contrast to the usual one of the direction of maximum increase of the cost functional for a unit change in the actuation at a location and time instance.
Our reason for the alternative definition of the gradient is that in the first iteration of the nonlinear conjugate gradient procedure, the search direction is opposite to that of the gradient (the direction of maximum increase).
Thus, the gradients presented in this section indicate the actuation (with a unit step size) to yield a reduction in the cost when the initial guess for the actuation is zero; i.e., the flow is unactuated. 

Figure \ref{fig:ClCdGrad} shows for the unactuated flow, gradient information at some of the body points (solid lines in figure \ref{fig:gradients}) and snapshots at time instances of one vortex-shedding cycle (figure \ref{fig:BLSnaps}).
Throughout the manuscript, the temporal variation of the various quantities ($C_p$, actuation velocity and gradients of the cost functionals) at the points on the airfoil surface have been colored according to the markers in the schematic shown in figure \ref{fig:airfoilsc}.
In the schematic, $s_{max}$ is the location of the maximum $y$-coordinate of the airfoil and is independent of the quantity being plotted. The location
$s_{vmax}$ on the other hand, is where the time-averaged quantity of interest has the maximum value on the suction surface and is thus quantity dependent (for most quantities the variation of $s_{vmax}$ is minor).
In figure \ref{fig:airfoilsc}, $s_{vmax}$ is at the location of the maximum time-averaged gradient of $\mathcal{J}_{C_l}$.
The maximum time-averaged gradient of $\mathcal{J}_{C_d}$ 
lies one discretized point downstream of the location in figure \ref{fig:airfoilsc} and the gradient at $s_{vmax}$ in the right plot is at this downstream location.
Since the gradients are accurate up to a scaling factor (because the adjoint equations are linear), the gradients shown in the figure \ref{fig:gradients} have been scaled 
so that the maximum value is $0.03$. 
The gradients in figure \ref{fig:gradients} have not been smoothed and correspond to target values of zero to avoid their influence on the comparison between the gradients of the two costs.
Figure \ref{fig:gradients} also shows the temporal variation of $C_p$ on the airfoil surface at the same body points as the gradients (dashed lines with round markers).

The discussion in this section revolves around relating the gradients at $s_{vmax}$
and $s_{TE_p}$ (a point near the trailing edge on the pressure side of the airfoil) with the formation and advection of the leading-edge vortex, LEV, and the trailing-edge vortex, TEV. 
The instantaneous lift and drag values are closely related to the LEV and TEV because of the pressure gradients associated with these flow features. As such, understanding the relation of the gradients relative to these quantities can inform the optimal actuation profiles computed.
The locations $s_{vmax}$ and $s_{TE_p}$ 
are of interest because the gradient magnitudes at these locations are high, indicating the sensitivity of the cost functionals to actuation at these locations.

The start of the optimization window is such that $t=0$ coincides with a time instance slightly before the $C_l$ peak (near $t=0.23$, as seen in the unactuated $C_l$ distribution in figure \ref{fig:TH}).
From figure \ref{fig:gradients}, the pressure at $s_{TE_p}$ (dashed orange curve with markers) is maximum around $t=0.13$ which is after the TEV (seen downstream of the airfoil in the snapshot at $t=0.22$) advects away but the LEV is still upstream of the trailing edge.
During this fraction of the vortex-shedding cycle, which characterizes LEV growth and advection close to the trailing edge, the gradient of $\mathcal{J}_{C_l}$ near $s_{vmax}$ (solid gray curve in left subplot of figure \ref{fig:gradients}) is positive.
It increases as the pressure near $s_{vmax}$ drops (dashed gray curve with markers in figure \ref{fig:gradients}) and subsequently decreases as the pressure there increases. 
Over $t\in[0.22,1.24]$, the LEV advects past the trailing edge, the TEV rolls up and advects away while a new LEV approaches the trailing edge, leading up to the first snapshot in the periodic vortex-shedding process (see snapshots in figure \ref{fig:BLSnaps}). 
During this process, the pressure at $s_{TE_p}$ first decreases (during the advection of the LEV and roll-up of the TEV) and then increases (when the TEV advects away).

As the pressure at $s_{TE_p}$
increases from the minimum to the maximum value, the gradient there increases and reaches its peak before maximum pressure is experienced. After this time the gradient decreases.
At $s_{vmax}$, the pressure increases during the advection of the LEV downstream of the trailing edge (and roll up of the TEV), and decreases when the TEV advects away (and the new LEV rolls up and grows in size over the suction surface).
The gradient there is negative when the pressure increases and it reaches its negative peak (around $t = 1$) shortly before maximum pressure is experienced (around $t=1.11$).
The role of positive actuation during the appearance of the pressure minimum at $s_{vmax}$ is closely related to modification of the streamline curvature on the suction surface and the accompanying pressure drop, as also found in  \citet{thompson2022surface} for sub-optimal actuation.
The interplay between positive actuation at $s_{vmax}$ and the LEV formation process will be further discussed in section \ref{sec:ClVSBL}.

Both negative as well as positive portions of the gradient at $s_{vmax}$ are preceded in time  at the spatial location $s_6$ (see the variation of the gradient at $s_6$ over $t\in[0.47,1.34]$ where the gradient switches to negative and positive values before the gradient at $s_{vmax}$).
Between the positive peaks of the gradient at $s_6$ and $s_{vmax}$, a trough is seen in the gradient variation at $s_6$, which roughly coincides with the peak in the gradient at $s_{TE_p}$. 
This convection-like behavior of the gradient suggests that optimal actuation will have wave-like behavior, as will be explored in section \ref{sec:actuation}.

The local minimum in the gradient at $s_6$ also coincides with a local minimum in the gradient at $s_1$, which is located near the trailing edge on the suction surface of the airfoil.
At $s_1$, the gradient is mostly negative over the optimization window.
The coincidence of the positive peak at $s_{TE_p}$ and the negative peak of $s_1$ is relevant because these points lie on opposite sides of the trailing edge---positive actuation on the pressure side and negative actuation on the suction side near the trailing edge suggests that the streamlines near the trailing edge are being deflected downward as the pressure near the trailing edge is near-maximum.
The gradient at $s_1$ shows a local maximum (around $t=1.1$) before the negative peak, which roughly coincides with the pressure maximum there.

In the case of the drag-based cost, $\mathcal{J}_{C_d}$, the gradients have some opposite trends to those of $\mathcal{J}_{C_l}$---while the gradient near $s_{vmax}$ is positive near the the start of the window for $\mathcal{J}_{C_l}$, it is negative in case of $\mathcal{J}_{C_d}$. A similar trend is seen in the gradient at $s_{TE_p}$ where the maximum of the pressure near the trailing edge is preceded by a negative peak in the gradient instead of a positive peak. 
Generally speaking, the gradients for the two cost functionals at the locations shown here are mirrored about the $y=0$ line and shifted with a location-dependent shift in time.
While the gradient at $s_{TE_p}$ is mostly positive for $\mathcal{J}_{C_l}$, 
in case of $\mathcal{J}_{C_d}$, it is not strictly negative even though the substantial portions are all negative. 
The peaks in the pressure at $s_{TE_p}$ are preceded by negative peaks of the gradient in the case of $\mathcal{J}_{C_d}$, instead of the positive peaks seen for $\mathcal{J}_{C_l}$.
At $s_{vmax}$, the local maxima in pressure are preceded by positive peaks for $\mathcal{J}_{C_d}$ instead of the negative peaks seen for $\mathcal{J}_{C_l}$.
Similar observations can be made about the gradients at the other two locations.
Despite the differences in gradient for the different cost functionals, common trends for each gradient across the various locations can be seen: for example, the peaks in the gradient at $s_{TE_p}$ coincide with peaks in the gradient at $s_6$ and $s_1$.
Also, the maxima in the pressure variation at both the trailing edge and leading edge are preceded by peaks in the gradients at these locations for both costs.

This shared relationship between the pressure and gradient for both costs at $s_{vmax}$ and $s_{TE_p}$ suggests an interplay between the flow features and the optimal actuation. One might infer that actuation at these two locations---both of which have relatively high gradient magnitudes---plays an important role in altering the vortex-shedding process to ultimately provide performance improvements.
Other observations such as the coincidence of the gradient peaks at $s_6$ and $s_1$ with the peak in the gradient at $s_{TE_p}$, and the appearance of extrema at $s_6$ prior to those at $s_{vmax}$, are reminiscent of traveling-wave behavior. These features of the gradient will be used to understand the computed optimal actuation for both the lift- and drag-based cost functionals  in section \ref{sec:actuation}.

For both costs, there exist cycle-to-cycle variations in the gradients over the optimization window even though the forward solution over the optimization window is perfectly periodic. For $\mathcal{J}_{C_l}$, the variation in the gradient profile from cycle-to-cycle is not monotonic whereas for the drag-based cost, the gradient at the various locations is seen to decrease in magnitude over the window. This variation is most clearly seen in the time evolution at $s_{vmax}$ in figure \ref{fig:ClCdGrad}. For both costs, the gradients substantially decrease in magnitude over the last $\approx 0.5$ time units because of the ``starting'' transients of the adjoint solution (near the final time). These variations in the gradient motivate a possible pathway by which a non-periodic solution may be obtained in optimal actuation, even when the underlying unactuated dynamics are periodic.

\begin{figure}
    \centering
    \includegraphics[width=0.95\textwidth]{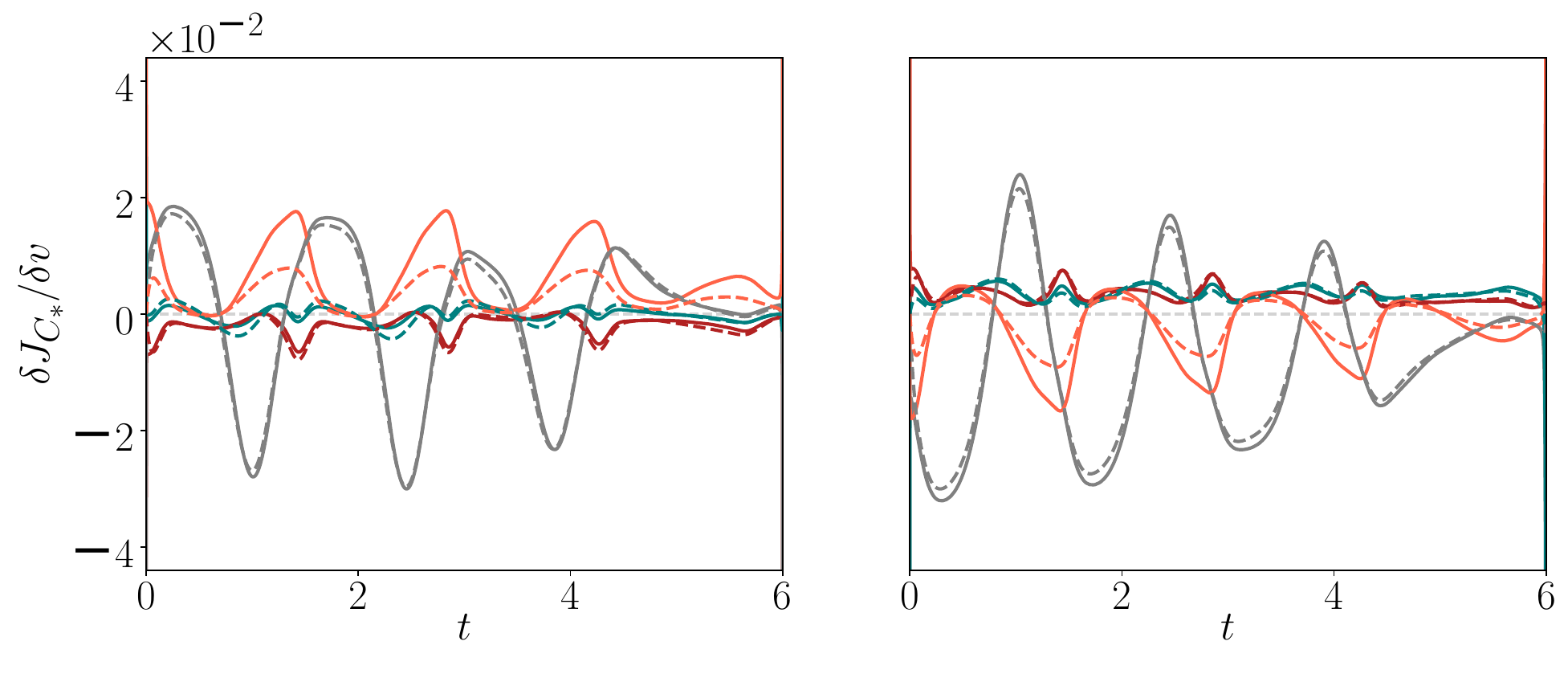} 
    \caption{Effect of smoothing and non-zero target values. 
    Solid lines represent non-smoothed gradient with target values set to zero, dashed lines represent smoothed gradients with non-zero target values.
    Left: 
    $\delta \mathcal{J}_{C_l}/ \delta v $, 
    right: 
    $\delta \mathcal{J}_{C_d}/ \delta v $. }
    \label{fig:gradientsmoothing}
\end{figure}

To depict the changes in the gradients arising from smoothing and non-zero target values, figure \ref{fig:gradientsmoothing} shows the smoothed gradients (dashed lines) alongside those without smoothing and with zero target values, from figure \ref{fig:ClCdGrad}.
The noticeable differences are that the smoothed gradients are zero at either end of the optimization window and that the gradient at $s_{TE_p}$ is much lower in magnitude due to spatial smoothing. Despite these differences, the figures demonstrate that smoothing has a small effect on the gradient over the majority of the time window.
%
%
\section{Properties of optimal actuation}
\label{sec:actuation}
In this section, we discuss the spatial and temporal properties of optimal actuation for each cost functional. 
To connect to prior work that assumed periodic 
(non-optimal)
actuation, which is known to yield aerodynamic performance improvements \citep{akbarzadeh2019numerical, akbarzadeh2020controlling, shukla2022hydrodynamics}, we draw similarities between optimal actuation and traveling-wave behavior where appropriate. 
It has already been established that actuation on the suction surface near $s_{max}$ is beneficial for performance \citep{kang2015effects, khan2017local, thompson2022surface, paris2023reinforcement}. 
We therefore comment on the spatial properties of actuation to identify key locations on the airfoil surface, 
and analyze the temporal variation of optimal actuation at these places.
From the gradient of the unactuated flow, 
the magnitude at $s_{TE_p}$ was found to be substantial for both costs. 
Further, for both costs, the variations at $s_1$ and $s_6$ were 
related to 
the gradient at $s_{TE_p}$ as well as $s_{vmax}$. 
It is thus of interest to determine whether actuation at these locations reflects a similar trend as the behavior of the gradient of the unactuated flow.

\begin{figure}
    \centering
    \hspace{0.05\textwidth}{\includegraphics[width=0.95\textwidth]{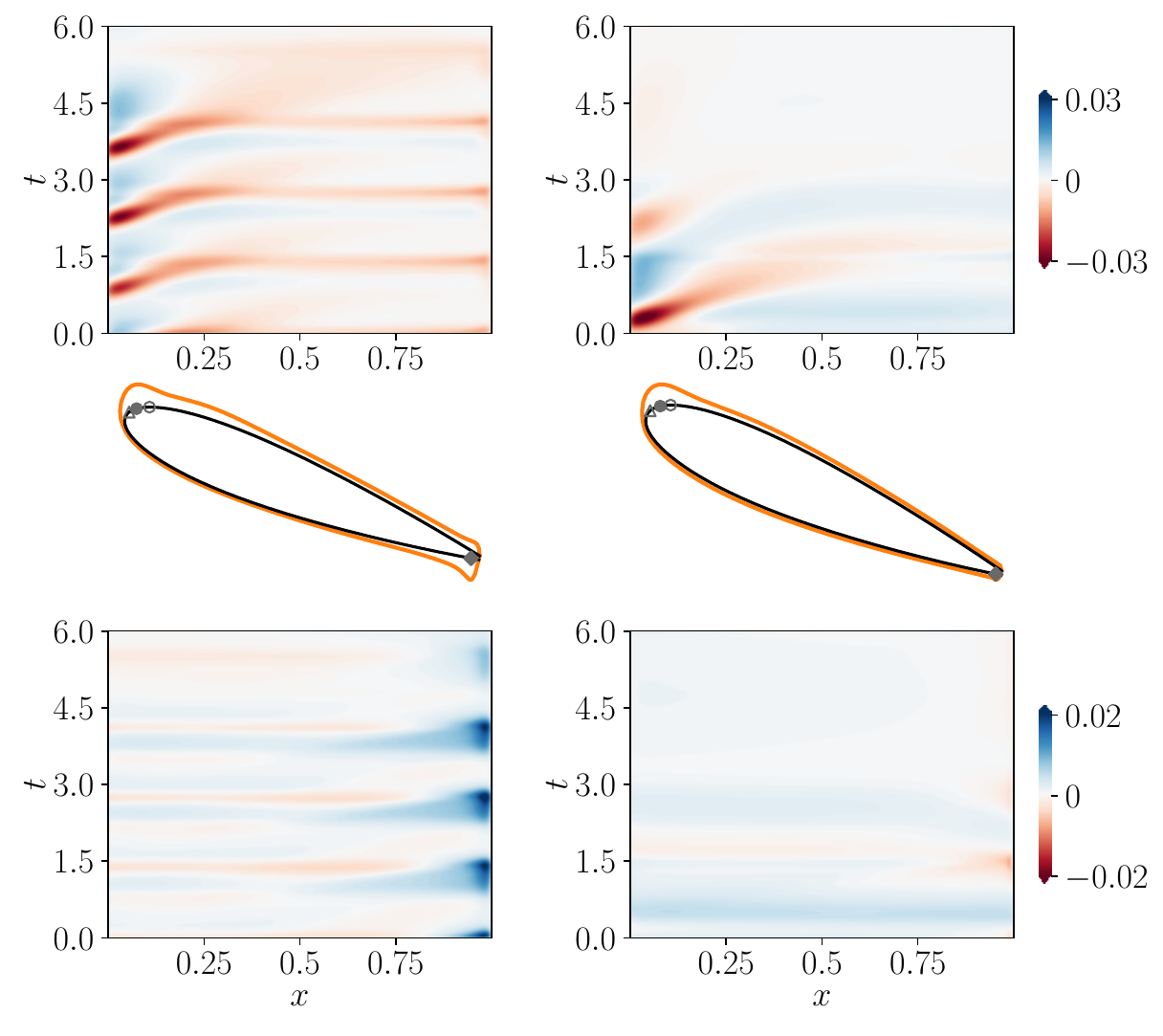}}
    \caption{Space-time contours of optimal actuation. Left column: $\mathcal{J}_{C_l}$, right column: $\mathcal{J}_{C_d}$.
    Top row: spatio-temporal variation on the suction surface; bottom row: spatio-temporal variation on the pressure surface; 
    Middle row: time-averaged actuation on the airfoil surface; hollow triangle: location of maximum time-averaged $C_p$, solid circle: location of maximum time-averaged actuation ($s_{vmax}$), hollow circle: $s_{max}$, solid diamond: location of maximum time-averaged actuation on pressure side.}
    \label{fig:vNSurf}
\end{figure}

Figure \ref{fig:vNSurf} details the space-time dependence of actuation as well as the time-averaged actuation profile.
From the space-time contours on the suction surface, patterns of positive and negative velocity exist for both cost functionals. 
For the lift-based cost, the streaks of negative actuation at $x<0.25$ are darker and narrower than the positive ones; implying large magnitude actuation for shorter durations. 
This observation is also true for the negative-actuation streaks in the drag-optimal actuation, although the magnitude of actuation in general reduces over the course of the optimization window. In the case of the lift-based cost, actuation is substantial not only near $s_{vmax}$ but also near $s_{TE_p}$.
For the drag-based cost however, the actuation at $s_{TE_p}$ is negligible.
The role of the actuation at the trailing edge for the lift-based cost is in line with the observations of section \ref{sec:gradients}. 
The discrepancy in the case of the drag-based cost could be an indication that a weakened vortex-shedding process is more easily achieved by actuating at $s_{vmax}$. 

The substantial actuation around $s_{max}$ is also reflected in the time-averaged plots in the second row. 
For both costs, the time-averaged profile shows a bump near $s_{max}$.
From the markers (see figure caption for description), the exact location of
$s_{vmax}$ is upstream of $s_{max}$ for both costs; 
with the location in the drag-optimal case being one discretized body point downstream of the lift-optimal case.
The location of $s_{vmax}$ for both costs is unchanged from that of the gradient of the unactuated flow (c.f., section \ref{sec:gradients}).
Also, for both costs $s_{vmax}$ is downstream of the location of the maximum time-averaged $C_p$.
Note that for both costs, the hollow markers are at the same locations.
The upstream location of $s_{vmax}$ with respect to $s_{max}$ is not surprising 
given the existence of a boundary layer that eventually separates.
On the pressure side, the time-averaged actuation is maximum near the trailing edge.

We now consider, for both cost functionals, actuation on the suction side, as actuation (exclusively) on this suction surface has been most commonly considered in prior investigations. 
For both cost functionals, there are time instances
when
the suction-surface actuation upstream of $x \approx 0.25$ is negative while the rest of the suction surface experiences positive actuation (in the case of drag-optimal actuation, see the start of the window and $t> 1.5$, for the lift-optimal case see the first few time instances when negative actuation occurs near the leading edge). 
Conversely, there are time instances when most of the suction surface experiences negative actuation while the airfoil surface near the leading edge has positive actuation. 
Additionally, as the negative actuation moves downstream, positive actuation appears near the leading edge. 

Moreover, while at a fixed time the spatial distribution of actuation is not obviously periodic, 
actuation exhibits 
traveling wave-like behavior as time evolves. 
To illustrate this behavior more clearly, consider the lift-optimal actuation over $t\in[2,3]$. 
The region of negative velocity of magnitude $\approx 0.03 $ seen close to the leading edge around $t=2$ moves downstream at later time instances while reducing in magnitude and spreading upstream and downstream. 
As this region of negative actuation moves downstream, positive actuation appears close to the leading edge ($t \approx 2.5$). Commensurate with this switch to positive actuation near the leading edge is an onset of (low-magnitude) negative velocity over the entire suction surface downstream of 
$x\approx 0.25$. 
This cyclic behavior is also seen to occur over $t \in [0.6, 1.5]$ and $t \in [3.2, 4.2]$. 
Similar traveling wave-like actuation is seen for the drag-based cost, even though actuation is negligible over the second half of the optimization window.

\begin{figure}
    \centering
    \includegraphics[width=0.96\textwidth]{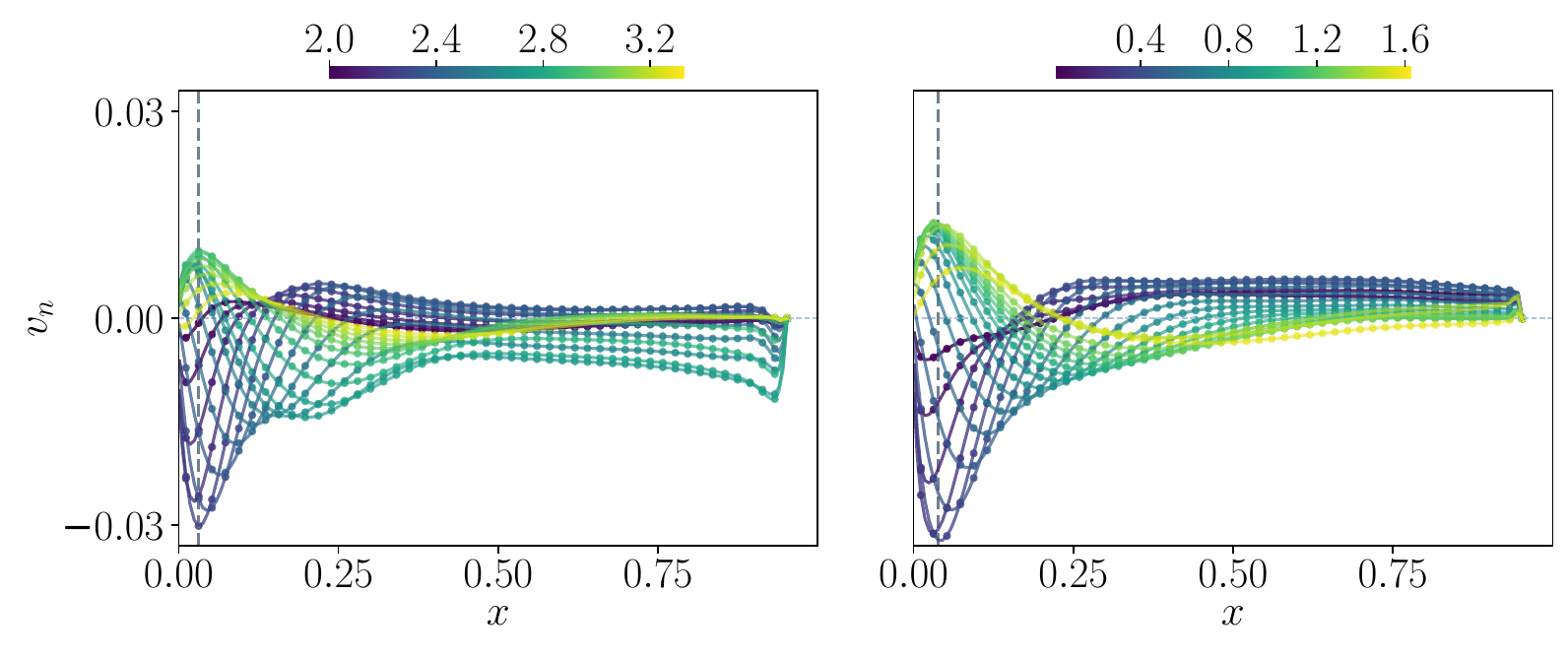} 
    \caption{Spatial profile of actuation on the suction surface for $\mathcal{J}_{C_l}$ (left) and $\mathcal{J}_{C_d}$ (right), for the time instances indicated by the colorbars. 
    The vertical dashed line in each subfigure represents the $x-$coordinate of the point on the suction surface with the maximum time-averaged value, $s_{vmax}$.
    }
    \label{fig:vNWave}
\end{figure}
To probe this traveling wave-like behavior further, figure \ref{fig:vNWave} shows the spatial variation of actuation on the suction surface at various time instances throughout a cycle for the two cost functionals.
In each plot, negative actuation is seen near the leading edge
at the first time instance. At the last time instance, the actuation at the leading edge is zero and negative actuation is about to reappear there. 
For both costs, at the initial time instances of negative actuation near the leading edge, the negative peak of actuation on the suction surface increases in magnitude as it moves downstream and reaches $s_{vmax}$.
Downstream of this negative peak near $s_{vmax}$, the actuation magnitude decreases and ultimately switches from negative to positive. 
At a later time when the negative peak of actuation travels sufficiently downstream of $s_{vmax}$,
positive actuation ceases to exist over the aft portion of the airfoil.
This behavior is more prominent for the lift-based cost where the actuation over the entire aft portion of the suction surface is negative for $t > 2.51$.
As the negative peak moves downstream, positive actuation appears near the leading edge.
For the lift-based cost, the actuation near the leading edge changes signs at 
$t= 2.49$
and for the drag-based cost, the switch occurs at $t= 0.41$. 
When positive actuation first appears at the leading edge, the maximum positive value is at the leading edge.
At later time instances, the peak positive actuation manifests as a local maximum at a location downstream of the leading edge. 

Between the two costs, the most noticeable differences are that the lift-based cost has a more pronounced negative peak around $x = 0.25$ and a larger actuation magnitude near the trailing edge. 
While in case of the drag-based cost, the magnitude of negative actuation peak is seen to decrease monotonically after the peak arrives at $s_{vmax}$, for the lift-based cost the peak magnitude first decreases and then slightly increases before monotonically decreasing in time again.

The traveling-wave behavior observed here has differences from the open-loop actuation previously studied \citep{akbarzadeh2019numerical, akbarzadeh2020controlling, shukla2022hydrodynamics, thompson2022surface}. In the current case, the spatial distribution of actuation contains at most one peak (characterized by a local maximum of positive actuation) or trough of actuation (characterized by local minimum of negative actuation) at a given time instance, while most of the suction surface has the opposite sign. 
Further, the length scales of the peak and trough vary with location.
This scale non-uniformity is evident from the spreading of the negative actuation of the lift optimal cost after $t=2$ as it travels from 
$x=0$ 
to $x = 0.25$. 
Note that the investigations of \citet{akbarzadeh2019numerical}, \citet{akbarzadeh2020controlling} and \citet{shukla2022hydrodynamics} involved surface displacements and it is possible that optimal actuation via surface displacements instead of the velocity boundary conditions considered here, could lead to differences in the optimal-actuation profile.
Further studies must be conducted to assess this point.

We now consider actuation on the pressure side. 
Figure \ref{fig:vNSurf} shows patterns of positive actuation near the trailing edge in the case of lift-optimal actuation. 
Each peak in positive actuation localized near the trailing edge is preceded in time by a small-magnitude but spatially extended region of positive actuation. As time evolves, that positive actuation advects to the trailing edge, shrinks in spatial extent and grows in magnitude.
Also seen are streaks of negative actuation over the upstream three-quarter portion of the pressure surface when the actuation near the trailing edge is maximum.
For the drag-based cost, the actuation on the pressure side is not substantial. 
Two faint streaks of positive actuation extending over the entire pressure surface are seen. 
The streaks are separated in time by a negative peak of actuation near the trailing edge and some time instances when negative actuation is seen over the entire pressure surface.
Before the actuation magnitude reduces to negligible values, negative actuation is seen at the trailing edge. Qualitatively, the results largely support prior studies' focus on suction-surface actuation, except for actuation near the trailing edge for the lift-based cost functional which will be explored further in section \ref{sec:ClVSBL}.

\begin{figure}
    \centering
    \includegraphics[width=0.99\textwidth]{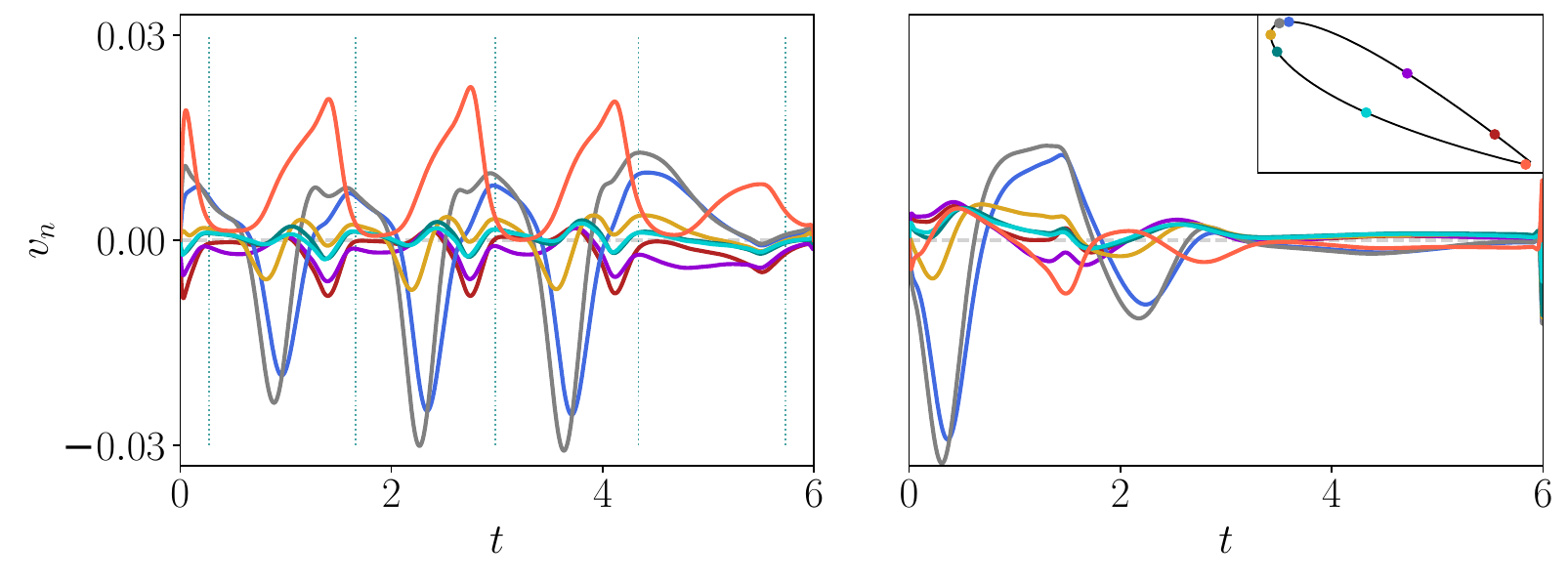}  
    \caption{
    Temporal variation of actuation at a subset of body points. Colors correspond to the markers in the inset (same as figure \ref{fig:airfoilsc}). Dashed vertical lines in the left subfigure represent time instances of lift peaks in top row of figure \ref{fig:TH}.
    }
    \label{fig:vNTemporal}
\end{figure}

To gain insights into the temporal variation of actuation at influential locations on the airfoil surface and to analyze the interplay between actuation at neighboring locations, figure \ref{fig:vNTemporal} shows the actuation profile at various points along the suction and pressure surfaces. For continuity with the analysis from earlier in this manuscript, these points include $s_{vmax}$ and $s_{TE_p}$, as well as a number of other locations for a more comprehensive representation of the full actuation behavior.

Observations from above in this manuscript are visible through this figure as well: the lift-optimal actuation comprises of traveling wave-like behavior with considerable magnitude throughout the optimization window whereas the drag-optimal actuation is of largest magnitude in the first half of the optimization window; 
for both costs, the negative actuation peaks are considerably larger in magnitude than the positive peaks; the actuation at the trailing edge is noticeably larger for the lift-based compared with the drag-based cost. 

For the lift-based cost, the positive phases of actuation at $s_{vmax}$ 
comprise of two local maxima (except at the first positive peak occurring over $t< 1$). 
The positive portion of the gradient in section \ref{sec:gradients} did not show the two peaks seen in the optimal actuation profile.
The two peaks are separated by a region of local minimum. 
Over the course of the three identifiable actuation cycles, the magnitude of the first peak relative to the second is seen to decrease.
The local minima of positive actuation at $s_{vmax}$ appear around the same instance as the actuation peaks at $s_{TE_p}$.
The occurrence of actuation extrema at $s_{vmax}$ 
and $s_{TE_p}$ is also seen for the drag-based cost functional, where the actuation at $s_{TE_p}$ attains its minimum around the same time instance as the actuation peak at 
$s_{vmax}$. 
The coincidence of the peaks (troughs) in the actuation at $s_{TE_p}$ along with troughs (peaks) at $s_1$ and $s_6$ for the lift-based (drag-based) cost
was also observed in the gradient of the unactuated flow
(section \ref{sec:gradients}).
This synchrony between the actuation at $s_{vmax}$ and the points on the pressure surface, as well as points near the trailing edge on the suction surface, will be further analyzed in section \ref{sec:flowfield}. 

Further connections between the actuation on the pressure side and that at 
$s_{vmax}$ 
can be observed for the lift-based cost: the occurrence of positive actuation is preceded in time by positive peaks on the pressure side. 
The actuation at the leading edge (represented by $s_{LE}$) reaches a positive peak shortly before the actuation at 
$s_{vmax}$ 
becomes positive. 
This peak follows the peak at downstream locations on the pressure side of the airfoil. 
This observation is also true for the drag-based cost: the actuation at the points on the pressure side of the airfoil reach positive peaks before the actuation at 
$s_{vmax}$ 
becomes positive. 
Further, it can be seen that the negative peak of actuation at 
$s_{vmax}$ 
appears after the local minimum in the actuation at the leading edge
and points near the leading edge on the pressure surface.

This section has highlighted the major features of actuation that is optimal for both maximizing mean lift and minimizing mean drag. 
For both costs, the time-averaged actuation on the airfoil surface indicated that the airfoil location upstream of $s_{vmax}$ on the suction surface is most effective for altering the vortex-shedding process.
For the lift-based cost, actuation at the pressure side of the trailing edge is also beneficial for modifying the flow.
Many of the attributes of the optimal actuation profile are similar to the gradient of the unactuated flow discussed in the prior section.
In the next section, we connect the optimal actuation behavior to the response of key flow structures, to mechanistically explain the reason behind the lift and drag performance benefits.  

\section{Flow physics of optimal actuation}
\label{sec:flowfield}
\begin{figure}
    \centering
    {\includegraphics[width=0.99\textwidth]{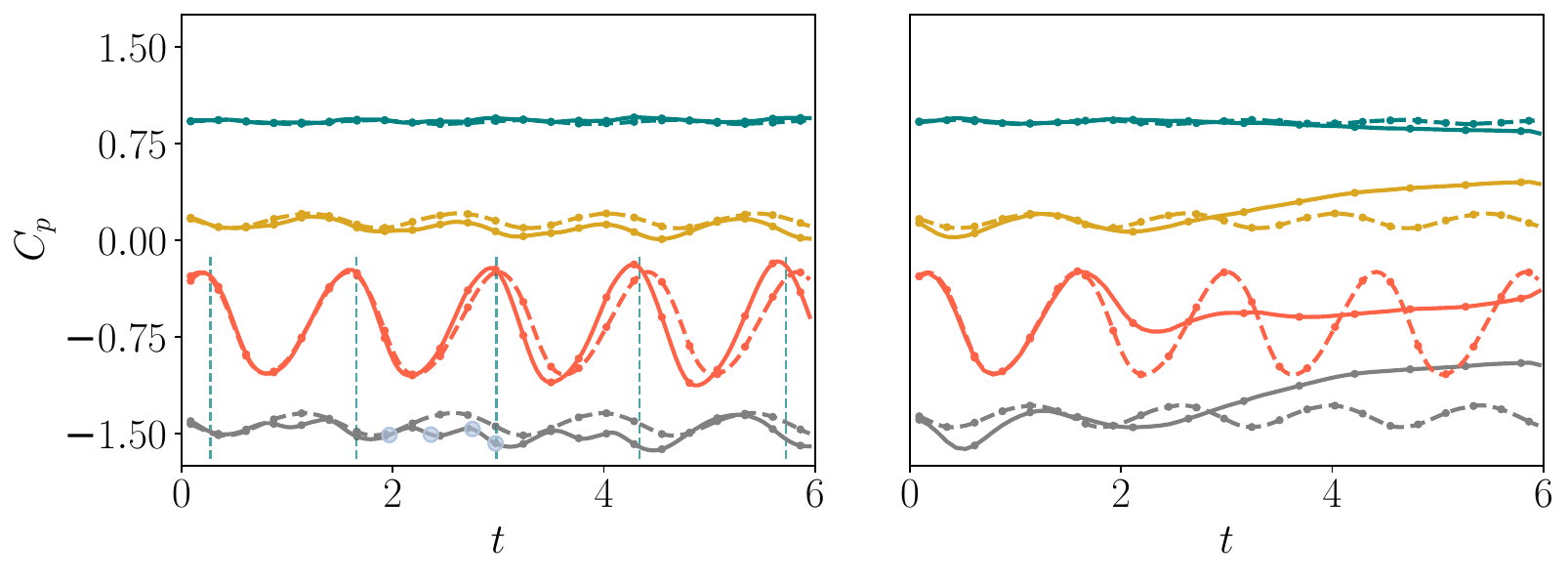}} 
    \caption{Temporal variation of $C_p$ at a subset of body points for the unactuated flow (dashed lines) and flow with optimal actuation (solid lines). 
    Colors correspond to the markers in figure \ref{fig:airfoilsc}.
    Left: $\mathcal{J}_{C_l}$, Right: $\mathcal{J}_{C_d}$. For the lift-based cost functional: time instances coinciding with the local maxima of lift are indicated by the dashed teal lines; the blue markers indicate the time instances of the snapshots in figure \ref{fig:ClAP}.
    }
    \label{fig:LECp}
\end{figure}

We now draw connections between the actuation and the flow phenomena it alters and induces. 
Prior to a detailed analysis of the flow features, we make comparisons between the temporal variation of the coefficient of pressure, $C_p$, on the surface of the airfoil with and without actuation. 
Figure \ref{fig:LECp} shows the variation of $C_p$ for the lift- and drag-based cost functionals, at various spatial locations. 
For the lift-based cost functional, $C_p$ at $s_{vmax}$ (gray curve) is altered so as to counter the lift detriments of the trailing-edge vortex formation. 
Without actuation, the pressure at $s_{vmax}$ is maximum shortly after the pressure near the trailing edge is minimum (as discussed in section \ref{sec:gradients}).
However, in the presence of actuation, a local minimum in the temporal variation (at $s_{vmax}$) is seen shortly after the time of the pressure minimum at the trailing edge. 
In the current discussion this local minimum in the pressure at $s_{vmax}$ will be referred to as the first local minimum (per cycle), to distinguish it from the 
second local minimum (per cycle),
which is analogous to the local minimum in the unactuated case and coincides with the roll-up of the LEV near the leading edge.
The first local minimum is absent towards the end of the optimization window. 
Thus, the first minimum can be attributed to actuation near the leading edge, since actuation 
there is negligible towards the end of the optimization window. 

The other characteristics of the curves are as follows: for the lift-based cost functional, the oscillation frequency (associated with vortex shedding) increases relative to the unactuated case (also evident from $\Delta t_{min}$ in table \ref{tab:JCltemp}), 
and the pressure near $s_{vmax}$ decreases over the optimization window, an observation that is in line with the increasing lift.
For the drag-based cost, the pressure 
at $s_{vmax}$, $s_{TE_p}$ and $s_{LE}$ increases while that at $s_{6}$ slightly decreases. 
Since the lift and drag-optimal actuation profiles and the aerodynamic coefficients and $C_p$ variations induced by them are substantially different, we discuss the flow changes for the two cases separately. 
\subsection{Analysis for Lift-Optimal Actuation}
\label{sec:ClVSBL}
\begin{figure}
    \centering
    {\includegraphics[width=0.9\textwidth]{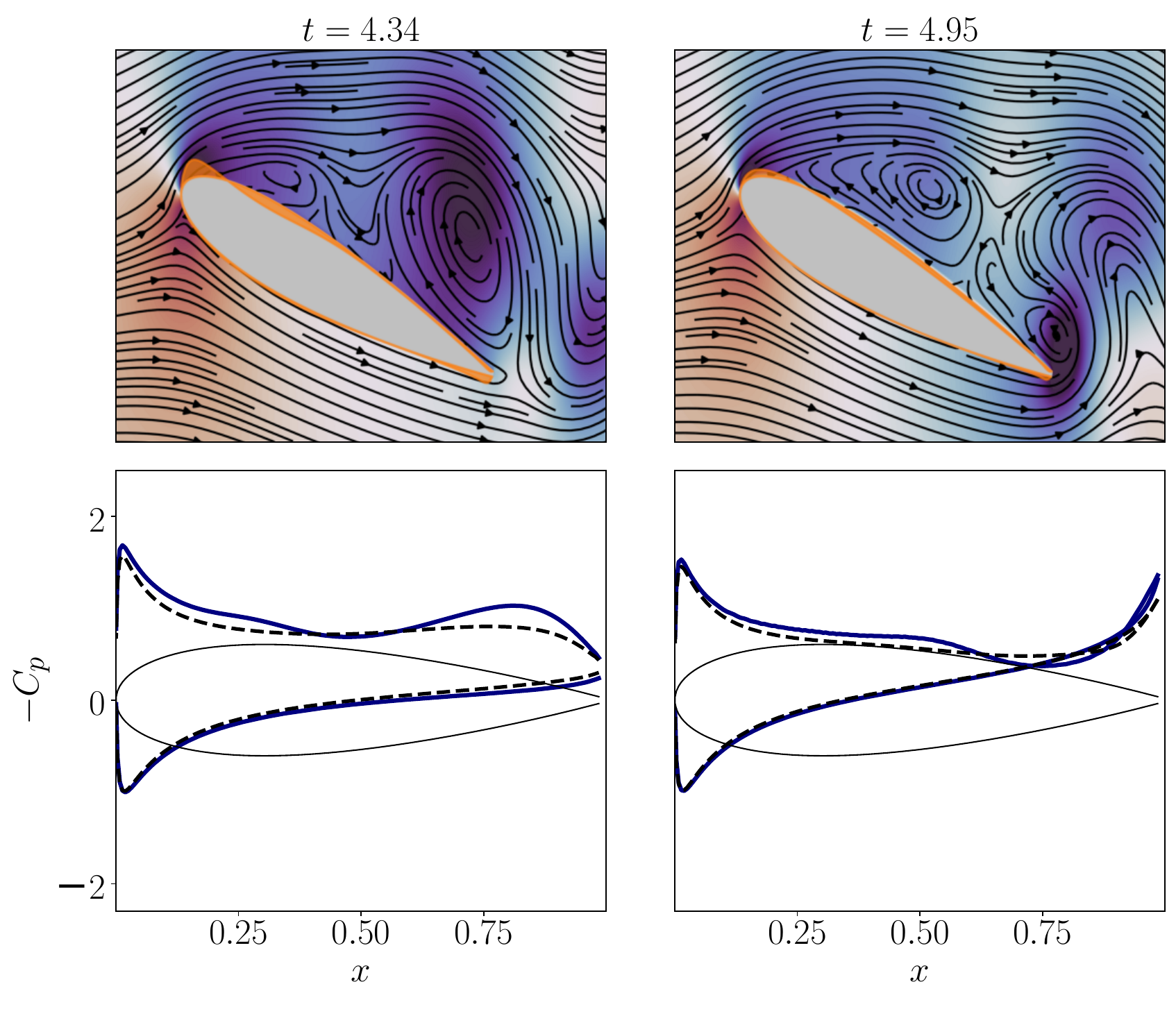}} \hspace{0.5em}   
    \caption{
    $C_p$ contours and streamlines, along with the surface actuation scaled for easy visualization, for the flow with lift-optimal actuation (top row; contour limits are the same as figure \ref{fig:BLSnaps}) and surface distribution of $C_p$ (bottom row; the dashed line represents unactuated flow and the solid line represents actuated flow).
    The left column shows contours at a time instance close to maximum lift and the right column represents flow field close to the time instance of minimum lift. The relevant contour plots for comparison in the unactuated flow are in figure \ref{fig:BLSnaps} ($t=0.22$ and $t=0.9$, respectively).
    }
    \label{fig:PMinMax}
\end{figure}
%
As seen in section \ref{sec:Ap}, actuation for lift optimization alters the lift dynamics, with an increase in the mean lift from one cycle to the next. 
To connect variation of aerodynamic forces to the flow field, snapshots of the flow and actuation behavior,
at time instances of maximum and minimum lift in a vortex-shedding cycle,
are shown in
figure \ref{fig:PMinMax}.
The analogous snapshots of the unactuated flow correspond to $t=0.22$ and $t=0.90$ in figure \ref{fig:BLSnaps}.

Though the two time instances represent similar phenomena in the vortex-shedding cycle, differences in pressure magnitude and phase between the LEV and TEV due to actuation are apparent. 
At the time instance of maximum $C_l$, compared to the unactuated case (c.f., figure \ref{fig:BLSnaps} at $t=0.22$), the LEV near the trailing edge on the suction side has a greater pressure magnitude (reflected by the darker contours)  and a new partially rolled-up LEV is seen near the leading edge.
The stronger LEV near the trailing edge and the partial roll-up of the new LEV near the leading edge contribute to higher pressure magnitude over most of the suction surface in the presence of actuation, as seen from the surface $C_p$ distribution.

At the time instance of minimum $C_l$, which coincides with TEV formation, the new LEV near the leading edge is more fully formed in the presence of actuation.
Thus, the $C_p$ on the suction surface near the leading edge has a higher magnitude.
Note that in the unactuated case the TEV is more developed than in the actuated case, suggesting that actuation changes the phase between the formation of the LEV and TEV.
The change in phase in terms of the accelerated formation of the LEV implies that the LEV is stronger when the TEV pinches off, leading to a smaller drop in lift as compared to the unactuated flow. 
The mitigation of the detrimental effects of TEV formation on lift is also reflected in the higher value of minimum $C_l$ in 
figure \ref{fig:TH}.
At both time instances, the higher magnitude of pressure occurs along with a larger curvature of the streamlines in the presence of actuation.  
\begin{figure}
    \centering
    \includegraphics[width=0.8\textwidth]{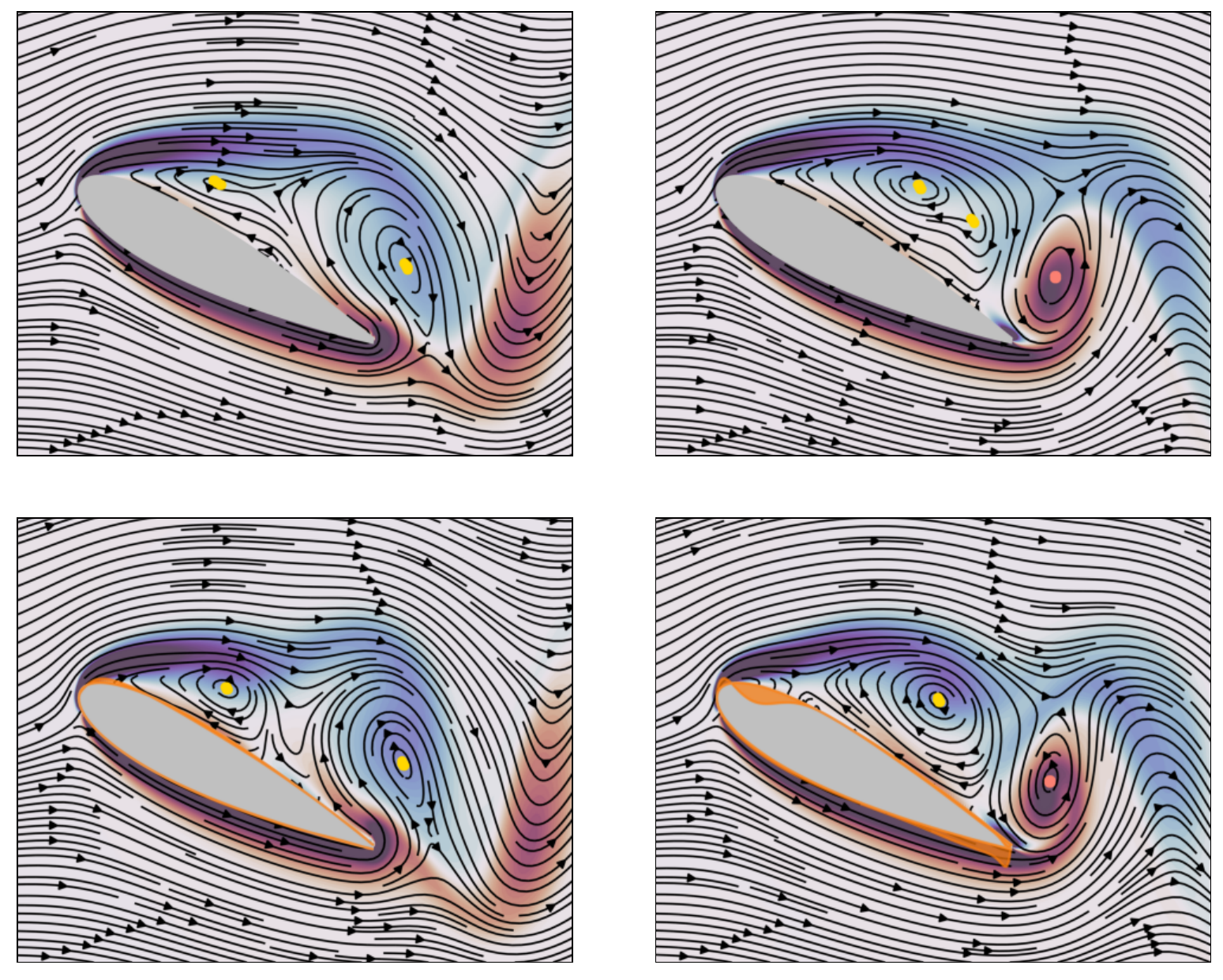}
    \caption{
    Vorticity contours and streamlines for unactuated flow(top row) and actuated flow with lift-optimal actuation (bottom row). 
    Gold markers represent center of clockwise vortices, salmon markers represent center of anti-clockwise vortices.
    }
    \label{fig:VortMinMax}
\end{figure}
%
\begin{figure}
    \centering
    \includegraphics[width=0.99\textwidth]{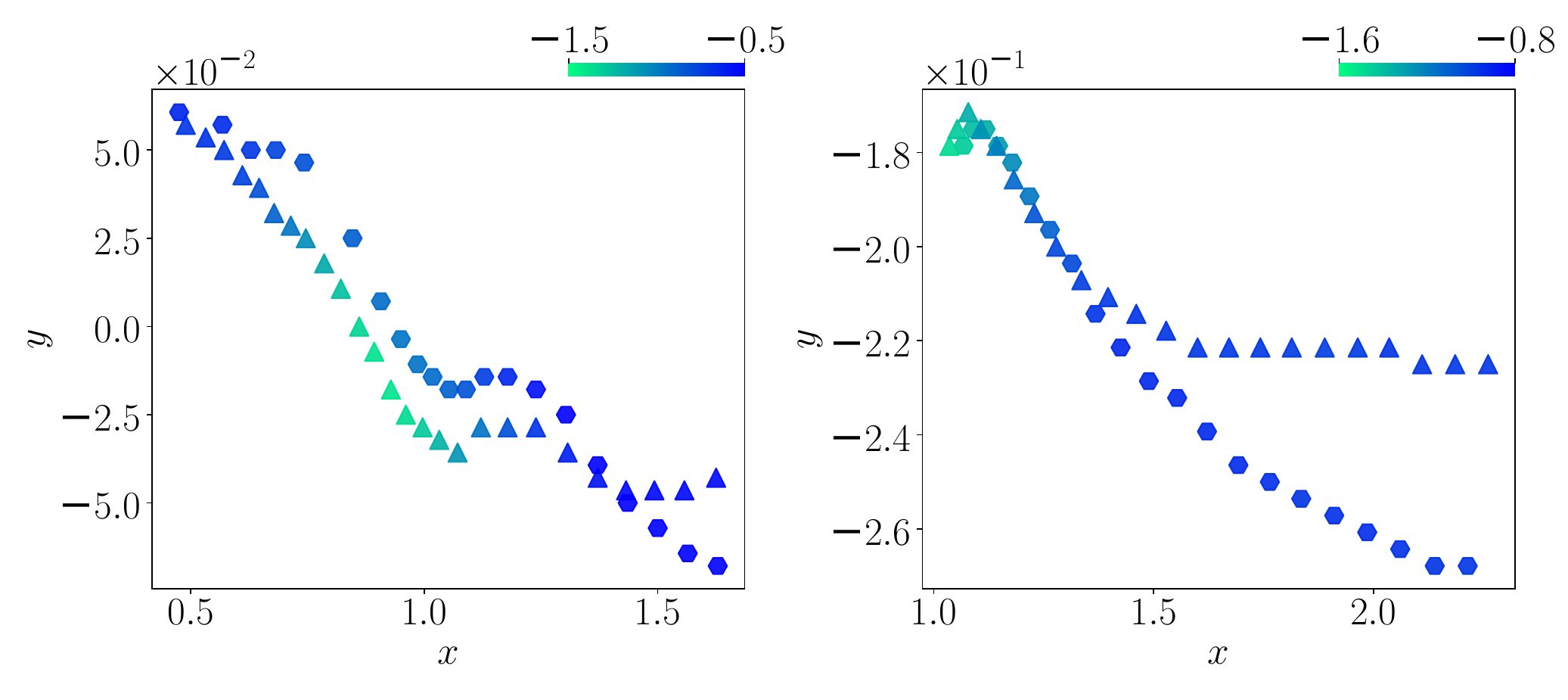}
    \caption{
    Trajectory of the pressure minimum associated with the leading-edge vortex (left) and trailing-edge vortex (right) for the unactuated flow (hexagonal markers) and flow with lift-optimal actuation (triangular markers).
    Markers are colored according to pressure. 
    }
    \label{fig:PminBL}
\end{figure}

The change in phase between LEV and TEV formation is also reflected 
in the distance between successive LEVs, as well as the LEV and the TEV from the earlier vortex-shedding cycle. 
Figure \ref{fig:VortMinMax} shows vorticity contours for the unactuated flow in the top row and the actuated flow in the bottom row. The time instances for the plots are chosen such that an LEV
is seen close to the trailing edge in the left column and a 
TEV
is seen near the trailing edge in the second column. Locations of representative vortex centers are indicated in yellow (LEV) and salmon (TEV) markers. 
The yellow markers indicate locations where vorticity is negative and the $x$- and $y$-components of velocity flip signs along the vertical and horizontal respectively. 
The salmon markers are at locations of positive vorticity with similar variations of the velocity components. This criteria yields multiple markers associated within a nominal vortex. For analysis purposes, the marker taken to represent each vortex within the set of markers associated with that vortex is the left-most and right-most marker for the downstream and upstream vortex, respectively.
The nondimensional streamwise distance between the LEVs is 0.58 for the actuated flow and 0.63 for the unactuated flow. The shorter distance between the vortices is also true for the separation between the LEV and TEV (0.36 for the actuated flow and 0.45 for the unactuated flow). 

Having analyzed the changes in the pressure magnitude and vortex-shedding characteristics caused by actuation, we investigate variations in the trajectories of the vortices, if any, that arise due to actuation. 
Figure \ref{fig:PminBL} shows trajectories of the center of the LEV (left plot) and the TEV (right plot), identified based on the 
grid point
of the pressure minimum in the vortices. 
The markers are colored by the magnitude of pressure as indicated by the colorbars.
There is cycle-to-cycle variation in the trajectories of the LEV and TEV in the actuated case. 
The trajectories shown in the figure are determined from flow snapshots between $t \approx 3.2$ and $t \approx 5.7$ (cycle-to-cycle variations are discussed next).
In the actuated case, the LEV is closer to the airfoil surface as it advects towards the trailing edge (the trailing edge is at $x\approx 0.95$).
As already discussed, the pressure magnitude of the LEV is higher in the case of the actuated flow.
Generally, increased proximity of the LEV to the airfoil surface yields a smaller pressure gradient between the center of the vortex and the airfoil surface. 
The higher pressure magnitude of the LEV in the presence of actuation, along with its closer proximity to the airfoil surface, both contribute to higher lift.
Downstream of the airfoil, the LEV in the presence of actuation advects at a higher $y$-coordinate value (see $x>1.5$).
The TEV also advects at a higher $y$-coordinate value in the presence of actuation (right subfigure).
The advection of the TEV at a higher $y$-coordinate affects the pressure on the pressure side of the airfoil: the pressure would be higher if the TEV advects closer to the suction surface than the pressure surface.
The higher $y$-coordinate of the TEV center in the actuated case is also evident close to $x=1$.
\begin{figure}
    \centering
    \includegraphics[width=0.99\textwidth]{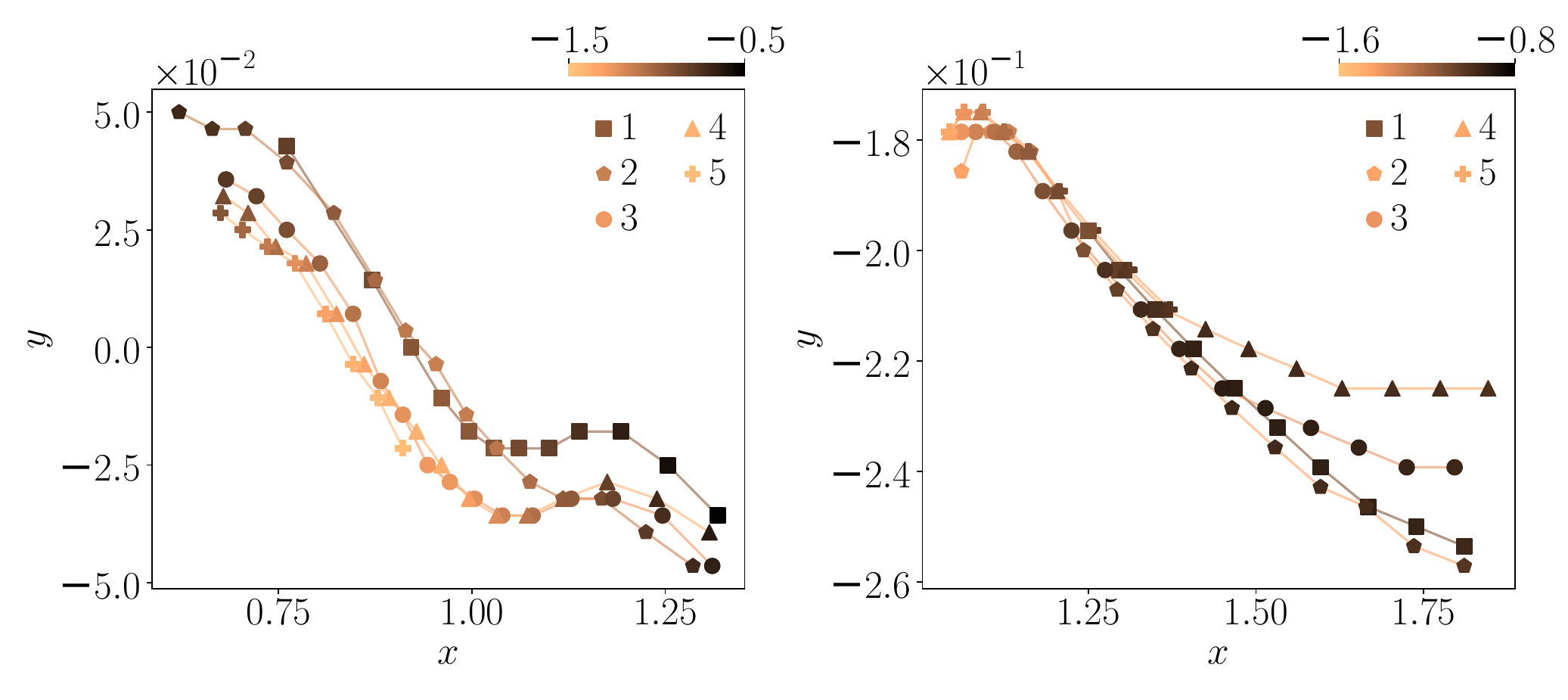}
    \caption{
    Trajectory of the pressure minimum associated with the LEV (left) and TEV (right). 
    The various markers represent the different vortex-shedding cycles indicated by the legends. 
    The markers are colored by the pressure at the location of the vortex center.
    The markers in the legends are colored as per the minimum pressure over the cycle they represent.
    }
    \label{fig:LEVCycle}
\end{figure}

The cycle-to-cycle variation in the trajectory of the LEV over the suction surface and the TEV in the wake are shown in figure \ref{fig:LEVCycle}.
The end of each ``cycle" is characterized by the advection of the LEV downstream of $x=1.3$ (the trailing edge of the airfoil is at $x \approx 0.95$). 
The LEV gets closer to the airfoil surface during the later cycles.
Also evident is the higher pressure magnitude of the LEV over the suction surface as the cycle number increases, as seen from the colors of the markers representing the trajectory of the LEV as well as the color of the markers in the legend.
The difference in the trajectory of the TEV, across the different actuation cycles, is slight except that
the TEV downstream of $x=1.4$ in the fourth cycle is at higher $y$-coordinate values as compared to the first cycle.
The $y$-coordinate values of the last few markers of the third cycle are also higher than that of the earlier cycles.
The higher $y$-values of the fifth cycle are also noticeable even though the cycle is not entirely captured because of the limited optimization window.

\begin{figure}
    \centering
    {\includegraphics[width=0.74\textwidth]{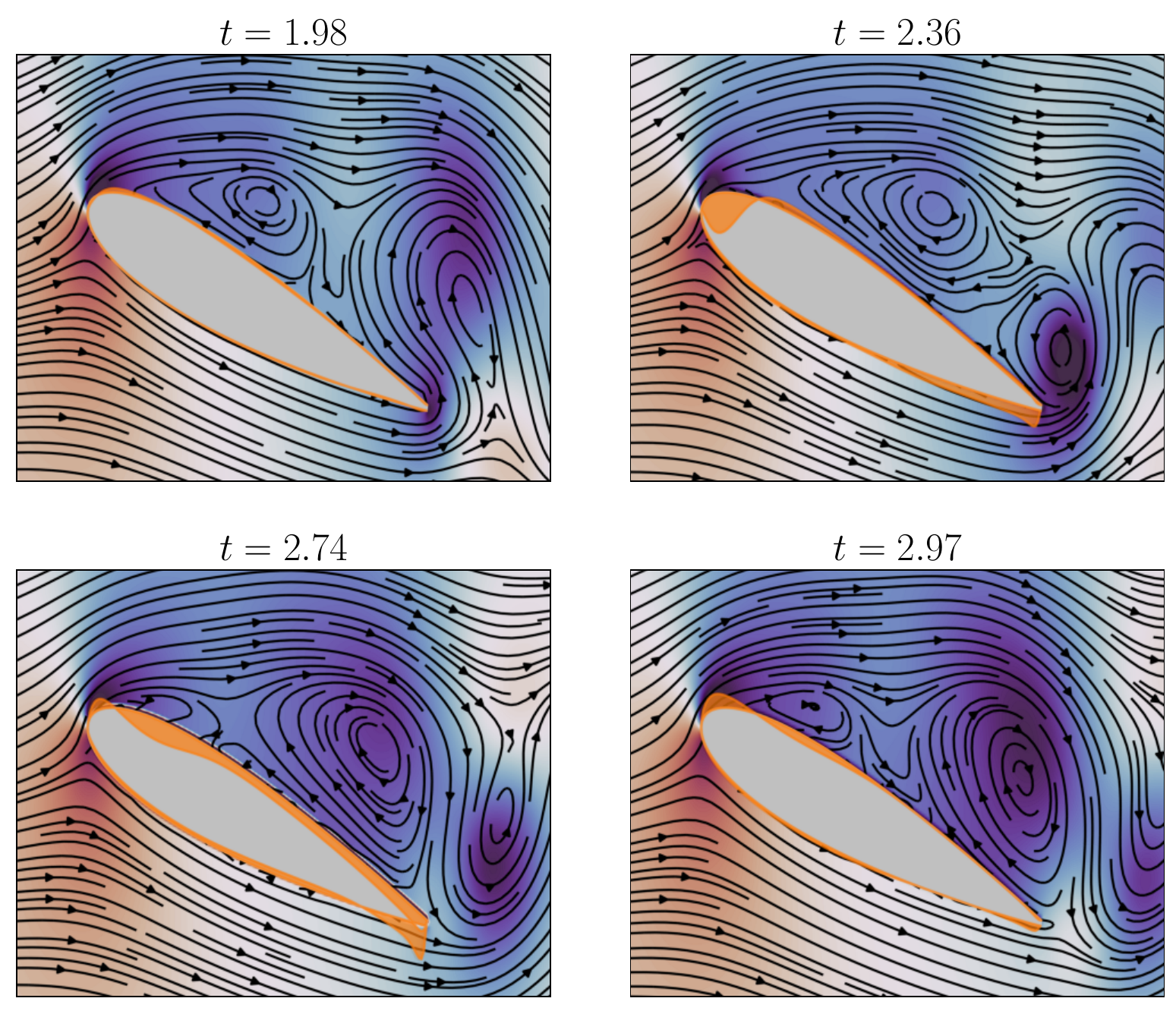}} 
    \caption{ 
    Flow snapshots and lift-optimal actuation over an actuation cycle. 
    Each snapshot comprises of pressure contours, flow streamlines and actuation on the airfoil surface (scaled for easy visualization).
    Contour limits are the same as in figure \ref{fig:BLSnaps}.
    }
    \label{fig:ClAP}
\end{figure}

We next discuss the effect of the velocity boundary condition imposed by actuation. 
In our earlier work, the role of non-optimal periodic, traveling-wave actuation in modifying the curvature of the streamlines near $s_{max}$ and the ensuing changes in the 
long-term vortex-shedding process was discussed  \citep{thompson2022surface}. 
However, there was no discussion of the effect of actuation near the trailing edge.
Likewise, changes in flow features due to actuation near the leading edge on the pressure side were not discussed since actuation was restricted to the suction surface.
In the rest of this section, we discuss the role of actuation near the leading edge on the suction and pressure sides, and close to the trailing edge on the suction and pressure sides separately.
Since actuation has the same qualitative attributes in all actuation cycles (see the temporal variation of actuation in figure \ref{fig:vNTemporal}),
we only analyze the flow between $t\approx 2.0$ and $t=3.0$. 
The time window is chosen such that the magnitude and dynamics of actuation as well as the resulting changes in the vortex-shedding process are substantial.
%

To investigate the interaction between actuation and the vortex-shedding process, figure \ref{fig:ClAP} shows contours of the coefficient of pressure, $C_p$, and the streamlines at four time instances within $t\in[1.9,3.0]$. 
The $C_l$ values at the time instances of the snapshots are indicated in figure \ref{fig:TH}, as well as in the temporal variation of $C_p$ at $s_{vmax}$ in figure \ref{fig:LECp}.
Of the four snapshots, $t=2.36$ is close to the first trough in the temporal variation of $C_p$ at $s_{vmax}$, 
$t=2.74$ is when the actuation on the aft portion of the suction surface has the lowest actuation velocity, and $t=2.97$ coincides with the peak of positive actuation at $s_{vmax}$ (see figure \ref{fig:vNTemporal}).

At $t=1.98$ when the LEV from  the current cycle is around the mid-chord, negative actuation starts to appear near the leading edge on the suction surface. 
The negative actuation near $s_{vmax}$ is pronounced at the next time instance ($t =2.36$). 
Similar to the earlier snapshot, the region of negative actuation is followed by a downstream region of positive actuation (also see the spatial profile of actuation in figure \ref{fig:vNWave}). 
This time instance coincides with the first trough in the temporal variation of $C_p$ at $s_{vmax}$.
The trough could be attributed to the larger curvature of the streamlines as the flow curls around the leading edge:
since the forward stagnation point (indicated by maximum pressure) is on the pressure side of the airfoil, the oncoming flow has to curl around the leading edge as it flows on to the suction surface.
The boundary condition, imposed by negative actuation on the suction surface, forces the flow to curl through a greater angle which is accompanied by a drop in pressure.
The actuation around $s_{vmax}$ eventually becomes positive again, to aid the formation of the new LEV.

At $t=2.74$, positive actuation appears near the leading edge. 
The positive actuation, in conjunction with the negative actuation downstream, results in a region of negative vorticity and promotes formation of the new LEV.
Between the regions of clockwise streamlines associated with the old and new LEVs is a region of counter-clockwise streamlines close to the airfoil surface.
Away from the airfoil surface
is a region of reduced pressure magnitude and concave streamlines, which is due to the negative actuation around $s_{vmax}$ at earlier time instances.
In the next snapshot at $t = 2.97$, the positive actuation near $s_{vmax}$ increases in magnitude and aids the roll-up of the new LEV.
The streamlines on the upstream side of the LEV point along the outward normal of the airfoil surface in this region while those on the downstream side point towards the airfoil surface (consistent with the direction of the actuation).
After $t=2.97$, the LEV continues to grow as the actuation at $s_{vmax}$ decreases.
The second trough in the $C_p$ variation occurs shortly after $t=2.97$ when the new LEV grows in strength.

As discussed in section \ref{sec:actuation}, 
substantial actuation is seen near the trailing edge in the case of lift-optimal actuation.
At $t = 2.36$, positive actuation is seen near the trailing edge. 
At $t = 2.74$, the region of positive actuation shifts towards the trailing edge while an increase in the spatial maximum occurs. 
The variation of positive actuation near the trailing edge between $t=2.36$ and $t=2.74$ can be traced to the contours of actuation on the pressure side in figure \ref{fig:vNSurf}.
The time instances of positive actuation near the trailing edge coincide with the roll-up and advection of the TEV. 
At $t=2.97$ when the TEV advects downstream of the airfoil
and the LEV reaches near the trailing edge, the actuation there becomes negligible.

\begin{figure}
    \centering
    {\includegraphics[width=0.92\textwidth]{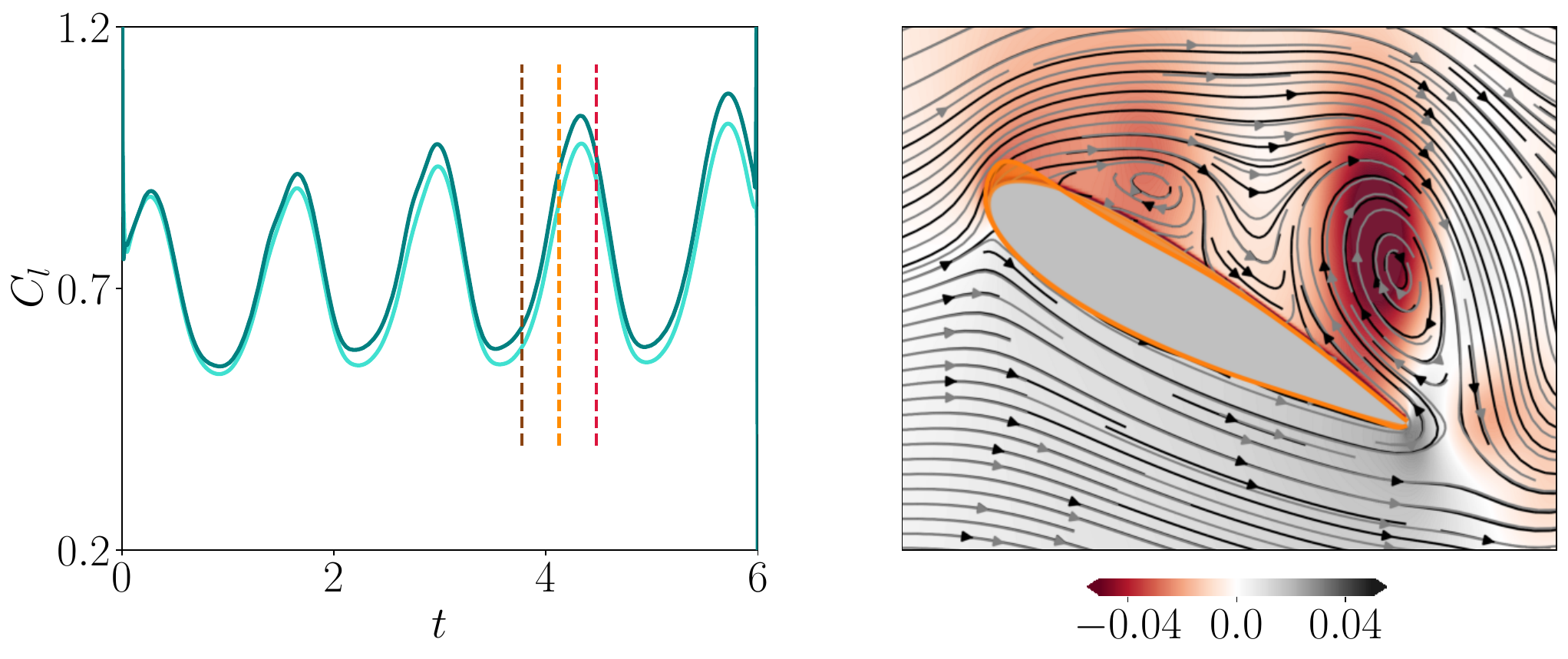}} 
    \caption{
    Influence of actuation near the trailing edge. (a): $C_l$ variation with optimal actuation (teal) and with the optimal actuation, except for a zeroed-out portion over the aft $30\%$ of the suction and pressure sides of the airfoil surface (turquoise).
    (b): Contours of the difference in coefficient of pressure between fully optimal actuation and actuation zeroed out near trailing edge. The snapshot is at $t = 4.5$ (indicated by the crimson dashed line in the left plot). Streamlines of the flow with fully optimal (black) and with TE-zeroed optimal actuation (gray) are superimposed. 
    }
    \label{fig:NoTEI}
\end{figure}
\begin{figure}
    \centering
    \hspace{0.05\textwidth}
    {\includegraphics[width=0.87\textwidth]{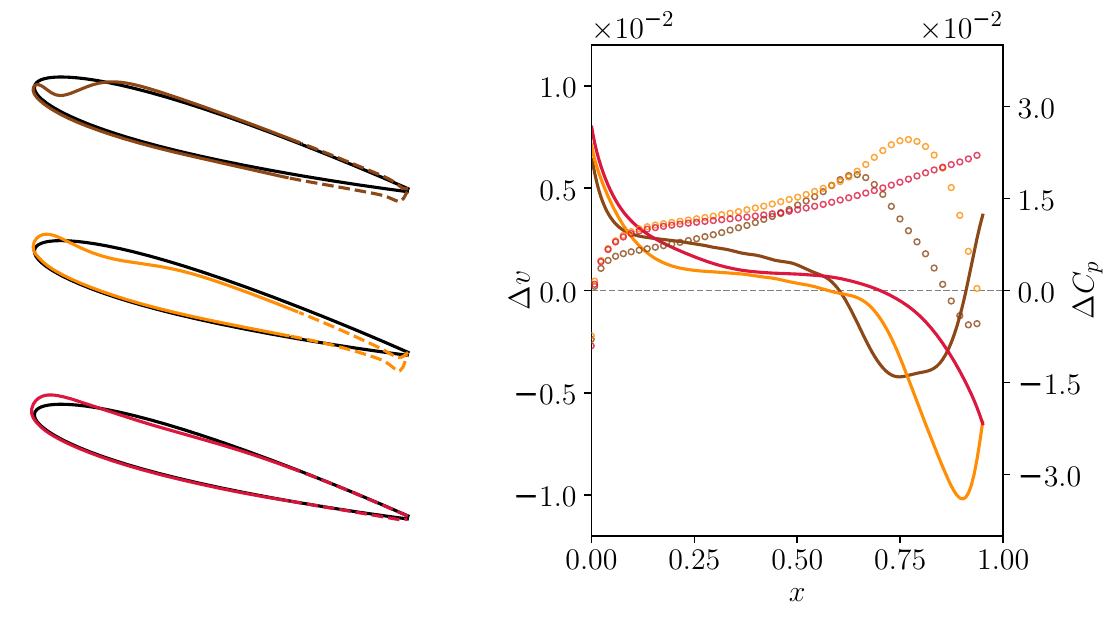}}
    \caption{
    Left: schematics of actuation at the time instances of the vertical lines in figure \ref{fig:NoTEI} (dashed lines indicate the region for which actuation is zeroed out in the modified actuation case). The time instances are indicated by the matching color in figure \ref{fig:NoTEI}. 
    Right: At the same time instances as in the left plot, difference in the 
    $y$-component of velocity (lines) and pressure (markers) between fully optimal and TE-zeroed actuation (explained in the main text). The pressure and velocity values are taken approximately $0.05$ dimensionless length units away from the pressure surface. 
    }
    \label{fig:dVNoTE}
\end{figure}
To investigate the effect of actuation near the trailing edge, we compare the flow field and the aerodynamic performance with and without actuation near the trailing edge. 
For the case without actuation at the trailing edge, the optimal actuation over the aft $30 \%$ of the airfoil surface on both the suction and pressure sides (measured from the trailing edge) is zeroed out.
Actuation on both the suction and pressure sides simultaneously is nullified since the actuation in these regions is related
(see section \ref{sec:actuation}), 
implying that actuation on either surface is meant to elicit the same change in the flow.
Similarities between the gradient on the suction and pressure sides of the airfoil near the trailing edge were also observed in section \ref{sec:gradients}.
Zeroing actuation on both surfaces simultaneously therefore facilitates easier identification of the influence of trailing-edge actuation. 

Figure \ref{fig:NoTEI} compares results between fully optimal actuation and optimal actuation with the trailing-edge effect zeroed out.
Zeroing out the actuation near the trailing edge reduces the mean lift by roughly 
$4\%$. 
The right plot shows contours of the pressure field of the modified actuation subtracted from the one with optimal actuation. 
The fully optimal actuation case is associated with higher pressure on the pressure side and lower pressure on the suction side.

To further probe the effect of trailing-edge actuation on the streamline curvature and the ensuing pressure on the pressure side, we show in figure \ref{fig:dVNoTE} the difference in the $y$-component of the velocity (indicated here as $v$) 
and pressure between modified and fully optimal actuation.
At the first of the three time instances (brown curves), the $v$-velocity at the trailing edge is higher for the optimal actuation case.
Slightly upstream of the trailing edge, a trough of negative $\Delta v$ is seen. 
The trough could be linked to positive actuation (along negative $y$).
From the $\Delta C_p$ at this time instance, the pressure in the presence of full actuation is higher everywhere except very close to the trailing edge and very close to the leading edge.
At the next time instance shown in the figure (orange curves), the trough of negative velocity 
advects downstream 
and has a higher magnitude. 
The higher negative trough implies flatter streamlines, accompanied by an increased pressure with full actuation as corroborated by the $\Delta C_p$ plot.
At the final time instance (crimson curves), the actuation near the trailing edge is near zero (third plot, left column) while the region of negative $v$-velocity is seen to advect further downstream. 
Similar to the earlier time instances, $\Delta C_p$ is higher with fully optimal actuation.
The $\Delta C_p$ variation at this time instance is a reflection of the contour plot in figure \ref{fig:NoTEI}.

The effect of actuation near the trailing edge can be interpreted as a shift in the rearward stagnation point towards a lower $y$-coordinate value (c.f., the different streamlines in figure \ref{fig:NoTEI}). 
This shift in the stagnation point in turn influences the streamline curvature and pressure distribution on both the suction and pressure surfaces. 
On the suction surface, the influence of a lower rear stagnation point is a larger curvature of the streamlines between the leading edge and the trailing edge, 
while the opposite is true for the pressure side.
The lower pressure on the suction surface is evident from the contours in figure \ref{fig:NoTEI}.
The higher pressure on the pressure surface with trailing-edge actuation is evident from figure \ref{fig:dVNoTE}.

\begin{figure}
    \centering
    {\includegraphics[width=0.75\textwidth]{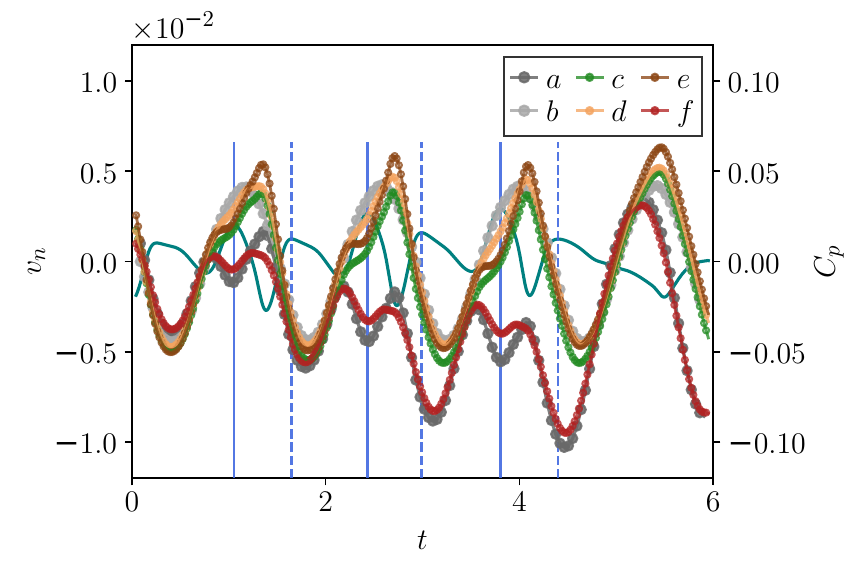}}
    \caption{
    Temporal variation of $C_p$ near $s_{vmax}$ (shifted along $y$-axis by $+0.7$ for visualization purposes).
    $a$: with full actuation,
    $b$: unactuated,
    $c$: with actuation over $50\%$ of the suction surface zeroed out,
    $d$: case $c$ with $30\%$ on either side of the trailing edge additionally zeroed out, 
    $e$: same as $d$, with actuation scaled by two,
    $f$: actuation over $50 \%$ of the suction surface and $30\%$ of the suction and pressure surfaces towards the trailing edge (inverse of case $d$). The additional
    teal line shows actuation at $s_6$.
    Blue vertical lines indicate relevant peaks in the actuation at $s_6$, solid (dashed): first (second) peak in actuation during an actuation/vortex-shedding cycle. 
    }
    \label{fig:LEp}
\end{figure}

We now turn our attention to actuation on the pressure side near the leading edge.
It was discussed as part of the analysis of the actuation profile in section \ref{sec:actuation} that
similarities exist
between the actuation on the pressure side of the leading edge and that around $s_{vmax}$. 
Additionally, the actuation over the aft portion of the suction surface was found to be similar to the actuation near the leading edge on the pressure side.
Among the snapshots in figure \ref{fig:ClAP}, $t=2.36$ and $t=2.74$ are time instances when the actuation at $s_6$ and $s_2$ have near-maximum and near-minimum values respectively (also see figure \ref{fig:vNTemporal}).
At $t=2.36$, concavity of streamlines is attained around $s_{vmax}$. 
However, the role of the actuation at $s_6$ at this time instance is not obvious.
Similarly, at $t=2.74$ the possible synergy between the actuation at $s_6$ and $s_{vmax}$ in the formation of the new LEV is not clear.

To isolate the effect of actuation over the fore portion of the pressure surface and the aft portion of the suction surface from the large-magnitude actuation around $s_{vmax}$ and $s_{TE_p}$, four cases with modified actuation are considered, and the variation of $C_p$ at $s_{vmax}$ for each case in addition to the unactuated flow, and flow with unmodified actuation are plotted in figure \ref{fig:LEp} (see figure caption for description of cases).
Actuation $e$, which is the same as case $d$ with the actuation scaled by two, is considered to study the effect of actuation at the less influential locations when the actuation magnitude there is no longer negligible.
With full actuation (case $a$), the peaks in the $C_p$ variation roughly align with the troughs in the actuation at $s_6$ and the troughs in the $C_p$ variation roughly align with the peaks in the actuation (indicated by blue vertical lines), suggesting that the actuation on the pressure side of the leading edge plays a role in the pressure variation and the related vortex formation process.

The results when actuation near $s_{vmax}$ is zeroed out (cases $c$--$e$) reinforce the prior results that actuation near $s_{vmax}$ is the dominant contributor to lift benefits: with actuation near this region zeroed out, the pressure signal is much closer to the unactuated profile (case $b$). We focus here on the role of actuation at these secondary locations. For cases $c$--$e$, an increase in pressure relative to the fully optimal case (case $a$) occurs because of the much smaller actuation magnitude at the other locations
and the smaller influence of the actuation there.
In case $c$ where actuation at the trailing edge still exists, a drop in $C_p$ relative to the unactuated case (case $b$) occurs in response to the actuation peaks. This drop is particularly apparent near the peaks and troughs of the pressure signals. 
The drop in pressure is reminiscent of negative actuation at $s_{vmax}$ which leads to regions of higher pressure between consecutive leading-edge vortices (see $t= 2.36, 2.74$ and $2.97$ in figure \ref{fig:ClAP}).
The $C_p$ peak of the unactuated flow is shifted near the time instance of an actuation trough (between the two actuation cycle peaks indicated by the blue lines).
The peak decreases in magnitude over the course of the window due to the increasing strength of the LEV.

In case $d$ where the actuation at the trailing edge is also zeroed out, similar qualitative behavior as case $c$ is seen but to a much smaller extent; i.e., the difference between the actuated case and the unactuated case is smaller.
In case $e$, the drop in $C_p$ associated with the first actuation peak as well as the increase in $C_p$ due to the trough in actuation (between the solid and dashed lines) are more evident than for case $d$.
This demonstrates that an increased actuation amplitude only serves to amplify many of the effects seen in actuation at these less important surface locations. 
However, the local minimum in $C_p$ near the second actuation peak is not as pronounced as in case $c$.
Similarly, the cycle-to-cycle decrease in the $C_p$ peak due to the strengthening of the LEV is negligible. 

For cases $c$--$e$, the correspondence between actuation troughs and local maxima in $C_p$ is evident.  
When the actuation over the fore portion of the pressure surface and part of the aft portion of the suction surface are zeroed out (case $f$), the peaks and troughs seen in the fully actuated case are less pronounced, in line with the observations in cases $c$--$e$. These observations demonstrate that the secondary benefits caused by actuation at locations away from $s_{vmax}$ are nonetheless synergistic with actuation near $s_{vmax}$, providing important benefits to the overall lift behavior. 
\subsection{Analysis for Drag-Optimal Actuation}
\label{sec:CdAnalysis}
\begin{figure}
    \centering
 {\includegraphics[width=0.9\textwidth]{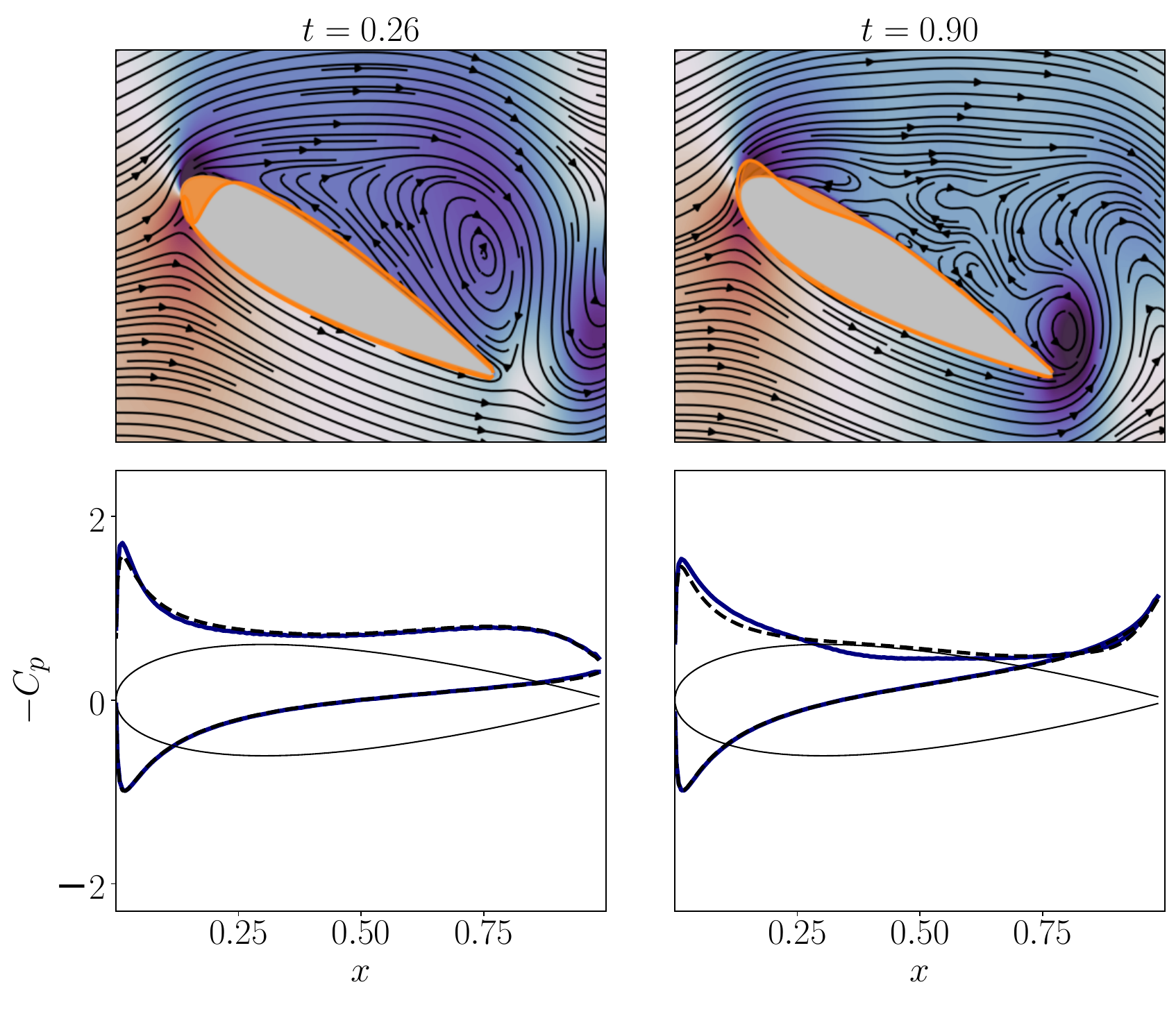}} \hspace{0.5em} 
    \caption{Analog of figure \ref{fig:PMinMax}, with drag-optimal actuation. }
    \label{fig:CdPMinMax}
\end{figure}

Unlike the lift-optimal case where 
vortex shedding continues to exist, figure \ref{fig:TH} suggests that drag-optimal actuation 
yields a flow field with reduced unsteadiness than in the unactuated case, reminiscent of the known unstable equilibrium state that is latent within the dynamical system.
From figure \ref{fig:vNTemporal}, the drag-optimal actuation has noticeable frequency content only in the first half of the optimization window which is also the time duration over which the undulations in the aerodynamic coefficients decay to small values (figure \ref{fig:TH}).
In this section, we discuss the interplay between the actuation and the vortex-shedding process over the first half of the optimization window which transitions the flow to a state conducive to a steady decrease in drag over the rest of the window.

Before analyzing the flow phenomena leading up to the substantial weakening of the vortex-shedding process,
analogous to figure \ref{fig:PMinMax}, figure \ref{fig:CdPMinMax} 
shows snapshots in the first row, and the surface pressure distribution with and without drag-optimal actuation at time instances of maximum and minimum $C_l$ (corresponding to the first peak and trough of $C_l$ in figure \ref{fig:TH}) in the second row.
The analogous snapshots of the unactuated flow correspond to $t=0.22$ and $t=0.90$ in figure \ref{fig:BLSnaps}.
At the time instance of the figure on the left, the strong negative actuation near $s_{vmax}$ influences the nearby velocity distribution and, through the curvature of the streamlines, the pressure distribution such that the formation of a new LEV is opposed.
At a similar time instance of the vortex-shedding cycle in the unactuated case ($t=0.22$ in figure \ref{fig:BLSnaps}), the initial stages of LEV formation in the form of an inflection in the streamlines appears on the suction surface near the leading edge. 
With drag-optimal actuation, 
the inflection appears at a downstream location (figure \ref{fig:CdPMinMax}).
From the surface $C_p$ distribution, the pressure close to the leading edge is higher in the presence of actuation while the pressure magnitude over most of the suction surface is smaller (albeit marginally). 
The higher pressure magnitude at the leading edge could be attributed to the larger angle by which the oncoming streamlines have to turn through while navigating around the leading edge, under the influence of the negative actuation on the suction surface.

At the time instance of the snapshot on the right, the actuation near $s_{vmax}$ initiates the formation of a new LEV. 
The role of this LEV and the subsequent changes in the vortex-shedding characteristics will be discussed in the 
subsequent paragraphs.
The LEV near mid-chord of the suction side seen in the unactuated case ($t=0.90$ in figure \ref{fig:BLSnaps}),
is far less pronounced in the actuated case due to the weakening of the LEV formation process caused by actuation at earlier time instances. 
The $C_p$ distribution on the suction surface has a lower magnitude over the mid-chord because of the mitigation of LEV formation. 
The pressure close to the leading edge is higher in the actuated case because a new LEV is being initiated by actuation.

\begin{figure}
    \centering
    \includegraphics[width=0.9\textwidth]{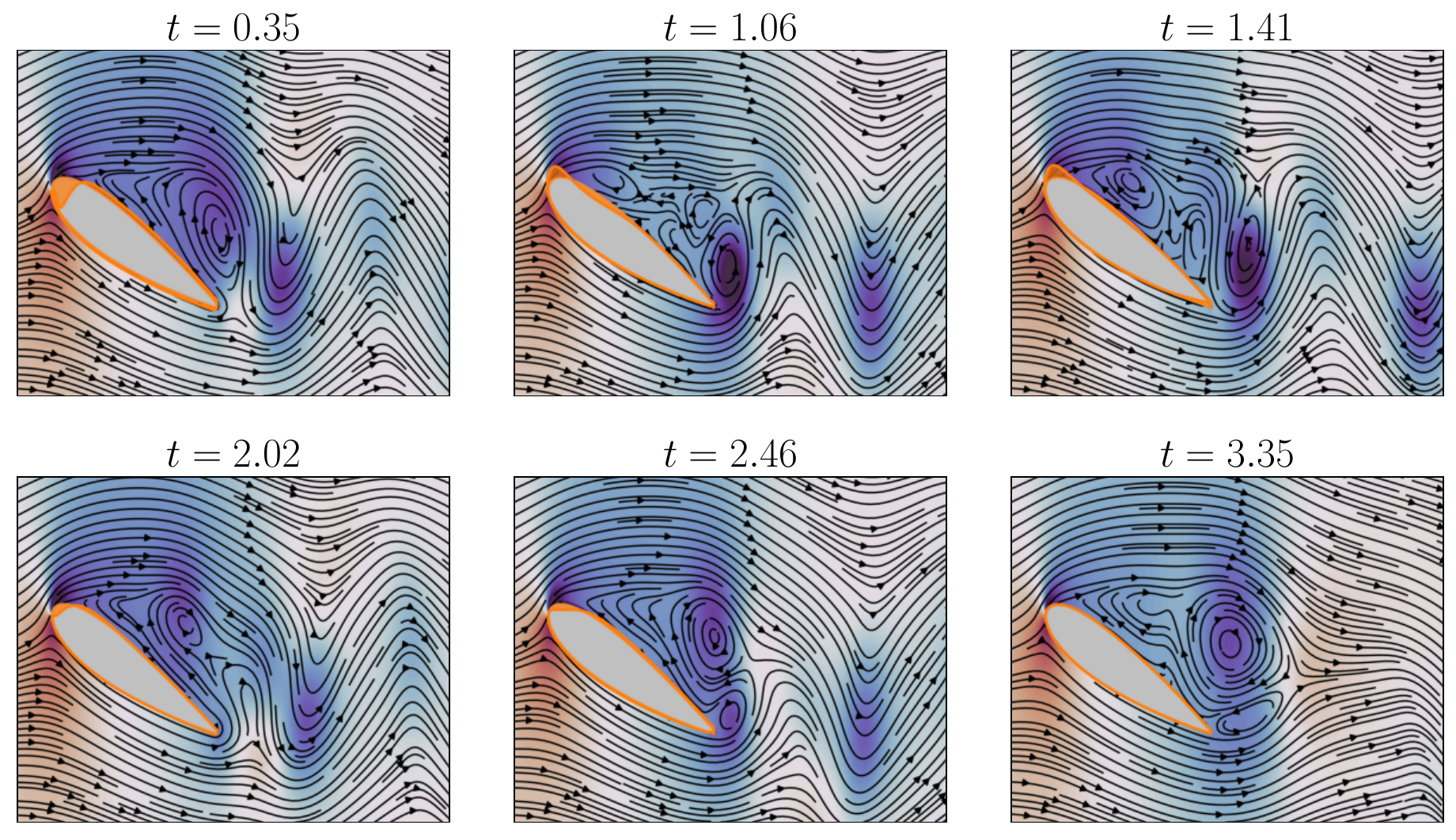}
    \caption{Flow snapshots depicting formation of LEV-TEV pair near the trailing edge. Each snapshot shows contours of coefficient of pressure, streamlines and drag-optimal actuation on the airfoil surface. }
    \label{fig:DragFormation}
\end{figure}
\begin{figure}
    \centering
    \includegraphics[width=0.75\textwidth]{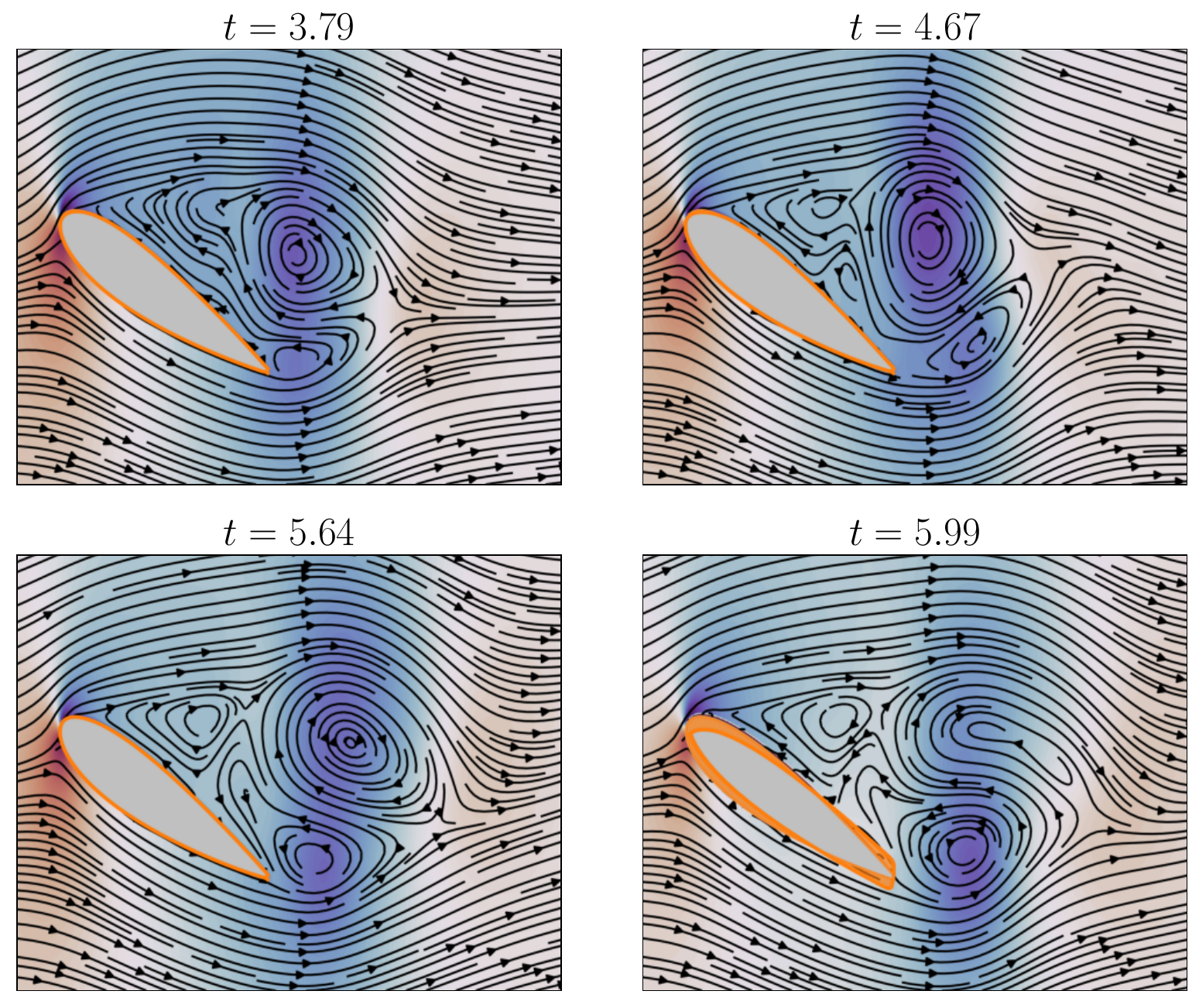}
    \caption{Advection of LEV-TEV pair and secondary vortex shedding on the suction surface.}
    \label{fig:DragAdvection}
\end{figure}

Figures \ref{fig:DragFormation}--\ref{fig:DragAdvection} demonstrate that this delayed LEV formation results in the LEV and TEV reaching the trailing edge around the same time and advecting downstream as a pair.
We will discuss the actuation profile and vortex interactions leading to the formation of the LEV-TEV pair in figure \ref{fig:DragFormation}, then the advection of this pair as well as additional vortex formation processes shown in figure \ref{fig:DragAdvection}. 
At $t=0.35$, the LEV from the unactuated flow
is seen close to the trailing edge. 
Note that this time instance is slightly later than the one corresponding to the first snapshot of the actuated flow in figure \ref{fig:CdPMinMax}.
At a similar time instance in the unactuated flow 
(see $t=0.22$ in figure \ref{fig:BLSnaps}),
an inflection in the shear layer is seen close to the leading edge which at later time instances evolves into a new LEV. 
In the current case however, the large magnitude of negative actuation near $s_{vmax}$ limits the curvature of the flow flowing past $s_{vmax}$.
The inflection in the shear layer and its interaction with the existing LEV occurs downstream of the region of negative actuation. 
It is also seen that positive actuation exists downstream of the negative actuation, like in the lift-optimal case ($t=2.36$ in figure \ref{fig:ClAP}). 

At a later time instance ($t=1.06$), the negative peak of actuation travels downstream of $s_{vmax}$ while reducing in magnitude.
Closer to the leading edge actuation is now positive which, in conjunction with the negative downstream actuation, aids the formation of a clockwise vortex. 
The actuation at this time instance results in a delayed LEV formation which has implications for the subsequent interaction of the LEV and TEV near the trailing edge. 
At this time instance, TEV formation occurs near the trailing edge. 
Between the LEV forming near the leading edge and the TEV at the trailing edge are regions of complex structures, most noticeable among which are regions of clockwise and anti-clockwise streamlines just upstream of the TEV.
As the TEV advects downstream at $t=1.41$, the clockwise and anti-clockwise streamlines from the previous snapshot resolve into a clockwise vortex. 
The LEV near the leading edge also grows in size and a region of anti-clockwise vorticity separates the two vortices.
As the downstream of the two clockwise vortices advects closer to the trailing edge, a trailing-edge vortex forms as seen at $t=2.02$ and $t=2.46$.
The formation of the TEV and the advection of the upstream LEV close to the trailing edge occur at similar time instances.
Following the occurrence of the LEV and TEV at the trailing edge, the two interact closely and move downstream as a vortex pair. 

The advection of the vortex pair and subsequent vortex formation processes are shown in figure  \ref{fig:DragAdvection}. 
Just as in the unactuated case, the formation of a clockwise vortex near the leading edge (see $t=4.67 - 5.99$) and an anti-clockwise vortex near the trailing edge (see $t=4.67$) occurs. 
However, the formation of the anti-clockwise TEV occurs on the suction surface 
instead of the wake of the airfoil.
At $t=3.35$ (in figure \ref{fig:DragFormation}), an incomplete clockwise vortical structure is seen near the leading edge. 
In the case of the unactuated flow, the advection of the LEV past the trailing edge is followed by TEV roll-up and advection. 
It is the formation of the TEV that mitigates the coalescence of the partially rolled-up LEV near the leading edge and the LEV downstream of the airfoil.
In the current case, the LEV near the trailing edge and the partially rolled-up structure near the leading edge coalesce, as seen at $t=3.79$.

At a later time instance, $t=4.67$, the clockwise vortex grows in size and strength (indicated by the larger pressure magnitude). The LEV also moves slightly upwards as it advects away. 
The upward motion of the LEV is beneficial to drag in that the lower pressure of the vortex has less of an effect on the surface pressure. 
Also seen at $t=4.67$ are the partial formations of a new clockwise vortex near the leading edge and an anti-clockwise vortex near the trailing edge.
The distinction between the secondary vortex-shedding in the current case and the usual vortex-shedding in the unactuated case, as already mentioned, is the formation of the anti-clockwise vortex over the aft portion of the suction surface instead of the pressure side of the trailing edge. 
At the next time instance, $t=5.64$, the secondary anti-clockwise vortex coalesces with the primary one. 
At this time instance, the primary LEV reduces in pressure magnitude while the primary TEV becomes stronger. 
The LEV also appears downstream of the TEV, in contrast to the configuration at $t=4.67$.
At the next time instance, $t=5.99$, the primary LEV is seen to break-up and the primary TEV grows in size. 
The abrupt change in actuation magnitude on the airfoil surface is due to the starting transients of the adjoint solve at the end of the optimization window.

The snapshots of figures \ref{fig:DragFormation} and \ref{fig:DragAdvection} can be used to explain the $C_d$ variation seen in figure \ref{fig:TH}.
The round markers in the drag plot correspond to the time instances of the snapshots. 
At $t=0.35$, the drag is slightly lower than that of the unactuated flow due to the effect of the negative actuation near $s_{vmax}$ in reducing streamline curvature and the consequent pressure magnitude. 
At the next time instance, the absence of a leading-edge vortex on the suction surface close to the trailing edge results in a drop in drag relative to the unactuated flow. 
Over the next two time instances, the drag is seen to increase and then decrease, in line with the growth of the LEV. 
At $t=2.46$, when the TEV grows, it reduces the pressure on the pressure side which in turn has a drag mitigating effect. 
Between $t=3.35$ to $t=5.99$ as the LEV-TEV pair advect further away from the airfoil, the pressure on the suction surface has a low magnitude. 
This is in part because the LEV-TEV pair force the streamlines over most of the suction surface to be flatter, which in turn reduces the pressure gradients there. 

\section{Conclusions}
In this article, we used numerical optimization to determine optimal surface actuation for the flow past a NACA0012 airfoil at a low Reynolds number of $Re=1000$ and a post-stall angle of attack of $\alpha= 15^{\circ}$. 
The iterative gradient-based optimization procedure involved minimizing a cost functional using the gradient of the cost with respect to the design-variable vector. 
The design variable was taken as the normal actuation at every discretized point on the airfoil surface, at every time instance of the optimization window.
The gradient of the cost was computed via the adjoint of the governing equations.
Optimization for lift- and drag-based costs was carried out separately.
The actuation magnitude was limited to values similar to our earlier work \citep{thompson2022surface}, distinguishing the optimal actuation profiles  from those of \citet{paris2023reinforcement} where flow reattachment with simultaneous improvements in lift and drag was possible with a higher actuation magnitude.

At the actuation magnitude and optimization window of interest, lift and drag benefits were not concurrently achieved.
Further, the time history of the aerodynamic coefficients with lift- and drag-optimal actuation suggested distinct mechanisms by which the two performance goals were realized: with lift-optimal actuation, vortex shedding persisted (albeit non-periodic) while in the drag-optimal case a considerable decrease in undulations accompanying an overall reduction in the values of $C_l$ and $C_d$ was seen. 
While the lift-optimal actuation led to an increase in drag, the time-averaged $C_l/C_d$ increased compared to the unactuated case.
Also seen were time instances where the drag was lower than the unactuated case (figure \ref{fig:TH}).
On the other hand, with drag-optimal actuation, both time-averaged $C_l$ as well as $C_l/C_d$ were lower than for the uanctuated case.
With lift-optimal actuation, the time-averaged lift improved by roughly $10 \%$ while drag-optimal actuation led to a drop of approximately $13\%$ 
in time-averaged drag.

Given the substantial distinction in the flow phenomena resulting from
the two optimal actuation 
profiles, the gradients of the unactuated flow for the two costs were compared to motivate the distinct actuation behavior and associated flow changes.
For both costs, the gradients were largest upstream of the location of the maximum $y$-coordinate of the airfoil ($s_{max}$) and on the pressure side of the trailing edge.
The gradients at $s_{vmax}$ and $s_{TE_p}$ (see figure \ref{fig:ClCdGrad}) were related to the formation and advection of the leading-edge vortex (LEV) and trailing-edge vortex (TEV) in the absence of actuation.
The temporal variation of the gradient at $s_{vmax}$ in relation to the pressure there had opposite characteristics for the two costs.
Similarly, the gradient at $s_{TE_p}$ for the two costs had opposite attributes. 
In general, the gradients for drag optimization were opposite in sign and shifted with a time- and location-dependent shift. 
Although the gradients for the two costs were roughly opposite to one another, the gradients at the various spatial locations were related to each other in a similar way.
For both costs, the gradients at $s_1$ ad $s_6$ had extrema that aligned with those at $s_{TE_p}$.
Also, traveling-wave behavior was evident from a comparison between the gradient at $s_{6}$ and $s_{vmax}$.
For both costs, the negative peaks at $s_{vmax}$ had larger magnitude than the positive ones. 
Cycle-to-cycle variation was observed even though the unactuated flow exhibits limit-cycle oscillation.

The lift-optimal actuation profile exhibited many of the same features as the gradient of the unactuated flow: three local minima indicating three actuation cycles were seen, the negative peaks had larger magnitude and shorter duration than the positive peaks and substantial time-averaged actuation magnitude existed near $s_{TE_p}$.
The drag-optimal actuation had negligible magnitude over the second half of the optimization window, including much smaller time-averaged actuation magnitude near $s_{TE_p}$ than for the lift-based cost.
For both costs, the actuation near $s_{max}$ shows attributes of traveling-wave behavior, though with distinct features from the periodic actuation considered in the literature \citep{shukla2022hydrodynamics, akbarzadeh2019reducing, thompson2022surface}. Broadly, the spatial distribution of the optimal actuation profile is non-periodic and exploits the nonuniform sensitivity of the flow to forcing along the airfoil surface as the vortex-shedding process evolves. Moreover, the time evolution of the actuation behavior is not perfectly periodic, also reflected in the 
gradient field of the unactuated flow. 
From the temporal variation of the optimal actuation at key locations on the airfoil surface, non-intuitive characteristics were noticed particularly for the lift-based cost.
The positive peaks of actuation at $s_{vmax}$ showed two local maxima separated by a local minimum which roughly aligned with local minima in the actuation at the various locations on the suction and pressure sides of the airfoil, and the peak of the actuation at $s_{TE_p}$.
Further, the magnitude of the first local maximum relative to the second decreased over the 3 actuation cycles.

The substantial differences in the flow arising from the two actuation strategies were reflected in the different temporal variations of $C_p$ at key locations on the airfoil surface.
In the drag-optimal case, the $C_p$ at 
$s_{vmax}$ and $s_{TE_p}$ 
increased, in line with the decreasing drag.
In the lift-optimal case, the $C_p$ at $s_{vmax}$ and $s_{LE}$ decreased over the course of the optimization window, consistent with the increasing lift and the stronger LEV.
Additionally, the $C_p$ at $s_{vmax}$ showed local minima around time instances when the $C_p$ at $s_{TE_p}$ is the lowest.
The different flow phenomena for the two costs prompted separate investigations of the interplay between the actuation profiles and the flow features.

A comparison with the unactuated flow showed a stronger LEV and a phase shift between the LEV and TEV when lift-optimal actuation is present. 
The trajectory of the LEV is closer to the airfoil surface and the trajectory of the TEV is at higher $y$-coordinate values with lift-optimal actuation.
The variation of these quantities over the various vortex-shedding cycles in the optimization window was reported.
An analysis of the flow snapshots and streamlines over one actuation cycle revealed that the lift-optimal actuation results in a stronger LEV by aiding its roll-up near $s_{max}$. 
The positive actuation (along the outward normal) appears in the vicinity of streamlines of the LEV pointing in a similar direction.
Similarly, the negative actuation appears near the streamlines of the LEV pointing towards the airfoil surface.
The large-magnitude negative actuation of short duration that appears at $s_{vmax}$ affects the flow streamlines such that two LEVs are separated by a region of lower pressure magnitude. 
Shortly after, positive actuation reappears there and promotes formation of a new LEV (figure \ref{fig:ClAP}).
As the LEV advects past the trailing edge, positive actuation occurs on the pressure side and negative actuation occurs on the suction side.
The effect of the actuation near the trailing edge is studied by analyzing the $C_p$ contours and the $v$-velocity and surface $C_p$ on the pressure side with actuation over the aft $30\%$ of the airfoil surface on both the suction and pressure surfaces zeroed out.
Zeroing out the actuation near the trailing edge leads to a weaker LEV (figure \ref{fig:NoTEI}) and lower pressure on the pressure surface (figure \ref{fig:dVNoTE}).
To study the effect of the actuation over the less significant locations, 
simulations without actuation around $s_{max}$ and additionally without actuation near the trailing edge were run. 
Using the temporal variation of $C_p$ at $s_{vmax}$ as a point of comparison, all cases without actuation near $s_{max}$ showed a decrease in $C_p$ (compared to the unactuated case) in response to the first peak in the actuation at $s_6$ and an increase in $C_p$ in response to the trough in the actuation between the first and second peaks (figure \ref{fig:LEp}).
The troughs and peaks in the $C_p$ profile with full actuation also aligned with the peaks and troughs of the actuation at $s_6$, suggesting that the actuation at $s_6$ (and $s_{TE_p}$) aids the large-magnitude actuation at $s_{vmax}$ in influencing the flow field near the leading edge.

The drag-optimal actuation changes the vortex-shedding processes such that the LEV and the TEV eventually approach the trailing edge at the same time and advect in tandem thereafter. 
The interplay between the actuation and the LEV formation during the various cycles was discussed in terms of the actuation at $s_{vmax}$. 
The changes in the flow field resulting from actuation at $s_{vmax}$ are similar to those in the lift-optimal case: the appearance of negative actuation over a region close to $s_{max}$ leads to flatter streamlines, an increase in pressure, and greater opposition to LEV formation. 
On the other hand, positive actuation there with negative actuation downstream leads to a region of negative vorticity and promotes LEV formation. 
The LEV formation is opposed at the start of the optimization 
window and later aided such that an LEV and TEV advect close to the trailing edge around the same time instance (figure \ref{fig:DragFormation}).
Following their advection to the trailing edge, the LEV-TEV pair travel downstream in tandem. 
Secondary vortex shedding with formation of new LEV and TEV is seen on the suction surface (figure \ref{fig:DragAdvection}).
The new vortices coalesce with the LEV-TEV pair in the wake of the airfoil.
The variation of the drag over the optimization window seen in figure \ref{fig:TH} was related to the various changes in the flow features.

Appendix \ref{sec:ASim} considers a suite of additional studies to augment the primary results of this article, which center on optimal actuation normal to the airfoil surface over the targeted time window. Three specific extensions are reported on in that appendix, with key conclusions summarized here for completeness.

First, optimization over a second time window (directly subsequent to the optimization window considered in the article) was performed. This computation was done to assess: (i) for the lift-based cost, whether the non-periodic nature of the lift-optimal actuation profile was due to transients in the adjoint gradient over the finite time window; (ii) for the drag-based cost, whether optimization over more time windows would move the flow closer to the unstable steady base state inherent to the unactuated airfoil system. The $C_l$ variation resulting from the lift-optimal actuation of the second window exhibited less cycle-to-cycle variation as compared to the first. 
However, perfect periodicity in neither the aerodynamic coefficients nor the actuation profile was achieved.
The drag in the second window had larger local minima than 
those in the first.
The lift-optimal actuation in the second window had many similarities to that in the first.
The drag-optimal actuation in the second window did not yield complete elimination of vortex shedding. 
This persistence of unsteadiness was due at least partly to the penalization of actuation magnitude; permitting larger actuation magnitude led to reduction in (but not complete elimination of)  the drag oscillations.
The drag-optimal actuation in the second window had similarities to the actuation profile from the first window but with a phase shift.

Second, optimal actuation with independent $x$- and $y$-components was computed for both costs. This optimization was performed to allow for a more general actuation profile with possible large-magnitude values at locations where normal actuation had insignificant variations, and to determine if actuation with a component along the local tangent could more efficiently yield performance benefits. 
No significant differences in the aerodynamic coefficients
as compared to the optimal normal case was found. 
When projected onto the local normal and tangential directions, the time averaged normal actuation was identical to the optimal normal actuation.
For the drag-based cost, the tangential actuation was insignificant everywhere but near $s_{max}$ while for the lift-based cost, non-negligible actuation was also seen very close to the trailing edge.
From the temporal variation of the tangential and normal components of the general actuation, the preferred inclination of the actuation at $s_{vmax}$ was slightly more inclined towards the oncoming flow as compared to the purely normal case.

Finally, the actuation from the current work was compared to that from our earlier work \citep{thompson2023optimal}, where optimal actuation over a shorter actuation window was computed.
The $C_l$ variation due to the lift-optimal actuation obtained over the shorter window did not lead to as much of an increase in the minimum lift value as compared to the actuation from the current work.
From the temporal variation of the actuation profiles, substantial differences were observed in the actuation at $s_{vmax}$.
At $s_{TE_p}$ however, similarities were seen away from the end of the shorter optimization window, where the starting transients of the adjoint solution necessitate smaller-magnitude actuation.
The drag-optimal actuation was less influenced by the optimization window. Identical temporal variations were observed at both $s_{vmax}$ as well as $s_{TE_p}$. 
Over the shorter time interval, the drag associated with the short window-optimal solution was smaller than for the solution optimized over the longer window for several time instances of the shorter window. The amplitude of oscillations was, however, more pronounced for this solution optimized over the shorter time window.

\section*{Acknowledgements}
E.T. and A.J.G. gratefully acknowledge support from the Air Force Office of Scientific Research under Award No.
FA9550-21-1-0182.

\section*{Declaration of interests}
The authors report no conflict of interest.

\appendix

\section{More details on numerical methodology}
\label{sec:NM_details}

The two-dimensional (2D) Navier-Stokes equations are solved in the vorticity-stream function formulation.
The dimensionless form of the equation for the vorticity along the $z$-direction and the equation enforcing the velocity boundary condition on the body are:
    \begin{align}
        \begin{split}
            \frac{\partial \omega }{\partial t} + \bm{u}\cdot \nabla \omega &= \frac{1}{Re}\nabla ^ 2 \omega + \nabla \times \int_{\Gamma} \bm{f}(\bm{\xi}) \delta (\bm{\xi} -\bm{x}) d\bm{\xi}, \\
            \int_{\Omega} \bm{u}(\bm{x}) \delta (\bm{x}-\bm{\xi}) d\bm{x} &=  \bm{u}_b(\bm{\xi}) = \bm{u}_{act} (\bm{\xi}) \bm{\hat{e}}_n,
        \end{split}
    \end{align}
where $\bm{u}$ is the flow velocity vector written in terms of the flow domain coordinate $\bm{x}\in \Omega$ ($\Omega$ denotes the flow domain). 
The surface force $\bm{f}$ which is computed at each coordinate on the airfoil, $\bm{\xi}\in\Gamma$ ($\Gamma$ represents the airfoil surface), 
ensures
that the flow velocity matches the surface velocity $\bm{u}_b$. 
The surface velocity at each point on the airfoil is determined from the normal actuation at the respective point, where $\bm{\hat{e}}_n$ is the local unit normal on the airfoil surface.
The grid parameters and time step are taken from our earlier work where the relevant convergence tests have been reported \citep{thompson2022surface}.
Our choice to recycle the grid and time step parameters is rooted in the observation from our earlier work that the most beneficial actuation parameters were of the same temporal and spatial scales as the flow features of the unactuated flow.
The suitability of the grid and time step parameters for the adjoint solution is demonstrated by quantifying the accuracy of the adjoint gradients at the end of this section.
The cost functionals in equation (\ref{eq:cost}) can be written as:
     \begin{align}
            \mathcal{J}_{C_*} &= \frac{1}{2}\int_0^{T} \int_{\Gamma} 
            J_{C_*} \text{  } d \xi dt,
    \end{align}
where $J_{C_*} = J(C_*, C_{*_{target}}, \phi, u_{act})$. 

The minimization of the cost functional is achieved using a gradient-based optimization procedure. 
The gradient used is that of the cost functional with respect to the design variables, which in the current work are the spatio-temporal values of the actuation at each body point and time step of the optimization window. 
Since the dimension of the vector of design variables is $n_b \times n_t$, where $n_b$ and $n_t$ are respectively the number of surface body points and time steps in the optimization window, the computation of the gradient vector by finite differences would require $n_b\times n_t +1$ simulations. 
The expensive computation of the gradient is circumvented by employing the adjoint of the forward equation. 
The derivation of the adjoint equation is the same as in \citet{flinois2015optimal}, with the exception that the actuation in the current work is normal to the body surface and is allowed to vary in space at every time instance. 
The adjoint equations solved here are the curl of the adjoint of the $x$ and $y$ momentum equations and are given by:
    \begin{align}
        \begin{split}
            -\frac{\partial \omega^+}{\partial t} - \bm{u^+}\cdot \nabla \omega + \nabla^2 (\bm{u^+} \times \bm{u})  &= \frac{1}{Re}\nabla ^ 2 \omega^+ + \nabla \times \int_{\Gamma} \bm{f}^+(\bm{\xi}) \delta (\bm{\xi} -\bm{x}) d\bm{\xi}, \\
            \int_{\Omega} \bm{u}^+(\bm{x}) \delta (\bm{x}-\bm{\xi}) d\bm{x} &=  \bm{u}^+_b(\bm{\xi}) = \frac{\partial J_{C_*}}{\partial \bm{f}}.
        \end{split}
        \label{eq:AdjNS}
    \end{align}
In the above equations, the superscript $+$ implies that the variables are the adjoint analogs of the respective forward variables. 
The adjoint equations are marched backwards in time as implied by the negative coefficient of the unsteady term. 
The ``initial'' condition for the adjoint variables at time $t=T$ is taken to be 0---consistent with the derivation of the adjoint equations. 
Since the adjoint equations have contributions from the forward variables, the forward field at every time instance of the optimization window is required for the computation of the adjoint field. 
To reduce the storage requirements, we store the forward solution 
every four time steps and use linear interpolation for the intermediate time steps.
To circumvent the generally large interpolation error at the start and end of the optimization window, the full forward field at each of the first and last twenty time steps are saved. 

Note that the dependence of the cost on $C_l$ and $C_d$ figures into the evolution of the adjoint equations (\ref{eq:AdjNS}) via the second equation. 
such that
\begin{align}
    \frac{\partial {J}_{C_d}}{\partial \bm{f}} = ( mC_d - C_{d_{target}}, 0 ),  \text{ }
    \frac{\partial {J}_{C_l}}{\partial \bm{f}} = (0, mC_l - C_{l_{target}} ).
\end{align}
The change in the cost functional due to a small change in the actuation is 
     \begin{align}
            \delta \mathcal{J} &= \frac{1}{2}\int_0^{T} \int_{\Gamma} 
            \frac{d J}{d \bm{u}_{act}} \delta \bm{u}_{act} \text{  } d \xi dt.
            \label{eq:gradI}
    \end{align}
For the form of actuation and penalty term, 
     \begin{align}
            \frac{d J}{d \bm{u}_{act}} = \bm{f}^+ \bm{\hat{e}}_{\bm{n}} + \phi \bm{u}_{act}.
            \label{eq:gradII}
    \end{align}
After discretizing the equations in time and space, the discrete gradient of the cost functional with respect to actuation is a vector of size $(n_b\times n_t)$ where, $n_b$ and $n_t$ are respectively the number of body points on the airfoil surface after spatial discretization and the number of time steps in the optimization window after temporal discretization. This discrete gradient is fed into a nonlinear conjugate gradient algorithm which iteratively updates the vector of design variables to converge to a local minimum of the cost functional as defined by the first-order sufficient condition.
In each iteration, the update of the design vector is along a search direction. 
The dependence of the search direction on the gradient is as per the Polak-Ribiere formula \citep{polak1971computational}. The update of the design vector can be written as
     \begin{align}
        \bm{x}^{k+1} = \bm{x}^k + \alpha \bm{d}^{k+1}. 
    \end{align}
The step size of the update along the search direction is computed using Brent's line search \citep{press2007numerical}. 
The line search procedure involves iteratively solving for the step size that satisfies within a tolerance: $\bm{g}(\bm{x}^k + \alpha \bm{d}^{k+1})^T\bm{d}^{k+1} =0$; i.e., the first-order sufficient condition of the optimality of the cost functional along the search direction $\bm{d}^{k+1}$. 
Though the gradient in the continuous space is given by equation (\ref{eq:gradI}), the discretized gradients are not guaranteed to be smooth, which in turn could lead to non-smooth 
or very small 
updates to the actuation. 
A lack of smoothness in the actuation profile, particularly in time, results in large acceleration magnitudes which keeps the optimizer from converging. 
To avoid sharp variations in the actuation, smoothness of the gradients is enforced by restricting the inner product definition in equation (\ref{eq:gradI}) to the space of twice-differentiable functions, as discussed in \cite{jameson2003aerodynamic} and \cite{bukshtynov2011optimal}.
The smoothed gradients are computed from the gradient in equation (\ref{eq:gradII}) by solving a Helmholtz equation \citep{jameson2003aerodynamic, bukshtynov2011optimal}.
The boundary conditions for the smoothed gradients are taken to be zero. 
Smoothing is incorporated both in space and time which means that the actuation at the trailing edge on both the suction and pressure surfaces at all time instances
and the actuation at all body points at the first and last time steps of the optimization window are all zero. 
The smoothing parameters are such that at all iterations of the optimization procedure, the gradient varies smoothly in space and time.
To keep the smoothing parameters as low as possible (resulting in less altered gradients), the smoothing in time is increased only when kinks appear in the gradient---noticeably near the start and the end of the window and near the trailing edge.
When the smoothing parameter is changed, the nonlinear conjugate gradient iteration is restarted along the direction of steepest descent, with the actuation from the previous iteration effectively being used as a starting guess.


The accuracy of the gradients computed by the adjoint approach is tested against finite difference calculations. 
The spatially and temporally discretized gradients arising from equations (\ref{eq:gradI}) and (\ref{eq:gradII}) are compared with those from finite difference computations.
The gradient computation is at an actuation velocity given by: 
\begin{align}
    v_n(s, t) = 0.01 \sin \left( \frac{\pi s}{l_s} \right) \sin \left( \frac{\pi t}{T} \right)
\end{align}
The above equation gives the normal velocity at any spatial location, $s$, on the airfoil, at time, $t$. The variables $l_s$ and $T$ are respectively the length of the airfoil surface and the length of the optimization window. 
For validating the gradients, both the penalty weight $\phi$ as well as the target values in equation (\ref{eq:cost}) are taken to be zero. 
While it is possible to compute the adjoint gradients at each body point and time instance, such a calculation with finite differences is prohibitively expensive.
Hence, we limit the computation of the finite-difference gradients to six spatial locations and ten time instances. 
The finite-difference gradient of the cost functional at a time instance and a spatial location is computed by evaluating the difference in the cost functional $\mathcal{J}$ in response to a perturbation in the actuation value at that point.
The perturbation in the actuation value is taken to be $\epsilon = 10^{-7}$, based on a convergence test. 
\begin{figure}
    \centering
    \includegraphics[width=0.96\textwidth]{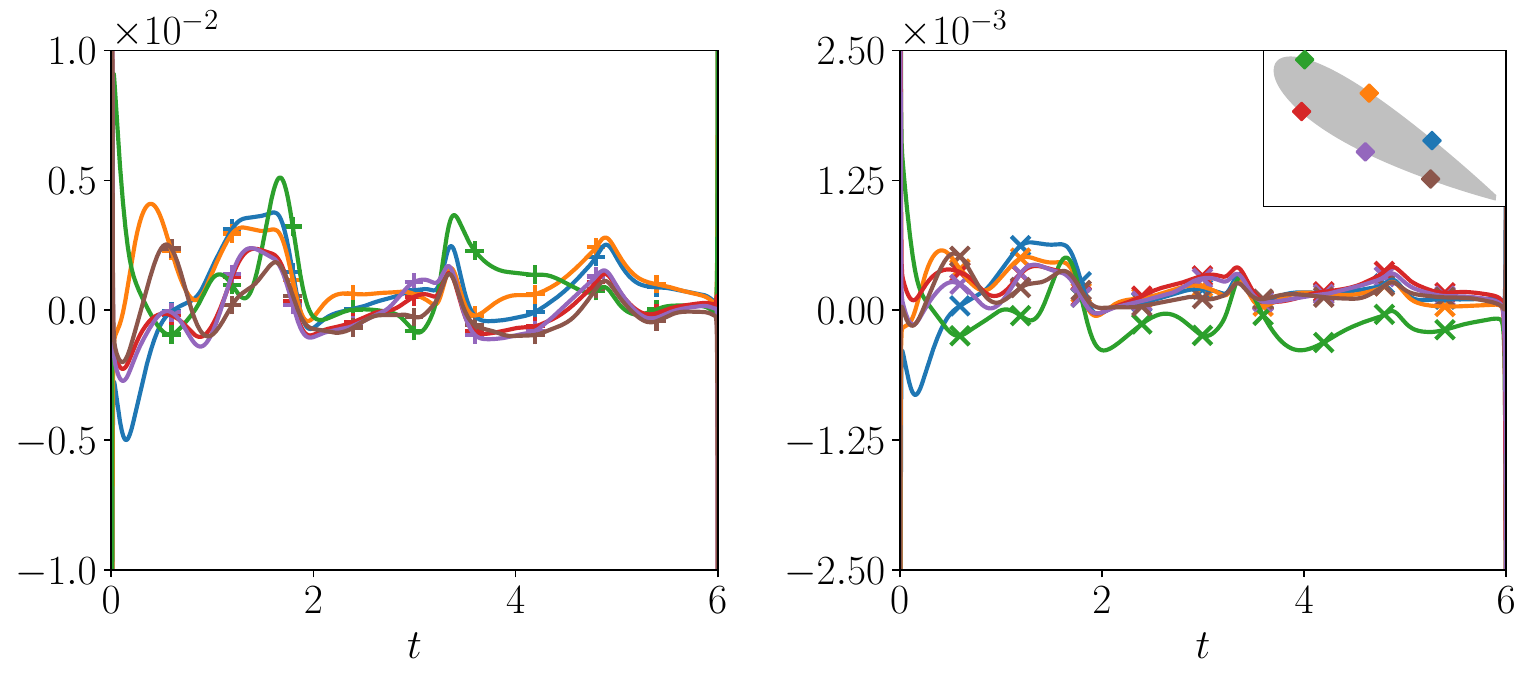}
    \caption{Gradient validation: comparison between adjoint (solid line) and finite difference (markers) gradients at a subset of body points. Left:
    $\delta \mathcal{J}_{C_l}/ \delta v $, 
    right:
    $\delta \mathcal{J}_{C_d}/ \delta v $; colors are according to the schematic in the inset of the right figure.}
    \label{fig:GradComp}
\end{figure}
The finite difference and adjoint gradients for both costs are plotted in figure \ref{fig:GradComp}. 
The inset in the right subfigure shows the airfoil and the spatial points for the comparison. The error in the adjoint gradients is $1.3\%$ for $\mathcal{J}_{C_l}$ and $0.6\%$ for $\mathcal{J}_{C_d}$.
The agreement between the adjoint simulation and the finite difference approximation indicates both a correct implementation of the adjoint solver and a sufficiently fine grid resolution.

\section{Additional simulations}
\label{sec:ASim}

The findings in the main section of this article prompt three further questions that are addressed in this appendix 
by way of additional optimization runs.
First, as discussed in sections \ref{sec:Ap} and \ref{sec:actuation} 
(and evident from figures \ref{fig:TH} and \ref{fig:vNTemporal}),
the aerodynamic coefficients and the lift-optimal actuation show non-periodic behavior even though qualitative similarities between the various cycles exist. 
On the other hand, the drag-optimal actuation causes the flow to approach a steady state that one may naturally associate with the unstable equilibrium of the unactuated flow.
To determine whether a periodic lift-optimal actuation profile and full stabilization of the flow with drag-optimal actuation can be achieved, optimization over a subsequent window of the same length is repeated.
Second, this study has only considered surface normal actuation, and it is unclear whether allowing for general actuation in the $x$- and $y$-directions could yield different benefits and actuation profiles (e.g., without actuation so focused near $s_{max}$). 
Third, one of our prior efforts considered a smaller actuation window \citep{thompson2023optimal}, and it is worth quantifying the differences in results across that window size and the one employed here, to more systematically characterize the effect of that parameter.

\subsection{Beyond the first optimization window} 
\label{sec:wII}
\begin{figure}
    \centering
    {\includegraphics[width=0.92\textwidth]{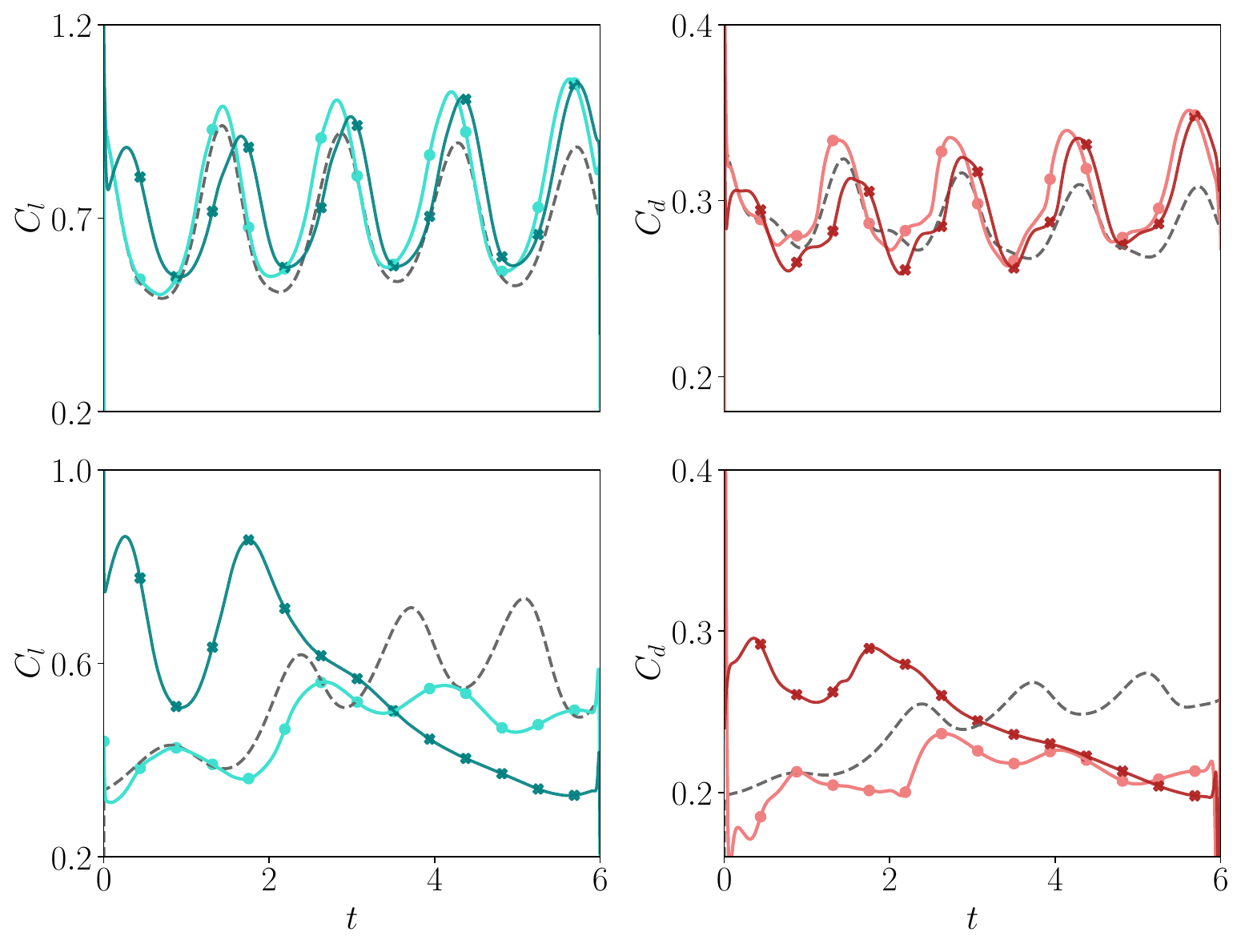}} 
    \caption{
    Coefficient of lift (left column) and drag (right column) for lift- and drag-based optimization (top and bottom rows, respectively). 
    The gray dashed lines represent the aerodynamic coefficients of the unactuated flow over the second optimization window. 
    In each subfigure, the brighter curve represents the aerodynamic coefficient over the second optimization window and the darker curve represents the coefficient over the first optimization window.
    }
    \label{fig:TH_WII}
\end{figure}
We first report the result of optimizing over a subsequent window of the same length. 
For both costs, the first time instance of the second window coincides with $t=5.9$ in the first 
optimization window.
This time instance was chosen so as to exclude the large temporal variations in the actuation profile (due to the adjoint field) towards the end of the first window. 
The penalty weight for the optimization procedure was chosen to limit the maximum actuation magnitude to roughly $0.03$, as was done in the first optimization window.  
Figure \ref{fig:TH_WII} shows the aerodynamic coefficients over the relevant optimization window.  
For ease of comparison, the results for both windows are plotted using the same $x$-axis, even though the start of the second optimization window coincides with $t=5.9$ in the first optimization window. (That is, the solution for the second window is appropriately time shifted so that the time axis $t\in[0,6]$ is appropriate).
For both costs, in the absence of actuation the flow restarted from the end of the first optimization window approaches the limit-cycle oscillation of the unactuated flow (but does not fully reach the limit cycle because of the limited window length).
\begin{table}
  \begin{center}
\def~{\hphantom{0}}
  \begin{tabular}{lccccccccc}
    \multirow{2}{7em}{Aerodynamic coefficient} & cycle & min & $t_{min}$ & $\Delta t_{min}$ & max & $t_{max}$ & $\Delta t_{max}$ & avg. \\
    \\
    \hline\hline 
    \hspace{2em} 
    \multirow{5}{2em}{$C_l$} 
     & 1 & 0.503 &0.679 & - & 0.989 & 1.436 & - & 0.736 \\
     & 2 &0.550 & 2.021 & 1.342 & 1.005 & 2.817 & 1.380 & 0.765 \\
     & 3 &0.573& 3.421 & 1.400 & 1.027 & 4.201 & 1.384 & 0.766 \\
     & 4 &0.562& 4.842 & 1.421 & 1.060& 5.631 & 1.431 & - \\
    \hline
    \hspace{2em}\multirow{5}{2em}{$C_d$} 
    & 1 & 0.275 & 0.642 & - & 0.335 & 1.351 & - & 0.299 \\
     & 2 &0.272 & 2.013 & 1.370 & 0.336 & 2.726 & 1.375 & 0.297 \\
     & 3 &0.263& 3.426 & 1.414 & 0.340 & 4.142 & 1.416 & 0.304 \\
     & 4 &0.277& 4.694 & 1.267 & 0.351 & 5.619 & 1.477 & - \\
    \end{tabular}
\caption{Properties of $C_l, C_d$ variation over the second optimization window with lift-optimal actuation, restarted from $t=5.9$ (near the end of the lift-optimized solution for the first time window). Properties are not shown for the flow restarted from the same time instance without actuation.}
  \label{tab:JCltempII}
    \end{center}
\end{table}

For the lift-optimal case, table  \ref{tab:JCltempII} shows information analogous to table \ref{tab:JCltemp}. 
The start of a cycle is characterized by the occurrence of a local minimum. 
Like in the first optimization window, the lift variation from one cycle to the next increases (see avg. $C_l$ in the last column of table \ref{tab:JCltempII}). 
However, the variation is not as pronounced as in the first optimization window, where the flow starts from a limit-cycle oscillation of the unactuated flow. 
The minima (particularly the first two) associated with the $C_l$ dynamics for the optimal solution are higher over the first optimization window than the second, 
while the peaks of $C_l$ (noticeably the first two) are higher over the second window.
These discrepancies can be attributed to the slight variation in the maximum actuation magnitude and the number of vortex-shedding cycles in the optimization window; while in the first optimization window five peaks in the $C_l$ variation are seen, in the second only four peaks exist.
The actuation towards the end of the window must have a small magnitude due to the starting transients of the adjoint equation and the extent to which performance improvements can be attained depends on the flow phenomena occurring during the time instances with substantial actuation magnitude.
From the $C_d$ variation, the drop in drag below the unactuated case still exists at some time instances.

Turning our attention to the drag-optimal solutions, optimizing over the second window does not yield a flow that reaches the unstable base state, as there are evident oscillations in both lift and drag.
While the drag drops significantly over the optimal solution for the first optimization window, it increases after the second cycle in the second optimization window.
The inability of actuation to eliminate vortex-shedding over the second window could be due to the limitation on the magnitude of actuation. We also note that while this second optimization suggests the magnitude-penalized actuation will not reach the unstable base state, it is possible that optimizing over a single large window (rather than multiple smaller ones) could affect these results as in \citet{flinois2015optimal}. 
An investigation into the various effects of optimization window selection are indicated in a later section of this Appendix, though further study is left to future work. 

\begin{figure}
    \centering
    \includegraphics[width=0.95\textwidth]{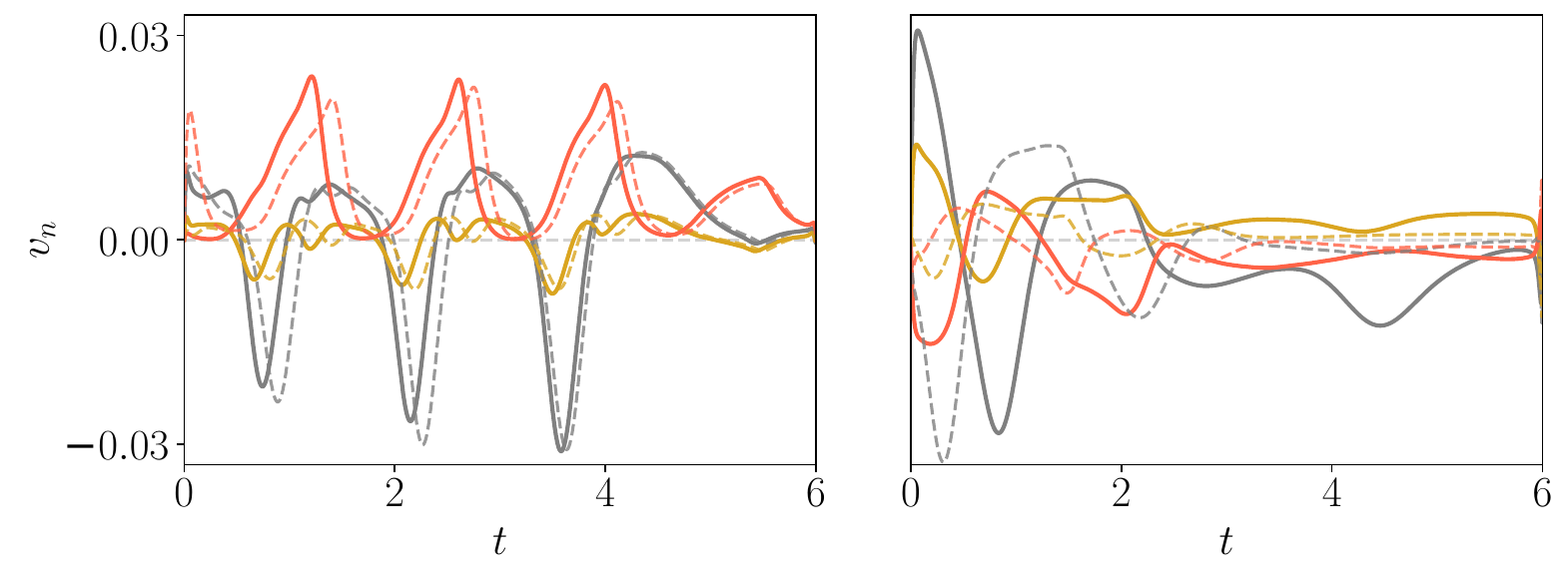} \hspace{0.03\textwidth}
    \caption{
    Optimal actuation for the lift-based (left) and drag-based (right) cost functionals. Solid lines: optimal actuation over the second window, dashed lines: optimal actuation over the first window. Colors are according to the markers in figure \ref{fig:airfoilsc}.
    }
    \label{fig:vNwII}
\end{figure}
\begin{figure}
    \centering
    \hspace{0.008\textwidth}
    \includegraphics[width=0.95\textwidth]{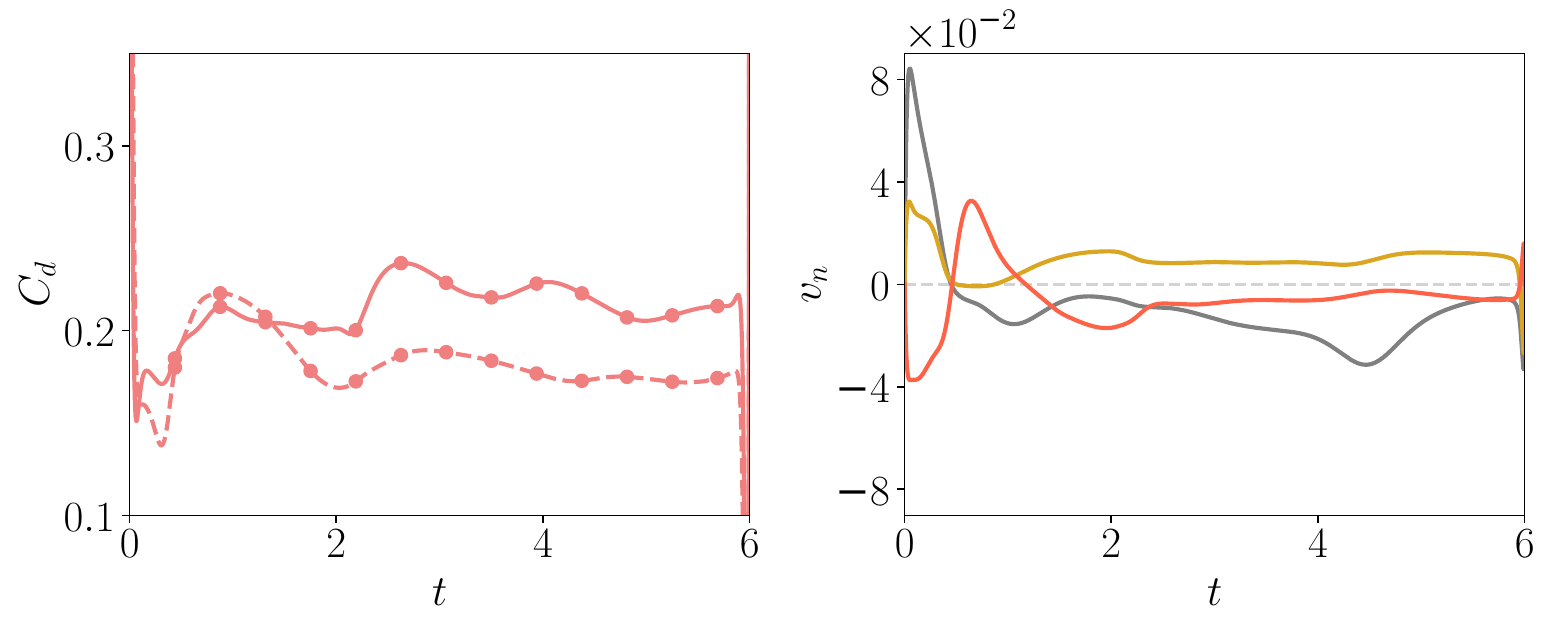}
    \caption{Drag variation (left) and actuation (right) over second optimization window with smaller penalty weight. (The solid line in the left subfigure is the drag with the larger penalty, 
    reproduced from figure \ref{fig:TH_WII} for ease of comparison). }
    \label{fig:CdwIIpII}
\end{figure}

Figure \ref{fig:vNwII} compares the actuation from the two optimization windows at a subset of the body points. 
The dashed lines represent actuation from the first window.
The location of $s_{vmax}$ remains unchanged for $\mathcal{J}_{C_l}$ ($x = 0.025$ measured from the leading edge along the chord) but shifts downstream for $\mathcal{J}_{C_d}$ ($x=0.038$ in the second window; $x=0.032$ in the first window, measured from the leading edge).
The lift-optimal actuation profiles have several similarities across both windows.
Both negative and positive peaks at $s_{vmax}$ are preceded by corresponding peaks 
at $s_{LE}$, characteristic of traveling-wave behavior. 
For both time windows of the optimal solution, the time regions of positive actuation at $s_{vmax}$ contain two peaks, the first of which becomes less prominent over later cycles. 
The troughs separating the peaks roughly coincide with the troughs in the actuation at $s_{LE}$ and the peaks in the actuation at $s_{TE_p}$. 
The negative peaks increase in magnitude with each cycle, with the difference in the magnitude of the negative peaks being more prominent over the second optimization window. 

While the drag-optimal actuation over the second window is noticeably different from that of the first window, features of the actuation profile are emblematic of phase-shifted behavior. For $s_{vmax}$, over the first window a time interval of positive actuation is preceded and superseded by negative actuation. 
This behavior occurs at $s_{vmax}$ for the second window as well, apart from the initial positive actuation. 
The actuation profiles at $s_{LE}$ also have similarities, apart from  the initial positive peak over the second window.
In the first window, the actuation at $s_{LE}$ is negative at the start of the window, followed by positive actuation. 
Such variation is also observed in the second window after the initial positive actuation.
The above observations suggest that the time instance of the start of the second window could be varied to get better vortex suppression. 
However, such an exploration is beyond the current scope.

For completeness, figure \ref{fig:CdwIIpII} shows the drag and actuation over the second window with a smaller penalty weight. The drag behavior with larger actuation magnitude has lower variation over time, particularly towards the end of the optimization window, indicating that large actuation may indeed stabilize the flow. 
\subsection{Independent $x$ and $y$ components of actuation}
\begin{figure}
    \centering
    {\includegraphics[width=0.92\textwidth]{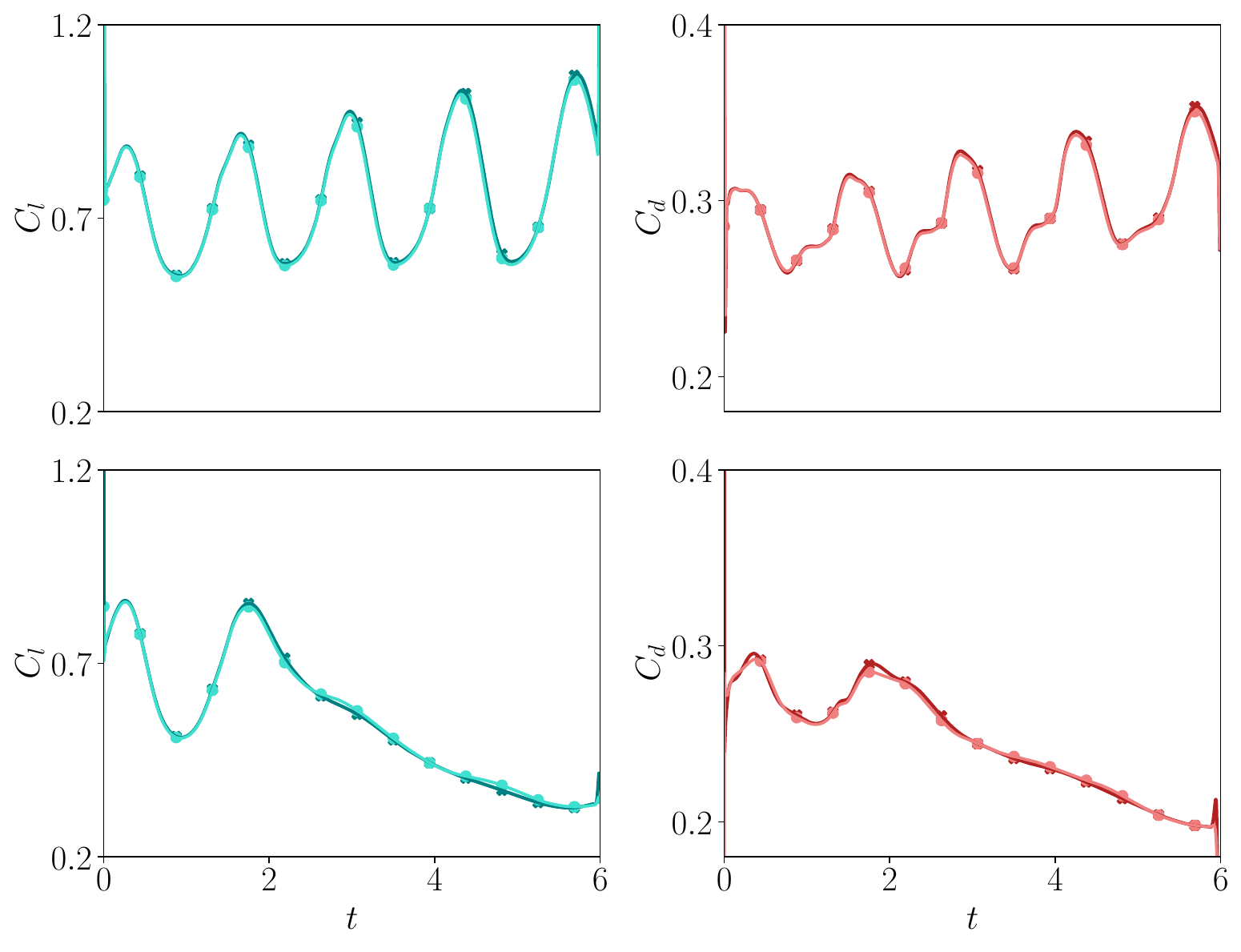}} 
    \caption{Analog of figure \ref{fig:TH} with independent $(x, y)$ components of actuation. 
    Within each subfigure, the brighter curve corresponds to actuation with independent $(x,y)$ components while the darker one is the same as in figure \ref{fig:TH}. }
    \label{fig:TH_VxVy}
\end{figure}
%
\begin{figure}
    \centering
    \subcaptionbox*{}
    {\includegraphics[width=0.495\textwidth]{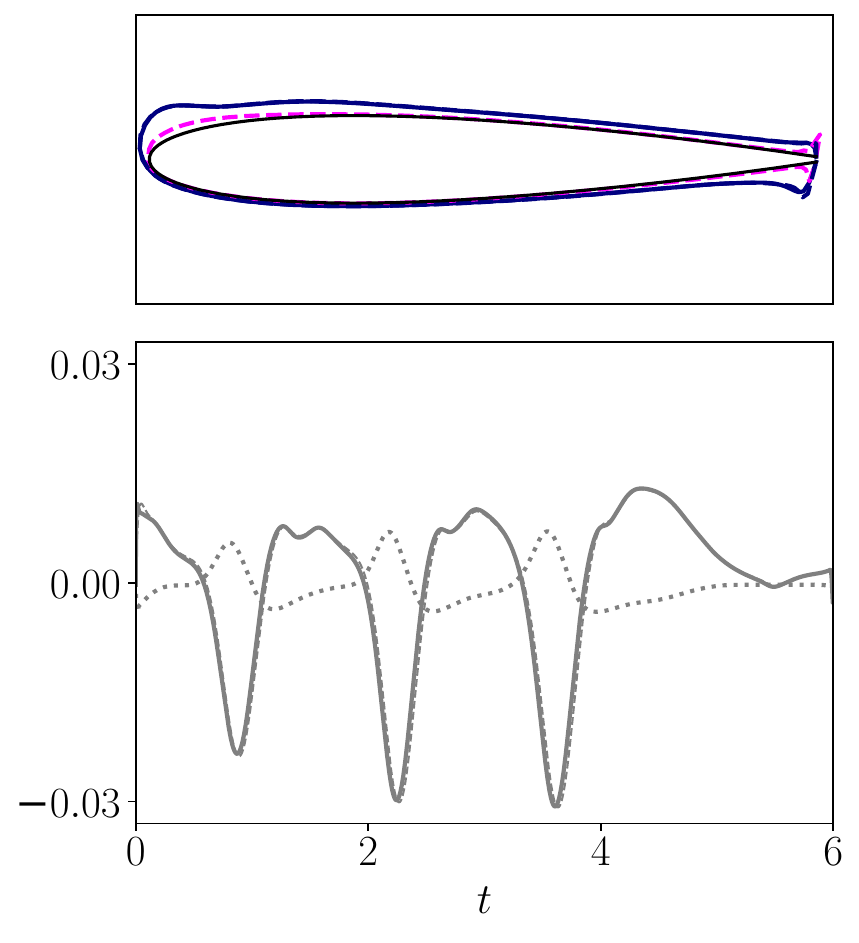}} 
    \subcaptionbox*{}
    {\includegraphics[width=0.495\textwidth]{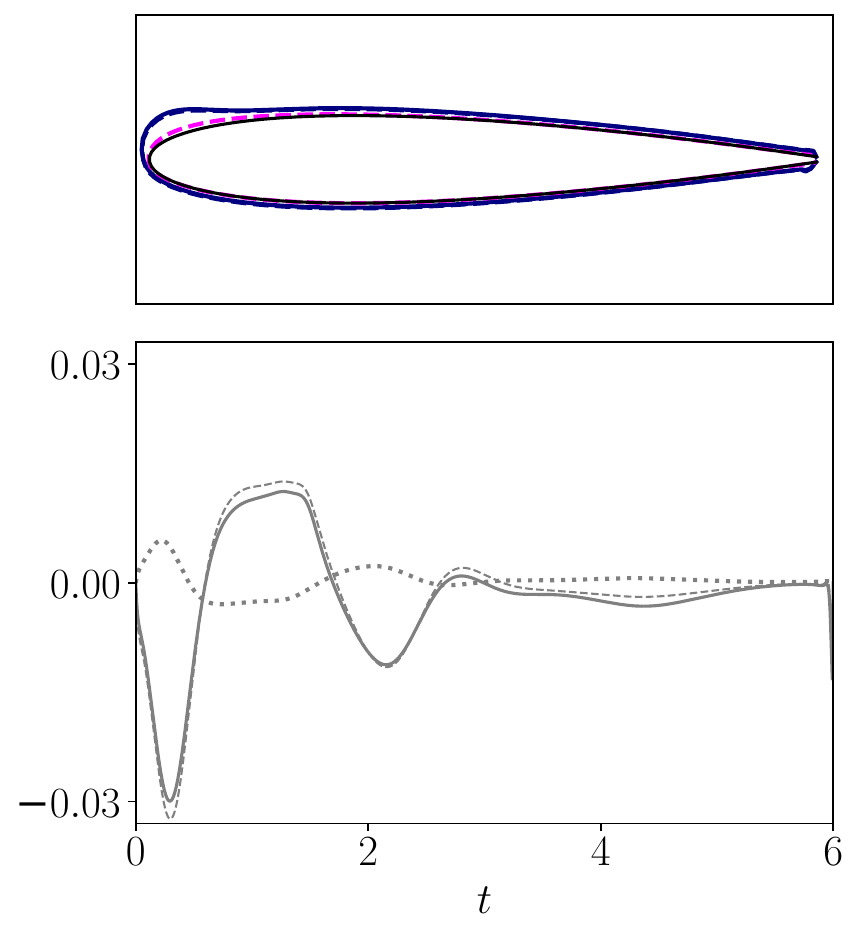}}
    \caption{
    Normal and tangential components of actuation, (left): $\mathcal{J}_{C_l}$, (right): $\mathcal{J}_{C_d}$.
    Top row: time-averaged actuation. For the actuation with $x$- and $y$-actuation, the tangential component is indicated by the dashed magenta line and the normal component by the dashed blue line. For normal-only case, the actuation profile is given by the solid blue line. 
    Second row: temporal variation of normal and tangential components of actuation at $s_{vmax}$.
    Dotted line: tangential component of general actuation, solid line: normal component of general actuation, dashed line: optimal-normal actuation.
    }
    \label{fig:VxVyAvg}
\end{figure}
To investigate whether the previously less influential locations of the airfoil play a larger role in affecting the flow field when a more general actuation strategy is allowed, we now consider optimization with actuation allowed to vary independently in the  $x$ and $y$ directions (rather than in the local normal direction along the airfoil). 
The design space is now twice as large as that where actuation is forced to be normal.
The penalty weight is adjusted such that the maximum magnitude of instantaneous actuation is roughly $ 0.03$.
A comparison between the aerodynamic coefficients corresponding to optimal normal actuation and actuation with independent $x$ and $y$ components is shown in figure \ref{fig:TH_VxVy} for both costs. 
The brighter curves correspond to general actuation.

We find that the general actuation strategy results in similar behaviour of the aerodynamic coefficients as normal actuation.
The slight differences in performance could be attributed to the variation in the maximum magnitude of actuation resulting from the choice of the penalty weight.
We next compare the time averaged spatial variation of the general actuation. 
To make a more direct comparison with the normal case, we project the $x$ and $y$ components onto the local normal and tangential directions of the airfoil surface. 
We take the normal to be outward and the tangential direction such that the tangent at $s_{max}$, the point with the maximum $y$-coordinate, points downstream 
(note that the tangential direction on the pressure side of the airfoil points upstream).

Figure \ref{fig:VxVyAvg} shows comparisons between the optimal-normal actuation and normal and tangential projections of the general actuation. 
The figure demonstrates that the normal actuation is nearly identical for  both cases, and even in the case where actuation is allowed to vary in two independent directions the tangential component is of much smaller magnitude than the normal component.
For both costs, the time-averaged value of the actuation on the suction surface in the normal direction coincides with $s_{vmax}$ of the optimal-normal actuation.
The time average of the projection in the tangential direction is maximum one discretized point upstream of $s_{max}$ for the drag-based cost. 
For the lift-based cost, the maximum value is near the trailing edge while upstream, it is maximum at the same location as in the drag-based cost (one discretized point upstream of $s_{max}$). 

To compare the variation of the optimal normal actuation and the projection of the general actuation onto the normal and tangential directions, these quantities at $s_{vmax}$ are shown in the second row. 
For both costs, the tangential actuation is seen to be positive when the normal actuation is negative and vice-versa. 
This counteracting behavior suggests that the actuation at $s_{vmax}$ is along a direction that has a larger angle with respect to the positive $x$-axis than the local normal. 
When compared to the purely normal actuation, the general actuation has a greater upstream pointing component when it is positive, thus decelerating the oncoming flow to a greater extent. 
Similarly, during negative actuation the general actuation accelerates the flow to a greater extent than the normal actuation. 
\subsection{Influence of optimization window}
\begin{figure}
    \centering
    {
\includegraphics[width=0.92\textwidth]{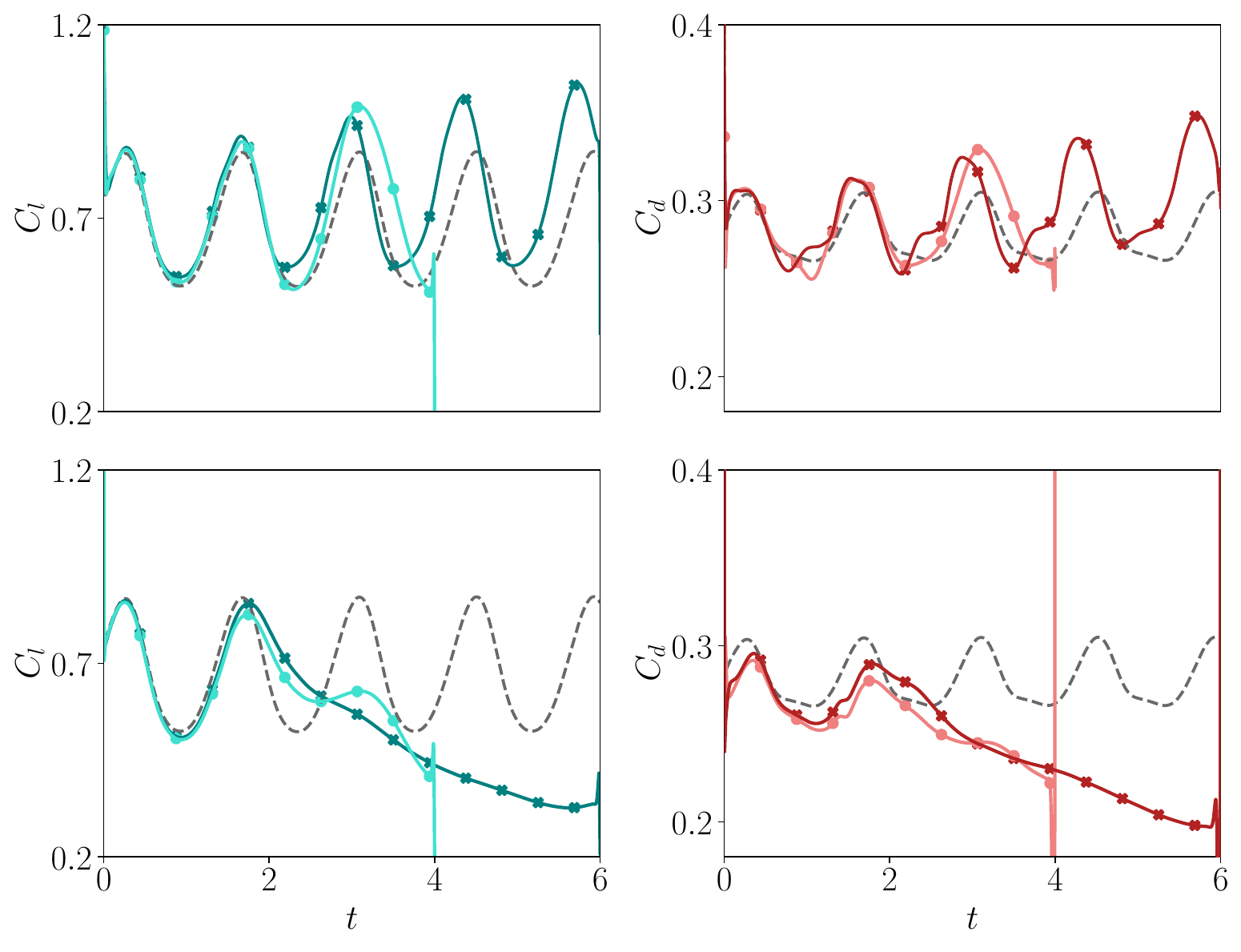}} 
    \caption{
    Analog of figure \ref{fig:TH} with brighter curves representing the aerodynamic coefficients for the case optimized over the shorter time window.
    }
    \label{fig:TH_4T}
\end{figure}
\begin{figure}
    \centering
    \includegraphics[width=0.95\textwidth]{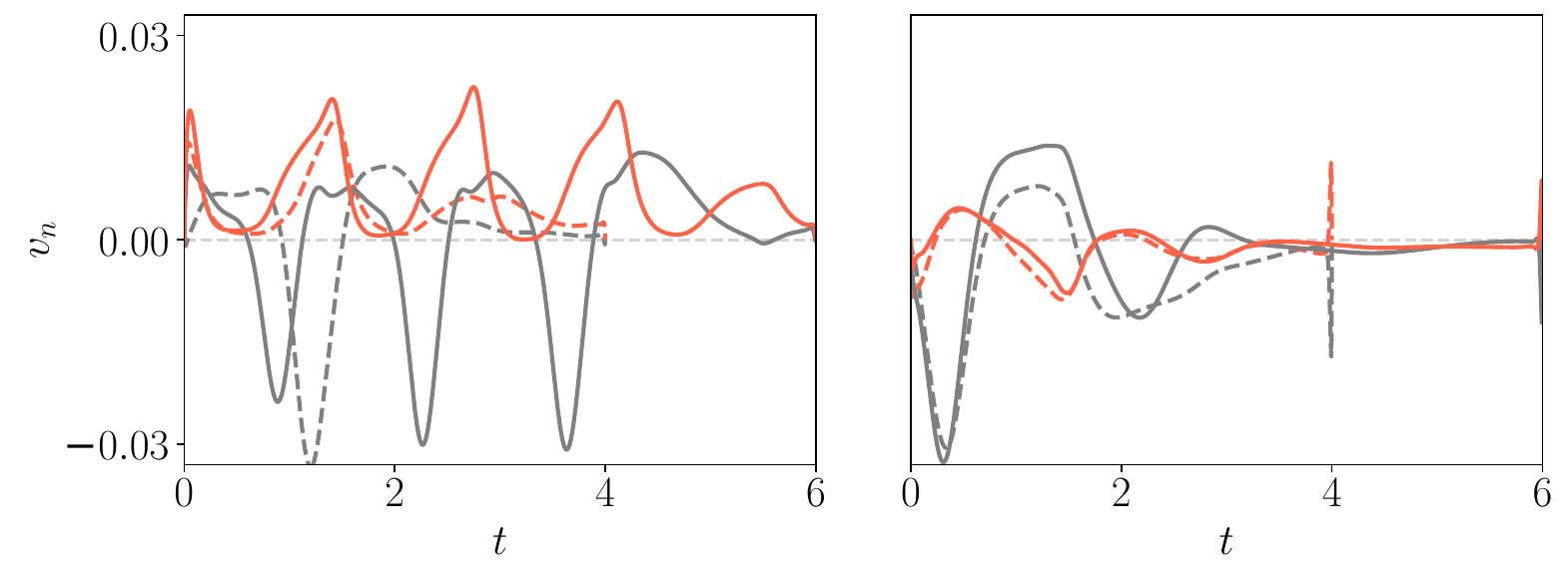} \hspace{0.03\textwidth}
    \caption{Influence of window length on the temporal variation of actuation; dashed curves correspond to actuation optimized over a shorter window.
    Left: $\mathcal{J}_{C_l}$, right: $\mathcal{J}_{C_d}$.
    }
    \label{fig:4TvN}
\end{figure}
Lastly, we compare the influence of the optimization window on the optimal actuation profile and the resulting aerodynamic performance. 
The length of the shorter window is $T=4$ which corresponds to less than 3 complete vortex-shedding cycles and is taken from our earlier work \citep{thompson2023optimal}. The lift and drag behavior for both time windows, along with the unactuated case for reference, is shown in figure \ref{fig:TH_4T}. For both costs, the frequency of the lift variation over the shorter window is almost unchanged as compared to the unactuated flow.
By contrast, for optimization over the longer window  the peaks shift to the left of the unactuated flow. 
In case of the shorter window, 
the lift minima of the vortex-shedding cycles are almost unchanged as compared to the unactuated flow whereas the lift minima increase in value for the solution optimized over the longer time window.
The peak lift however, increases with each cycle like in the case of the solution optimized over the longer window.

For the drag-based cost, the drag for the case optimized over the shorter window is lower over that time interval than for the solution optimized over the longer  window (aside form the large fluctuation at the end of the shorter window). 
The fluctuations in the aerodynamic coefficients are, however, more pronounced for the solution optimized over the shorter window.
Comparing the actuation profiles shown in figure \ref{fig:4TvN}, the lift-optimal actuation at $s_{vmax}$ is noticeably different. 
The starting transients of the adjoint solution, which in turn influence the gradients and consequently the actuation profile, force the actuation to have smaller magnitude near the end of the optimization window. 
This smaller actuation near the end of the window in turn affects the temporal behavior throughout the shorter optimization window. 
While the actuation at the very start of the window is positive for the case optimized over the longer window, the actuation for the case optimized over the shorter window is shifted such that the first positive peak occurs at a later time instance. 
The actuation near the trailing edge, however, seems to be similar to that for the case optimized over the longer window (except near the end, where the transients of the adjoint solution have a large influence). 

In the case of the drag-optimal actuation, the influence of the optimization window is less substantial. 
Apart from the end of the window where the actuation is forced to have small magnitude, the actuation profiles are qualitatively similar. 
The differences are possibly due to the differences in penalty weights, which yield slight differences in the targeted actuation magnitude. 
For both costs, the location of $s_{vmax}$ does not vary with the length of the optimization window.

\bibliography{References.bib}
\bibliographystyle{jfm}
\end{document}